\documentclass[usenatbib]{mnras}
\usepackage{graphicx}
\usepackage[space]{grffile}
\usepackage{latexsym}
\usepackage{textcomp}
\usepackage{longtable}
\usepackage{multirow,booktabs}
\usepackage{amsfonts,amsmath,amssymb}
\usepackage{natbib}
\usepackage{url}
\usepackage[utf8]{inputenc}
\usepackage[english]{babel}
\usepackage{times}
\usepackage{xspace}

\newcommand{\coone}{\mbox{\rm CO($1\text{--}0$)}} 
\newcommand{\cotwo}{\mbox{\rm CO($2\text{--}1$)}}

\newcommand{\numcloudhomogtotal}{5758\xspace}
\newcommand{\numcloudhomog}{4986\xspace}
\newcommand{\complim}{\ensuremath{4.7\! \times\! 10^{5}~M_\odot}}
\newcommand{\OSU}{Department of Astronomy, The Ohio State University, 140 West 18th Avenue, Columbus, OH 43210, USA}
\newcommand{\Alberta}{Department of Physics, University of Alberta, Edmonton, AB T6G 2E1, Canada}

\newcommand{\Caltech}{Infrared Processing and Analysis Center (IPAC), California Institute of Technology, Pasadena, CA 91125, USA}
\newcommand{\Carnegie}{Observatories of the Carnegie Institution for Science, 813 Santa Barbara Street, Pasadena, CA 91101, USA}

\newcommand{\CNRS}{CNRS, IRAP, 9 Av. du Colonel Roche, BP 44346, F-31028 Toulouse cedex 4, France}

\newcommand{\Heidelberg}{Astronomisches Rechen-Institut, Zentrum f\"{u}r Astronomie der Universit\"{a}t Heidelberg, M\"{o}nchhofstra\ss e 12-14, D-69120 Heidelberg, Germany}
\newcommand{\IRAM}{Institut de Radioastronomie Millim\'{e}trique (IRAM), 300 Rue de la Piscine, F-38406 Saint Martin d'H\`{e}res, France}
\newcommand{\ITA}{Universit\"{a}t Heidelberg, Zentrum f\"{u}r Astronomie, Institut f\"{u}r Theoretische Astrophysik, Albert-Ueberle-Str.\ 2, D-69120 Heidelberg, Germany}
\newcommand{\IWR}{Universit\"{a}t Heidelberg, Interdisziplin\"{a}res Zentrum f\"{u}r Wissenschaftliches Rechnen, Im Neuenheimer Feld 205, D-69120 Heidelberg, Germany}

\newcommand{\MPE}{Max-Planck-Institut f\"{u}r extraterrestrische Physik, Giessenbachstra{\ss}e 1, D-85748 Garching, Germany}
\newcommand{\MPIA}{Max-Planck-Institut f\"{u}r Astronomie, K\"{o}nigstuhl 17, D-69117, Heidelberg, Germany}

\newcommand{\OAN}{Observatorio Astron\'{o}mico Nacional (IGN), C/ Alfonso XII, 3, E-28014 Madrid, Spain}

\newcommand{\Toulouse}{Universit\'{e} de Toulouse, UPS-OMP, IRAP, F-31028 Toulouse cedex 4, France}
\newcommand{\UBonn}{Argelander-Institut f\"ur Astronomie, Universit\"at Bonn, Auf dem H\"ugel 71, 53121 Bonn, Germany}

\newcommand{\UGent}{Sterrenkundig Observatorium, Universiteit Gent, Krijgslaan 281 S9, B-9000 Gent, Belgium}

\newcommand{\UMass}{Department of Astronomy, University of Massachusetts Amherst, 710 North Pleasant Street, Amherst, MA 01003, USA}
\newcommand{\UWyoming}{Department of Physics and Astronomy, University of Wyoming, Laramie, WY 82071, USA}
\newcommand{\UToledo}{University of Toledo, 2801 W. Bancroft St., Mail Stop 111, Toledo, OH, 43606}
\newcommand{\JHU}{Department of Physics \& Astronomy, Bloomberg Center for Physics and Astronomy, Johns Hopkins University, 3400 N. Charles Street, Baltimore, MD 21218}

\hypersetup{draft}

\title[GMC Properties in PHANGS-ALMA]{Giant Molecular Cloud Catalogues for PHANGS-ALMA: Methods and Initial Results}

\author[Rosolowsky et al.]{Erik Rosolowsky$^{1}$,  Annie Hughes$^{2,3}$, Adam K. Leroy $^{4}$, Jiayi Sun$^{4}$, Miguel Querejeta$^{5}$,
\newauthor Andreas Schruba$^{6}$, Antonio Usero$^{5}$, Cinthya~N.~Herrera$^{7}$, Daizhong Liu$^{8}$, J\'er\^ome Pety$^{7}$,
\newauthor Toshiki Saito$^{8}$, Ivana Be\v{s}li\'c$^{9}$, Frank Bigiel$^{9}$, 
Guillermo Blanc$^{10}$,
M\'elanie Chevance$^{11}$,  
\newauthor 
Daniel A. Dale$^{12}$, 
Sinan Deger$^{13}$, 
Christopher M. Faesi$^{14}$, 
Simon C. O. Glover$^{15}$, 
\newauthor 
Jonathan D. Henshaw$^{8}$, 
Ralf S. Klessen$^{15,16}$, 
J. M. Diederik Kruijssen$^{11}$, Kirsten Larson$^{13}$, 
\newauthor Janice Lee$^{13}$, 
Sharon Meidt$^{17}$, Angus Mok$^{18}$, Eva Schinnerer$^{8}$, David A. Thilker$^{19}$, 
\newauthor 
Thomas G. Williams$^{8}$\\
$^{1}$\Alberta \\
$^{2}$\CNRS\\
$^{3}$\Toulouse\\
$^{4}$\OSU\\
$^{5}$\OAN\\
$^{6}$\MPE\\
$^{7}$\IRAM\\
$^{8}$\MPIA\\
$^{9}$\UBonn\\
$^{10}$\Carnegie\\
$^{11}$\Heidelberg\\
$^{12}$\UWyoming\\
$^{13}$\Caltech\\
$^{14}$\UMass\\
$^{15}$\ITA\\
$^{16}$\IWR\\
$^{17}$\UGent\\
$^{18}$\UToledo\\
$^{19}$\JHU
}

\begin{document}

\defcitealias{heyer_2009}{H09}
\defcitealias{cprops}{RL06}
\defcitealias{sun_18}{S18}
\defcitealias{srby87}{S87}

\date{\today}

\pagerange{\pageref{firstpage}--\pageref{lastpage}} \pubyear{2020}

\maketitle

\label{firstpage}

\begin{abstract}
We present improved methods for segmenting CO emission from galaxies into individual molecular clouds, providing an update to the \textsc{CPROPS} algorithms presented by \citet{cprops}. The new code enables both homogenization of the noise and spatial resolution among data, which allows for rigorous comparative analysis. The code also models the completeness of the data via false source injection and includes an updated segmentation approach to better deal with blended emission. These improved algorithms are implemented in a publicly available python package, \textsc{pycprops}. We apply these methods to ten of the nearest galaxies in the PHANGS-ALMA survey, cataloguing CO emission at a common 90~pc resolution and a matched noise level. We measure the properties of \numcloudhomog\ individual clouds identified in these targets. We investigate the scaling relations among cloud properties and the cloud mass distributions in each galaxy. The physical properties of clouds vary among galaxies, both as a function of galactocentric radius and as a function of dynamical environment. Overall, the clouds in our target galaxies are well-described by approximate energy equipartition, although clouds in stellar bars and galaxy centres show elevated line widths and virial parameters. The mass distribution of clouds in spiral arms has a typical mass scale that is $2.5 \times$ larger than interarm clouds and spiral arms clouds show slightly lower median virial parameters compared to interarm clouds (1.2 versus 1.4).
\end{abstract}%


\bibliographystyle{mnras}

\section{Introduction}
\label{sec:intro}

Star formation is one of the key processes by which galaxies evolve over time, building up their stellar mass and heavy elements. Feedback from star formation plays a major role in setting the structure of galaxy discs and returning material to the circumgalactic and intergalactic medium. In the local Universe, all star formation occurs in the molecular interstellar medium (ISM). Thus, the properties of the molecular ISM represent the immediate initial conditions for star formation and stellar feedback. Understanding how these properties change from galaxy to galaxy and within galaxies has been a longstanding goal of ISM studies, with direct implications for galaxy evolution theory.

Cataloguing {\it molecular clouds} is a well-established technique to describe changing conditions in the molecular ISM. In this approach, which originated in investigations of isolated dust extinction features \citep{heiles71}, gas in the molecular ISM is assigned to discrete structures. Then the macroscopic properties of each structure -- mass, line width, and radius -- are measured. The ensemble properties of clouds in a given galaxy or region capture the physical state of the molecular gas. Comparing the physical properties of different cloud populations reveals how conditions in the molecular ISM change within and among galaxies.  

Analyzing surveys of the Milky Way's Galactic plane in CO emission \citep{ss75, sss79, sss85, scoville87, srby87} established scaling relationships between the macroscopic properties of molecular clouds. Since these scalings resemble the correlations pointed out by \citet{larson}, they are frequently referred to as the ``Larson Laws.''  Within a limited range of galactic environments, these scaling relationships include a power-law relationship between cloud size, $R$, and spectral line width, $\sigma_v$, of the form $\sigma_v \propto R^\beta$, with $\beta\approx 0.5$. They also imply an approximate equipartition between the gravitational binding energy ($U_g$) and the kinetic energy ($K$) of the clouds, frequently phrased in terms of molecular clouds being virialized (i.e., where $2K=|U_g|$ and all other terms in the virial theorem being negligible).

In this study, we focus on observations of extragalactic {\it giant} molecular clouds (GMCs), because of the poor sensitivity and resolution of extragalactic observations compared to Milky Way clouds. GMCs are usually defined as molecular clouds with masses $M>10^5~M_\odot$ and radii $R>20~\mathrm{pc}$, which is an approximate minimum mass associated with O-star formation \citep{psp3}, and significant self-gravitation \citep{hc01}. Although early studies drew a sharp distinction between dark clouds and GMCs \citep[e.g.,][]{penzias75}, subsequent work showed that molecular cloud populations exhibit a continuous distribution in mass from small ($10^2~M_\odot$) to large clouds \citep[$>10^7~M_\odot$; e.g.,][]{srby87,hc01,rice2016, miville17, colombo19}.

Following the detection of extragalactic CO emission \citep{sol75}, molecular cloud populations have also been catalogued in other galaxies. Single dish telescopes can measure CO emission at the typical mass and size scales of individual GMCs in the Magellanic Clouds \citep{cohen88,rubio93,nanten-fukui,nanten,hughes-2010,magma} and the other two spiral galaxies in the Local Group \citep[M31 and M33;][A.~Schruba et al., in preparation]{iram-m31-aa,braine12, braine18}. Early millimetre interferometers achieved similar resolution and sensitivity in more distant Local Group galaxies and the nearest other galaxy groups \citep[e.g.,][]{vbb87,ws90,wr91,wilson_6822}. The first large surveys of Local Group galaxies concluded that their GMC populations exhibited similar scaling relationships to one another and the Milky Way \citep{nanten, eprb03, psp5, bolatto08, fukui-kawamura}.

The Local Group contains a limited range of environments. Once interferometric CO surveys were conducted in more extreme systems, it was discovered that molecular cloud populations in high density regions show markedly different properties than those in the Milky Way disc. Studies of the Galactic centre \citep{gmcs-galcen, shetty12, henshaw16} and molecule-rich regions of external galaxies \citep{gma-wilson,keto-m82,m64-gmcs,wei12} revealed clouds with line widths higher than expected from the Milky Way size--line width relationship. Clouds in these environments also showed high surface densities, such that they often still appeared to exhibit approximate virialization.

Continuing improvements in instrumentation allowed surveys of molecular clouds in more distant systems and with higher sensitivity and completeness \citep{rebolledo12,canon-gmcs,paws,rebolledo15}. With high completeness and careful homogenization among data sets \citep{hughes13}, variations in molecular cloud populations within galaxies also became clear. For example, the arms, interarm regions, and central disc of M51 show significantly different cloud populations \citep{hughes13,colombo14}. Meanwhile, \citet[][hereafter \citetalias{heyer_2009}]{heyer_2009} pointed out that variations of the size--line width relationship were also seen within the cloud population of the Milky Way disc.  

Along with observations that the mass distribution of GMCs varies between galaxies and with galactocentric radius \citep{mspec,gratier-m33}, these studies established that the properties of GMC populations vary with environment. Most of these studies found virialized GMCs to be a more universal result than the size--line width relation. Following the theoretical ideas of \citet{eg89-pressure} and more recently \citet{field11}, several observational works also highlighted that external pressure from a diffuse gas layer or an external gravitational potential may play an important role in regulating the properties of clouds in the Milky Way and other galaxies \citep{m64-gmcs,hughes-2010,hughes13,sun_18,schruba19}. The effect of external pressure is to raise line widths above the expectation for a self-gravitating cloud in energy balance.  Such clouds would still be considered ``bound'' once the set of external forces (gravitational potential, diffuse gas pressure) acting on the molecular gas is considered \citep{dale19, kruijssen19orb, sun_20}.

To build our knowledge of the molecular ISM in the local Universe beyond isolated case studies, the next step is to survey the CO emission and catalogue the cloud populations across a large sample of galaxies that is representative of star-forming galaxies in the local Universe. The Atacama Large Millimeter/submillimeter Array (ALMA) makes this possible by reducing the time required to survey the CO emission at cloud-scale resolution across the disc of single nearby galaxy to ${\sim}1$ hour each \citep[e.g.,][]{alma-antennae, leroy15,pan-m100,faesi-ngc300,hirota-m83,kruijssen19-ngc300,gmc-ngc1300}. ALMA also makes it possible to extend these studies beyond star-forming, disc galaxies to early-type galaxies \citep[][T.~Williams et al., in preparation; M.~Chevance et al., in preparation]{utomo15,wisdom-gmcs} and ultraluminous infrared galaxies \citep[][T.~Saito et al., in preparation]{ulirg-molgas}.

The Physics at High Angular resolution in Nearby Galaxies (PHANGS) project is a multi-facility campaign to observe the tracers of the star formation process at the scale of molecular clouds across different galactic environments.  The PHANGS-ALMA survey leverages the capabilities of ALMA to take this next step in extragalactic GMC studies. PHANGS-ALMA is a survey of the $^{12}$\cotwo\ emission from 74 nearby star-forming galaxies \citep{PHANGSSurvey}. The survey targets are massive ($10^{9} <M_\star/M_\odot < 10^{11}$), moderately inclined ($i<70^\circ$), nearby ($d < 17~\mathrm{Mpc}$), star-forming galaxies where an angular resolution of ${\sim}1''$ achieves a linear scale of $\lesssim 150$~pc. This allows the identification and characterization of high-mass GMCs across a large, statistically representative galaxy sample for the first time. 

This work presents the methods, software, and first results for the GMC catalogues constructed from the PHANGS-ALMA data. We adopt an approach built around careful data homogenization, which \citet{hughes13} demonstrated to be essential to comparative analysis of molecular cloud properties. Our input data and the homogenization procedure are described in Section~\ref{sec:data}.  In Section~\ref{sec:methods}, we present \textsc{pycprops}, a python implementation of the GMC cataloguing algorithm \textsc{cprops} \citep[][hereafter \citetalias{cprops}]{cprops}. We apply these methods to ten galaxies that are close enough to analyse the CO emission at a common linear resolution of $90$~pc. In Section~\ref{sec:scaling} we present the ``Larson Law'' GMC scaling relationships in these galaxies and in Section~\ref{sec:massdist} we present the mass distributions. Finally, Section~\ref{sec:variations} explores how physical properties of the GMC populations change with galactocentric radius and dynamical environment. A~companion paper (A.~Hughes et al., in preparation) will present the GMC catalogues for the full PHANGS-ALMA sample. 

Finally, we refer the reader to a parallel line of investigation that has measured the distributions of molecular gas surface density and line width at fixed spatial resolution for the PHANGS-ALMA galaxies \citep[][hereafter \citetalias{sun_18}]{sun_18} and \citet{sun_20b}. \citetalias{sun_18} measured the surface density and line width in $\sim 30\,000$ apertures at fixed size scales of 45, 60, 90, and 120~pc in 15 nearby galaxies. \citet{sun_20b} expands this analysis to $\sim 10^5$ apertures at 150~pc scales in $70$ PHANGS-ALMA targets. Both works found a wide range of internal conditions in the molecular ISM. While the gas was typically found near energy equipartition through a vast range of galactic environments, the internal pressure in the molecular gas ranged over $>6$ orders of magnitude. They found smaller, but still significant variations in surface density and line width. The ``GMC''-centred view of the molecular ISM adopted here and the non-parametric approach in \citetalias{sun_18} and \citet{sun_20b} are complementary \citep{leroy_16} and largely consistent, provided controlled comparisons are drawn.  We make frequent comparisons to their results throughout the text.

\section{Data}
\label{sec:data}

\begin{table*}
\begin{tabular}{l r r r l r r r}
\hline
Galaxy &  Distance & Incl. & Pos. Angle & Morph.  & $R_e$ & $M_\star$ & $M_\mathrm{mol}$ \\
(NGC~)  &  (Mpc)    & (${}^\circ$) & (${}^\circ$) & Type  & (kpc) & ($10^{9}~M_\odot$) & ($10^9~M_{\odot}$) \\ 
\hline
0628 &  9.8 &   9 &   21 & 	SA(s)c & 3.6 & 18.3 & 1.74 \\
1637 & 11.7 &  31 &   21 & SAB(rs)c & 2.6 & 7.7 & 0.67 \\
2903 & 10.0 &  67 &  204 & SAB(rs)bc & 4.1 & 28.9 & 3.81 \\
3521 & 13.2 &  69 &  343 & SAB(rs)bc & 3.7 & 66.3 & 6.61 \\
3621 &  7.1 &  66 &  344 & SA(s)d & 2.6 & 9.2 & 1.08 \\
3627 & 11.3 &  57 &  173 & SAB(s)b & 3.9 & 53.1 & 6.03 \\
4826 &  4.4 &  59 &  294 & (R)SA(rs)ab & 1.6 & 16.0 & 0.30 \\
5068 &  5.2 &  36 &  342 & SAB(rs)cd & 2.0 & 2.2 & 0.24 \\
5643 & 12.7 &  30 &  319 & SAB(rs)c & 3.8 & 18.2 & 2.26 \\
6300 & 11.6 &  50 &  105 & SB(rs)b & 3.6 & 29.2 & 1.59 \\
\hline                
\end{tabular}
\medskip

\caption{\label{tab:sample} Properties of galaxies presented in this work.  Distances are from the meta-analysis by \citet{anand20}.  Orientation parameters (inclination and position angle) are from \citet{lang2020}. Morphological type is from the NASA Extragalactic Database homogenized morphologies, which are based on \citet{rc3}. Stellar masses and $R_e$ are derived from the data in \citet{z0mgs}. The molecular mass, $M_\mathrm{mol}$, is calculated using a variable CO-to-H$_2$ conversion factor and refers to the mass in the ALMA field of view rather than an estimate for the entire galaxy.}
\end{table*}

We selected 10 galaxies from PHANGS-ALMA for our GMC cataloguing procedures. We list these targets and summarize their properties in Table~\ref{tab:sample}. We selected targets that were close enough to access a common resolution of $90$~pc across the subsample. For context, the coarsest linear resolution for the entire PHANGS-ALMA sample is ${\sim}150$~pc, and about half of the targets have linear resolution $<90$~pc. To determine these linear resolutions, we use the distances compiled by \citet{anand20} and reported in Table~\ref{tab:data}.  Resolution was our only strict selection criterion. Our selection includes galaxies with a range of inclination, mass, and molecular gas morphology as diverse test cases for the GMC cataloging methodology. The GMC catalogues for all PHANGS-ALMA galaxies will appear in A.~Hughes et al. (in preparation).

\begin{table*}
\begin{tabular}{@{\extracolsep{4pt}}l r c c r r r r@{}}
\hline
&  \multicolumn{5}{|c|}{Native} & \multicolumn{2}{|c|}{90 pc}\\
\cline{2-6} \cline{7-8}\\
Galaxy & Noise & Resolution & Linear Res. & $f_\mathrm{mask}$ & $f_\mathrm{cat}$   &  $f_\mathrm{mask}$ & $f_\mathrm{cat}$\\
(NGC~)   & (mK) &  (arcsec) & (pc) &  & \\ 
\hline
0628$^{\rm m}$ &  113 & 1.12 &  53 & 0.48 & 0.66 & 0.58 & 0.78 \\
1637 &   36 & 1.39 &  79 & 0.78 & 0.92 & 0.59 & 0.76 \\
2903$^{\rm m}$ &   70 & 1.46 &  71 & 0.77 & 0.91 & 0.73 & 0.87 \\
3521$^{\rm m}$ &   62 & 1.28 &  82 & 0.82 & 0.91 & 0.78 & 0.89 \\
3621$^{\rm m}$ &   39 & 1.82 &  62 & 0.83 & 0.94 & 0.68 & 0.85 \\
3627$^{\rm m}$ &   79 & 1.62 &  89 & 0.80 & 0.90 & 0.79 & 0.90 \\
4826 &   77 & 1.26 &  27 & 0.88 & 0.95 & 0.86 & 0.94 \\
5068$^{\rm m}$ &  208 & 1.04 &  26 & 0.30 & 0.44 & 0.33 & 0.50 \\
5643 &   76 & 1.30 &  80 & 0.69 & 0.82 & 0.66 & 0.81 \\
6300 &  114 & 1.08 &  60 & 0.66 & 0.80 & 0.69 & 0.85 \\
\hline                
\end{tabular}
\medskip
\caption{\label{tab:data} Properties of the PHANGS-ALMA data that we analyse in this work. We report the noise and resolution levels of the native data. Targets marked with $^{\rm m}$ were observed in multiple separate parts and linearly mosaicked together. The parts have matched resolution but initially distinct noise properties. The scalar value in the table is average noise value across these parts.
For the homogenized data, the resolution is $90$~pc and the noise level is $75$~mK.  We also report the fraction of flux in the signal-identification mask ($f_\mathrm{mask}$) and in the catalogued GMCs including clipping corrections ($f_\mathrm{cat}$), which are calculated from the homogenized data.  Both fractions are measured with respect to the total flux in the cube. We find $f_\mathrm{cat}>f_\mathrm{mask}$ because of the extrapolations applied to the measured cloud properties (Section \ref{sec:extrap}).}
\end{table*}

\subsection{Original Data}
\label{sec:origdata}

We began with $^{12}$\cotwo\ data cubes from PHANGS-ALMA internal data release version 3.4, which is nearly identical to the PHANGS public data release described in \citet{PHANGSSurvey}. These cubes combine data from ALMA's 12-m, 7-m, and total power telescopes and so should be sensitive to emission from all spatial scales. The observations and main steps of data reduction are described in \citet{PHANGSPipeline} and briefly summarized here. 

We observed each target using one or more large, multi-field mosaics. We jointly imaged the 12-m and 7-m data, using channels near, but not exactly equal to, $2.5$~km~s$^{-1}$ in width. The small variations in channel width occur since the data are made from fixed channels in topocentric frequency and the widths vary with the galaxy's recession velocity. We deconvolved the emission using a multi-scale clean followed by a deep single-scale clean. After imaging, we applied a primary beam correction to the data and convolved the cubes to have a round synthesized beam. Then we linearly mosaicked our galaxy targets that were observed in multiple parts, which are marked in Table~\ref{tab:data}. During this linear mosaicking procedure, we matched the resolution of the individual parts of the mosaic to the coarsest common beam.  The noise in different parts that make up the mosaics can vary by up to 30\% after convolution to a common beam and there can be a 20\% variation in the noise across the spectral bandpass.

In parallel, we reduced the total power data following the procedures described in \citet{herrera20}. We combined the total power data with the interferometric imaging using feathering\footnote{Feathering creates a final image by forming the weighted combination of the interferometric and total power data in the Fourier plane \citep[e.g.,][]{feather-vogel,lincom}.}. Finally, we convert the units of the cubes to Kelvin.

\subsection{Homogenized Data}
\label{sec:homogenized}

Table~\ref{tab:data} reports the native physical resolution and the root mean square (RMS) noise in Kelvin in each data cube at that resolution. The original data show a factor of $>2$ range in noise and physical resolution. The noise also varies within individual mosaics with the aforementioned $<30\%$ change in different mosaic blocks and $<20\%$ change across the bandpass.

Heterogeneous noise and resolution present a major problem for our analysis. Our signal identification algorithm is based on identifying significant emission, with significance defined relative to the local noise level. The {\sc cprops} decomposition algorithm that we use for identifying individual GMCs also divides structures up based on the significance of features relative to the local noise level. For algorithms like {\sc cprops} the physical resolution also strongly affects the size of derived structures \citep{pineda_09,hughes13,leroy_16}. As shown by \citet{hughes13}, homogenizing the resolution and noise levels in the data is essential to a robust comparative analysis.

To enable a fair comparison, we smooth all data to share a common resolution of $90$~pc and common surface brightness sensitivity of $75$~mK per ${\sim}2.5~\mathrm{km~s}^{-1}$ channel. This process also removes variations of the noise coming from different parts of multi-part mosaics.

To create a uniform noise level for all data cubes in the sample, we first empirically estimate the three dimensional noise distribution in each 90~pc resolution data cube. To do this, we follow the procedure described in \citep{PHANGSPipeline} for the PHANGS-ALMA data.  Briefly, the noise is estimated from the distribution of signal-free data assuming the spatial and spectral variations are independent and smooth.  The estimates of the noise scale are built up iteratively, while also refining the definition of the signal-free region. This creates an estimate $\sigma_{T,0}(x,y,v)$, of the noise amplitude in the original data (subscript of~0) as a function of position and velocity.

Next, we homogenize the noise to the target level of $\sigma_T = 75$~mK across each data cube. To do this, we generate a cube of random deviates drawn from a standard normal distribution, $N(x,y,v)$.  The cube has the same astrometric grid as the original data and the same spatial correlations as are expected for a 90~pc beam. Then, we scale the cube of deviates by a spatially and spectrally variable factor so that, when added to the original data, a uniform noise level of $\sigma_T=75$~mK is achieved.  The scaled cube of deviates has the form:
\begin{equation}
    N^*(x,y,v) = N(x,y,v)\sqrt{\sigma_T^2 - \sigma_{T,0}^2(x,y,v)}~.
\end{equation}
The effect of this scaling is to create a cube of noise with the opposite trends as are present in the original data; there is a high noise level in $N^*$ where the noise in the original cube is relatively low.  This common value of $\sigma_T = 75$~mK represents the limiting noise across our whole data set after convolving to 90~pc resolution. Note that the noise values quoted in Table~\ref{tab:data} refer to the data at their original resolution. For the closer galaxies, the noise will be reduced by the convolution to a common size scale, and the homogenization process will then add noise to achieve the common value.

\subsection{Environmental Masks}
\label{sec:envmasks}
A key variable for analysis in our work is the study of GMC properties as a function of galactic environment.  To enable this analysis, we use the PHANGS ``environment'' masks by M.~Querejeta et al. (in preparation) to divide the GMCs into distinct categories. The environment mask creation leverages the decompositions of S$^4$G \citep{s4g} carried out by \citet{herrera-endoqui15} and \citet{salo-s4ggalfit} with spiral arms identified on a log-spiral fit to near infrared data and checked by eye.

The environment masks provide a wealth of information about the different environments within a galaxy, and we define five sub-populations of GMCs based on these masks. 
\begin{itemize}
    \item {\bf Bar} -- Six of our 10 galaxies have stellar bars (NGC~1637, NGC~2903, NGC~3627, NGC~5068, NGC~5643, NGC~6300).  We include all clouds on lines of sight that project onto the bar in the ``Bar'' population.  
    \item {\bf Disc} -- For the six galaxies {\it with bars}, we define the disc clouds as those clouds at galactocentric radii larger than the maximum radial extent of the bar. Lines of sight that are at radii smaller than the extent of the bar but not projected on the bar are excluded.
    \item {\bf Spiral Arm} -- These clouds are located within a spiral arm of the galaxy with well defined arms (NGC~0628, NGC~1637, NGC~3627, NGC~5643).  In barred galaxies with spiral arms, we only include GMCs associated with spiral arms at galactocentric radii beyond the extent of the bar. GMCs at bar ends are not in the spiral arm population but are in the Bar population.
    \item {\bf Interarm} -- For the four galaxies with defined spiral arms, this population includes clouds not associated with the spiral arms.  In barred galaxies, we restrict this population to be at radii larger than the bar extent.
    \item {\bf Centre} -- We define a population ``Centre'' clouds that include all clouds that are in centre of the galaxy where the central region is distinct from the stellar disc.  This population includes all clouds that are associated with a stellar bar, as well as clouds in a compact nuclear feature or a stellar bulge as per \citet{sun_20}.
\end{itemize}
These sub-populations are designed to highlight specific contrasts and are not designed to contain all the GMCs in any comparison. The Bar and Disc clouds are mutually exclusive as are the Spiral Arm and Interarm clouds. For galaxies with both spiral arms and bars, the Spiral Arm and Interarm clouds comprise the Disc clouds for that galaxy.  GMCs from galaxies without arms or bars (NGC~3521, NGC~3621, NGC~4826) are not included in any of these sub-populations of GMCs.

For Centre clouds, we expect the ISM to respond to the non-disc-like stellar potential and exhibit different internal conditions.  We only use this Centre identification in our discussion of the scaling relationships between cloud properties (Section~\ref{sec:scaling}) to highlight which clouds are in these environments.

\section{GMC Identification and Characterization}
\label{sec:methods}

We identify and characterize molecular clouds in the CO data cubes using \textsc{pycprops}\footnote{\label{footnote:pycprops}\url{https://github.com/phangsteam/pycprops/}}, a \textsc{python} implementation of the {\sc cprops} algorithm originally described in \citetalias{cprops}.  The shift to \textsc{python} increases the cataloguing speed and moves the algorithm outside the proprietary \textsc{IDL} software environment. Compared to the original \textsc{cprops} implementation, \textsc{pycprops} also implements data homogenization (Section~\ref{sec:homogenized}) and assessment of completeness (Section~\ref{sec:completeness}) as core parts of the analysis. These additions make the revised version better suited for rigorous comparative analysis.

Following \citetalias{cprops}, we approach molecular cloud cataloging in two distinct phases.  First, we identify significant emission in the data cube and assign each pixel containing significant emission to  a distinct molecular cloud. Then we characterize the emission associated with each cloud to determine the properties of that cloud.

\subsection{Signal Identification}
\label{sec:segment}

We begin with a spectral line data cube that contains measurements of brightness temperature at each position-position-velocity pixel, $T(x,y,v)$. At this stage we construct a new three dimensional estimate of the RMS brightness temperature noise, $\sigma_T(x,y,v)$, following the same procedure as used in Section~\ref{sec:homogenized} \citep[described in ][]{PHANGSPipeline}. In practice, the process of homogenizing the data has already removed nearly all the spectral and spatial variation, so the cubes are well described by a single noise value at this stage.

From $T(x,y,v)$ and a local noise estimate, $\sigma_T(x,y,v)$, we construct a Boolean mask, $\mathcal{M}(x,y,v)$, that indicates which pixels are likely to contain significant emission. Following \citetalias{cprops}, we construct two masks: a high significance mask and a low significance mask. The high significance mask contains regions with two adjacent spectral channels with $T>4\sigma_T$. The low significance mask contains regions with two adjacent channels with $T>2\sigma_T$. We then reject all contiguous regions in the low significance mask that do not contain any pixels in the high significance mask. This pruned low significance map represents the final Boolean emission mask $\mathcal{M}(x,y,v)$ indicating pixels that are likely real emission. Cloud identification and decomposition is restricted to lie within this emission mask. 

We choose these specific thresholds since they avoid false positives, which we assess by applying the same algorithm to the data cube scaled by a factor of $-1$. Inverting the cube assesses whether the negative noise fluctuations in the data cube would be detected using the masking algorithm, and we find $\le 2$ detections of negative noise deviates in each of the cubes. Alternative masking thresholds are feasible, such as 2 adjacent channels with $T>5\sigma_T$, but we found our chosen combination finds most of the bright, visible emission in the cubes.  We test the impact of this choice through false source injection (Section~\ref{sec:completeness}).

\subsection{Cloud Decomposition Algorithm}
\label{sec:algorithm}
In {\sc pycprops}, each cloud is associated with a local maximum of emission in the spectral line data cube. The algorithm first identifies those local maxima. Then it checks whether those maxima are significant with respect to fluctuations that would be expected from noise in the data and whether they are distinct from neighbouring local maxima. Finally, it assigns emission to each local maximum.  Here, we describe the parameters and implementation details of the general algorithm and then describe the application to the PHANGS-ALMA data including parameter choices in Section~\ref{sec:phangsdecomp}. 

To identify local maxima, we take advantage of the fast dendrogram algorithm provided by the \textsc{astrodendro} package. This algorithm efficiently constructs a dendrogram representation \citep{dendrograms} of all emission in the mask. In the process, it identifies all local maxima in the data and also measures the intensity contour levels at which each maximum ``merges'' with each other maximum. The efficient implementation of \textsc{astrodendro} offers a major performance improvement compared to the original \textsc{cprops} implementation.  From the local maxima identified by \textsc{astrodendro}, we select only those likely to be significant. Following the original algorithm, \textsc{pycprops} implements features to reject local maxima based on four criteria: significance, number of associated pixels, separation, and uniqueness.

First, we require that the intensity at local maximum, $T_\mathrm{max}$, must be an interval $\delta$ above the highest contour of emission containing at least one other neighbour.  We define the {\it merge level}, $T_\mathrm{merge}$, as the highest value contour containing a given local maximum and one other neighbour. We then compare the brightness difference  between the local maximum and the merge level to the parameter $\delta$, requiring $T_\mathrm{max} - T_\mathrm{merge} > \delta$.  This criterion ensures that the local maximum is significant with respect to the local background. By setting $\delta$ to a multiple of the local noise, $\sigma_T$, we reject local maxima that are noise fluctuations.  As laid out in \citet{dendrograms}, the default value of $\delta = 2\sigma_T$ is a good compromise between sensitivity to cloud structure and robustness against noise.

Second, we require a minimum number of cube pixels (sometimes called ``voxels''), $N$, to be uniquely associated with the local maximum. For this test, we count only pixels contained in the isosurface above the merge level containing only that local maximum. By setting $N$ to some multiple of the beam size in pixels, we reject small, poorly defined objects.

Third, we require that local maxima are separated from each other by a minimum distance of $d_\mathrm{min}$ spatially and $v_{\min}$ spectrally.  Related to the previous point, this requirement ensures that maxima are reasonably resolved from one another. In the case of two maxima not well-separated from one another, we discard the fainter maximum from the list. 

Finally, we test whether local maxima are unique if their properties change significantly when merged with another object.  For a merging pair of leaves in the dendrogram, we consider both objects to be unique if the measured properties of each object changes by a factor of $>s$  when combined with one another. This is the {\tt SIGDISCONT} parameter in the original {\sc cprops} code.  This parameter can evaluate changes in flux and in size, where size is defined as the the second moment of the emission distribution along the coordinate axes.  In the case that a merger is not unique, we discard the fainter maximum from the list. 

After identifying a set of unique, significant local maxima, we then use these as seeds to assign emission to molecular clouds. To do this, we use a watershed algorithm that associates all pixels in the mask with a local maximum.  Some pixels are already uniquely associated with a single local maximum in the dendrogram. The remainder of the emission lies at an intensity level where it could be associated with multiple local maxima. The watershed algorithm assigns these contested pixels by growing the regions associated with the local maxima until all pixels are assigned to one of the regions.  In this way, pixels with ambiguous assignments are assigned to the closest unique object in position-position-velocity space.

This approach follows the ``seeded'' version of \textsc{clumpfind} \citep{clumpfind} adopted by \citet{m64-gmcs}. It is implemented as one of the options in the original version of \textsc{cprops}, but is not the default decomposition algorithm. The default approach in \citetalias{cprops} is to only catalogue emission uniquely associated with clouds. This often leaves a large amount of emission unassigned in the watershed \citep[e.g., see][]{colombo14}, particularly in crowded fields. That approach relies heavily on the extrapolation or clipping corrections described in \citetalias{cprops} and Section~\ref{sec:properties}. The approach adopted here is significantly less reliant on these clipping corrections, although we still apply them.

In detail, {\sc pycprops} uses the seeded watershed algorithm in \textsc{scikit-image} \citep{scikit-image}. This algorithm includes the compactness parameter introduced by \citet{watershed-compactness}. We adopt a value of $1000$, which leads the algorithm to seek the most compact possible structures. This leads to more natural boundaries between regions compared to the original {\sc clumpfind}. Note that when applying this algorithm, it is possible, though rare, for clouds to span disconnected regions of the signal identification mask (Section~\ref{sec:segment}). We visually inspected these occurrences, finding that there is often evidence for real emission below the mask sensitivity limit that connects the two apparently disjoint regions. We therefore retain these identifications as genuine clouds.  

The watershed algorithm assigns each pixel in the emission mask ($\mathcal{M}$) to a molecular cloud, generating a label cube $\mathcal{L}(x,y,v)$.  The label map has an integer value so that all pixels in the $j$th cloud are labelled with a value of $j$ in $\mathcal{L}$.  All pixels in $\mathcal{L}$ outside the emission mask (i.e., $\mathcal{M}=0$) are also set to zero.

\subsection{Cloud Decomposition in PHANGS-ALMA}
\label{sec:phangsdecomp}
 
\begin{figure}
    \centering
    \includegraphics[width=\columnwidth]{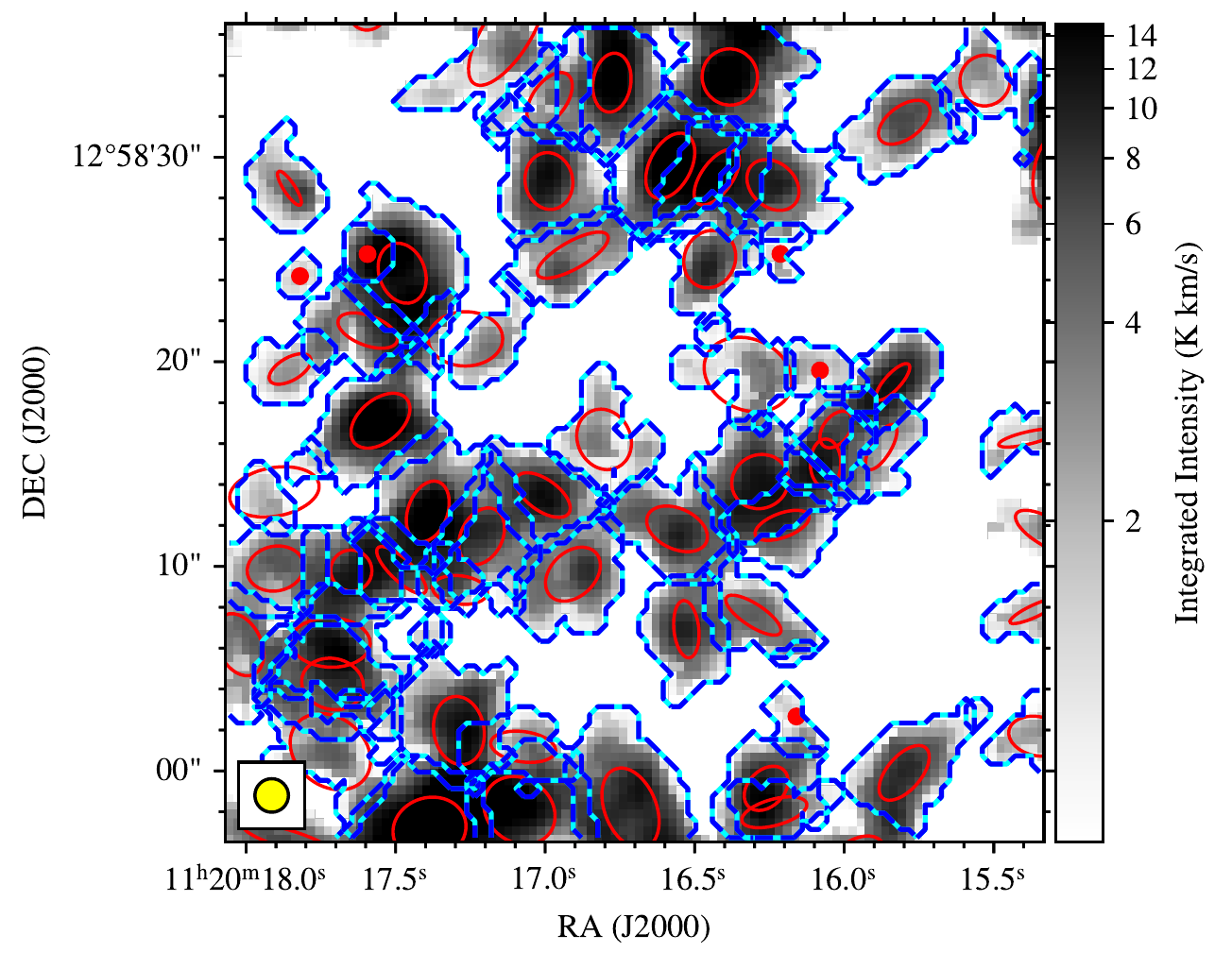}
    \caption{Cloud decomposition of a region in NGC~3627.  The greyscale background shows the masked \cotwo\ integrated intensity image by PHANGS-ALMA.  The blue dashed contours show the projected boundaries of the molecular clouds.  The red ellipses are elliptical approximations to the emission distribution.  The ellipses are centred on the emission centroid of the cloud and show the size and orientations of the molecular clouds as the deconvolved major and minor axes (FWHM; see Section~\ref{sec:derived} for details).  Clouds that cannot be deconvolved are shown as filled circles.  The circle in the bottom left-hand corner illustrates the beam size.}
    \label{fig:zoomfig}
\end{figure}
 
For the decomposition of the PHANGS-ALMA data, we only use the contrast and minimum volume criteria to select local maxima. Because our 90~pc beam size is comparable to the $\lesssim 100$~pc size of GMCs, we expect to select compact, almost beam-sized structures that may be crowded together. Therefore we impose no minimum separation, setting $d_\mathrm{min}=0$ and $v_\mathrm{min}=0$. We also set $s = 0$, and so do not merge peaks based on lack of uniqueness. These choices reflect a prior expectation on GMC size and that all emission in the mask can be decomposed into GMCs.  We note the mask does not necessarily contain all the emission in the data cube and this low lying emission will not be incorporated into the cloud catalogues.

We require all maxima to have $\delta = 0.15~\mathrm{K}$, or $2\sigma_T$, contrast against the local merge level. We also require $N>0.25\Omega_\mathrm{bm}/\Omega_\mathrm{pix}$ cube pixels uniquely associated with each local maximum, meaning those pixels above the contour level at which the object merges with another object. Here $\Omega_\mathrm{bm}$ and $\Omega_\mathrm{pix}$ are the solid angles of the resolution element and the pixel, respectively.  Since the pixellization of interferometer images is arbitrary, we have linked the decomposition parameter to the resolution of the image. While $N$ is smaller than a resolution element, the number of pixels in the resulting GMCs is much larger after the watershed algorithm is applied to the data.   This criterion ensures that each cloud in crowded, high signal-to-noise regions has a small neighbourhood around the local maximum that can be a stable seed for the decomposition.

\begin{figure}
    \centering
    \includegraphics[width=\columnwidth]{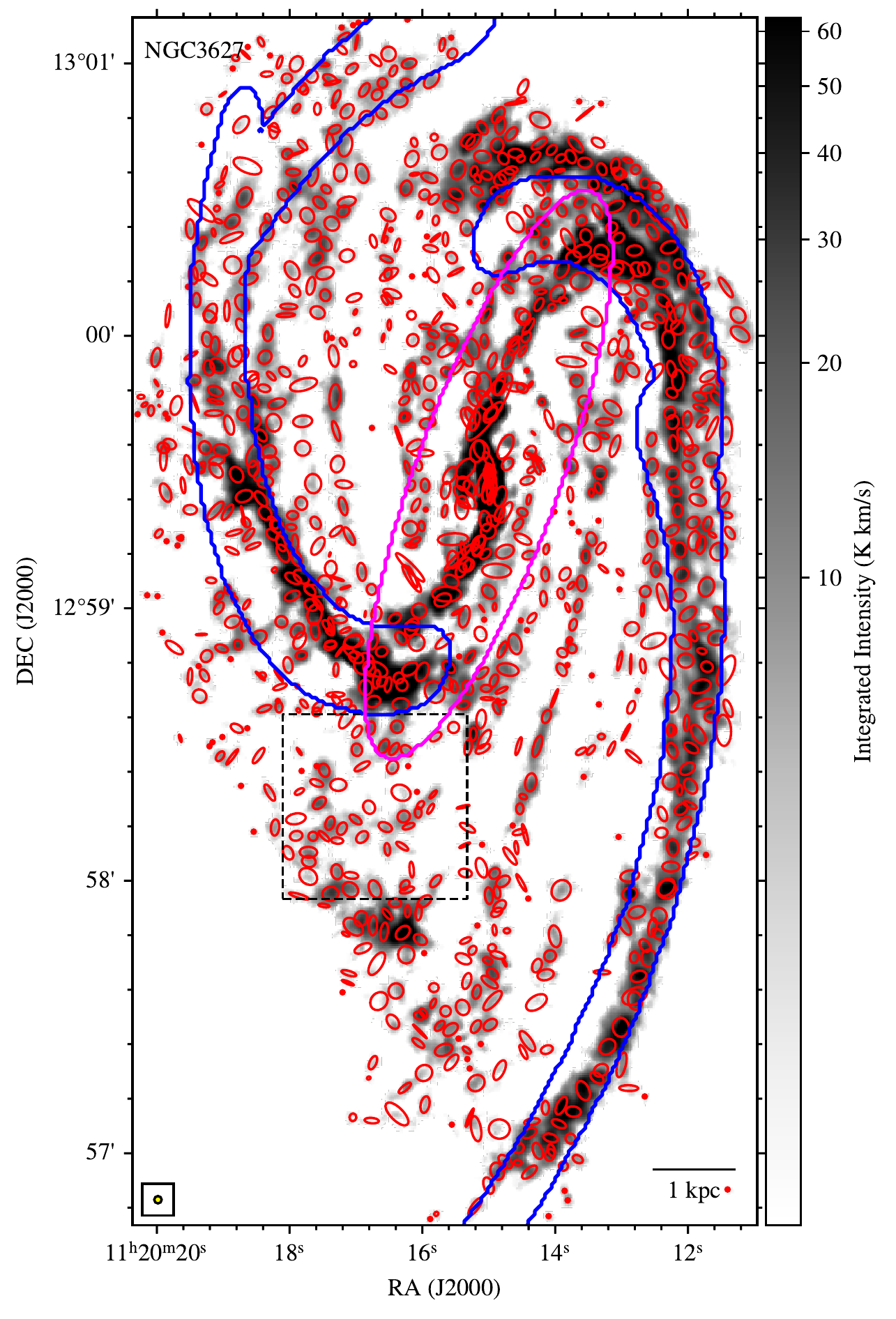}
    \caption{Integrated intensity map (``moment 0'') of $^{12}$\cotwo\ emission for NGC~3627 with locations of GMCs overlaid as red ellipses on an arcsinh color stretch.  The ellipses indicate the locations, the deconvolved major and minor axes (FWHM), and the position angles of the clouds as in Figure~\ref{fig:zoomfig}.  Unresolved clouds are shown as filled circles.  The beam size is indicated as the yellow circle in the lower-left. The magenta ellipse highlights the region identified as belonging to a bar and the solid blue contour indicates regions associated with spiral arms (Section~\ref{sec:envmasks}). The dashed black box indicates the region shown in Figure~\ref{fig:zoomfig}.  Similar figures for other targets are available online.}
    \label{fig:ngc3627atlas}
\end{figure}

Figure~\ref{fig:zoomfig} shows the results of this decomposition approach applied to a region in the PHANGS-ALMA data for NGC~3627. The projected boundaries of catalogued clouds are illustrated with blue contours. These show that the approach segments the emission into compact regions and that the boundaries between blended regions are approximately straight.  Often, the projected boundaries appear to cross one another, but this arises because the clouds have different velocities in the data cube.  The figure also illustrates how the current approach allocates all significant emission into clouds, unlike the default original \textsc{cprops} algorithm. Conforming to physical expectation, the clouds identified in the figure tend to be centrally concentrated and look like discrete clumps. The watershed algorithm does a good job of assigning extended fainter emission to the identified peaks in a natural way.

In Figure~\ref{fig:ngc3627atlas} we present the full-galaxy image for NGC~3627, highlighting the locations and orientations of each cloud.  We also illustrate the environmental regions described in Section~\ref{sec:envmasks}. Similar images for the remaining nine galaxies in our sample are available in an online supplement.

\subsection{Property Estimation}
\label{sec:properties}

After assigning emission to clouds, \textsc{cprops} calculates properties of the identified clouds. For this step, our \textsc{python} implementation largely follows \citetalias{cprops} with a few improvements. Property determinations are moment-based. We estimate the size and line width of the cloud based on the second moments of the emission distribution along the spatial and spectral axes. We estimate the flux based on the zeroth moment, i.e., the sum of the intensity.  We then correct the measured moments for the effects of sensitivity and the finite resolution of the data. Finally, we translate the moments into estimates of physical quantities. 

\subsubsection{Moment-based Property Estimators}

For these calculations, we consider the pixels in a cloud mask $\mathcal{C}$, which is just those pixels belonging to a single cloud in the label map $\mathcal{L}$ generated by the segmentation algorithm (Section~\ref{sec:segment}). We measure the luminosity of the cloud as 
\begin{equation}
L_{\mathrm{CO}} = A_{\mathrm{pix}} \sum_{i\in \mathcal{C}} T_i \Delta v~,
\end{equation}
where $A_{\mathrm{pix}}$ is the projected physical area of a cube pixel in pc$^2$, $\Delta v$ is the channel width in km~s$^{-1}$, and $T_i$ is the brightness of the cube pixels measured in K in the cloud mask $\mathcal{C}$. The resulting $L_{\mathrm{CO}}$ has units of K~km~s$^{-1}$~pc$^2$. We also record the equivalent integrated flux, $F_{\mathrm{CO}}$, in units of K~km~s$^{-1}$~arcsec$^2$, using the solid angle of a pixel instead of the physical area. 

We estimate cloud line widths (specifically, the velocity dispersions) by calculating the intensity-weighted variance in the spectral direction:
\begin{equation}
\sigma_{v,\mathrm{obs}}^2 = \frac{\sum_{i\in \mathcal{C}} T_i \left(v_i - \bar{v}\right)^2 }{\sum_{i\in \mathcal{C}} T_i}~,
\label{eq:sigv}
\end{equation}
where $\bar{v}$ is the intensity-weighted mean velocity calculated over the cloud mask. 

To measure the cloud size, we calculate the intensity-weighted second moments over the two spatial axes of the cube following a similar form as Equation~\eqref{eq:sigv}. This yields spatial variances $\sigma_x^2$ and $\sigma_y^2$. We also calculate an intensity-weighted covariance term $\sigma_{xy}$. Then we place these values in a variance-covariance matrix. We diagonalize the matrix to determine the major and minor axes of the emission distribution, $\sigma_{\mathrm{maj}}$ and $\sigma_{\mathrm{min}}$, as well as the position angle following \citetalias{cprops}.

\subsubsection{Extrapolation and Sensitivity Correction}
\label{sec:extrap}

Our masking strategy and calculation of moments only includes emission above an intensity threshold, which will bias these estimators. To account for this bias, we extrapolate from the actual measured cloud properties to those we would expect to measure given a 0~K contour threshold. This calculation corrects for the finite sensitivity of the data. \citetalias{cprops} describe this extrapolation in detail, and we briefly summarize it here. 

We sort the data ($T_i$) associated with the cloud in order of decreasing intensity. Then we repeatedly measure the moment values, each time including only data from the GMC above some intensity threshold (i.e., the cloud pixels with intensity $\ge T_i$). We repeat the calculation of moments, progressively lowering the threshold from the highest to the lowest intensity value in the clouds. As we do so, the moments increase in value as the threshold intensity level decreases. 

We fit the measured moment as a function of intensity threshold, using a linear form for the size and line width and a quadratic form for the luminosity. To carry out the fit of property versus intensity threshold, we use a robust least-squares regression with an arctan loss function, which is chosen for its robustness to outliers. Then the extrapolated property is equal to this fit evaluated at an intensity threshold of 0~K.

The fitting procedure and its interaction with the decomposition are the main differences from the original \citetalias{cprops} implementation. Our robust fitting improves on the original approach, which adopted the median of simple linear least-squares fits to all levels. The two methods mostly agree to better than $<5\%$ for luminosity, size, and line width, but in discrepant cases, the robust least-squares approach appears to provide a more reasonable extrapolation.

Our revised approach to decomposition also minimizes the impact of the extrapolation on the final measurements. The original \textsc{cprops} decomposition yielded high clipping levels and so relied heavily on the extrapolation. Our current approach incorporates much more low intensity emission into the cloud assignments, which reduces the dependence of the derived cloud properties on the extrapolation.

Other studies have adopted similar corrections using a Gaussian functional form \citep[e.g.,][]{m64-gmcs} or through directly fitting a Gaussian emission profile to the data \citep[e.g.,][]{canon-gmcs}. The extrapolation-based corrections to the brightness moments lead to nearly the same values as a Gaussian model for a bright, isolated cloud. For crowded regions, the local maxima in the emission distribution are less distinct and all approaches become less stable. We prefer our approach because it is non-parametric and tends to produce stable estimates of the moments.

To estimate a characteristic uncertainty in these properties, we inject false sources into signal-free regions of the data cube (see Section~\ref{sec:completeness}). These tests suggest that the cloud luminosity measurements typically have $\sim 30\%$ errors and a systematic bias to underestimate the true luminosity by $\sim 20\%$, even after the extrapolation correction.  The cloud size and line width measurements show typical errors of $\sim 30\%$ and no evidence of a systematic bias. These errors get larger for faint clouds with local maxima near the noise floor ($>50\%$ for peak signal-to-noise $<10$; see also \citetalias{cprops}). We adopt the extrapolation correction since the other options (Gaussian correction or direct fitting) show larger property errors in the low signal-to-noise case, though they show slightly smaller biases for luminosity estimates.

In Table~\ref{tab:data}, we report the fraction of emission found in catalogued objects, $f_\mathrm{mask}$, by comparing the sums of the unmasked data cube and the masked data cube so that $f_\mathrm{mask} \le 1$.  These show a wide range of values from $f_\mathrm{mask}=0.33$ for NGC~5068 to $f_\mathrm{mask} = 0.86$ for NGC~4826. We also report the fraction of emission found in GMCs after carrying out the extrapolation, $f_\mathrm{cat}$.  Because this extrapolation models the emission outside the mask, we could find $f_\mathrm{cat}>1$ though none of our targets show such high values.  The extrapolation should include all emission near the emission mask.  The remaining emission not included in the extrapolated flux values presumably corresponds to low surface brightness emission found elsewhere in the cube away from the boundary of the signal mask.

\subsubsection{Derived Physical Properties}
\label{sec:derived}

We use the measured size, line width, and luminosity to calculate several derived quantities. First, we convert from luminosity to a CO-based mass estimate. To do this, we scale the extrapolated luminosity by a CO-to-H$_2$ conversion factor, $\alpha_{\mathrm{CO}}$:
\begin{equation}
    M_\mathrm{CO} = \alpha_\mathrm{CO} L_\mathrm{CO}~.
\end{equation}
For PHANGS-ALMA, we adopt the $\alpha_\mathrm{CO}$ treatment of \citet{sun_20}. We designate $\alpha_\mathrm{CO}$ as the \cotwo-to-H$_2$ conversion factor
\begin{equation}
    \alpha_\mathrm{CO}^{(2-1)} = \frac{\alpha_\mathrm{CO}^{(1-0)} (Z)}{R_{21}}~,
\end{equation}
where $R_{21}$ is the \cotwo-to-\coone\ brightness temperature ratio and $\alpha_{\rm CO}^{(1-0)}$ refers to the \coone-to-H$_2$ conversion factor, which we allow to vary with metallicity. We adopt $R_{21} = 0.65$ based on \citet{Leroy_2013} and \citet{denBrok20}, measured at kpc scales. The ratio does vary, but these variations have magnitude $\pm 20\%$ in \citet{denBrok20} and we neglect them here.

Following \citet{sun_20}, we consider a metallicity dependence $\alpha_{\rm CO}^{(1-0)} \propto Z^{-1.6}$, following \citet{accurso2017} and in good agreement with a wide range of previous literature. This prescription is scaled to match the standard Galactic $\alpha_{\rm CO}^{(1-0)} = 4.35$ M$_\odot$~pc$^{-2}$ (K~km~s$^{-1}$)$^{-1}$ at solar metallicity \citep{Bolatto_2013}. In our targets, we estimate the metallicity locally as a function of galactocentric radius using the global mass-metallicity scaling relation of \citet{sanchez19} and the universal metallicity gradient of \citet{sanchez14}. A more complete discussion of this calibration as applied to the PHANGS-ALMA data is presented in \citet{sun_20}.

To estimate the line width of the emission, we deconvolve the line spread function from the extrapolated line width, $\sigma_{v,\mathrm{extrap}}$, to obtain a final line width measurement:
\begin{equation}
    \sigma_{v} = \sqrt{\sigma_{v,\mathrm{extrap}}^2 - \sigma_{v,\mathrm{chan}}^2}~,
\end{equation}
where $\sigma_{v,\mathrm{chan}}$ is the equivalent Gaussian width of a channel. \citetalias{cprops} equated the line spread function to a single top hat-shaped channel and subtracted the equivalent Gaussian width of a channel from the measured line width in quadrature. Here, we refine that approach to account for line spread functions broader than a single channel. We adopt the method of \citet{leroy_16}, who model the equivalent Gaussian width of a channel as:
\begin{equation}
\sigma_{v,\mathrm{chan}} = \frac{\Delta v}{\sqrt{2\pi}}\left(1 + 1.18k + 10.4k^2\right)~,
\end{equation}
where $\Delta v$ is the channel width and $k$ is a correction factor determined from the Pearson correlation coefficient, $r$, between noise in adjacent channels arising from line spread function in our data. As in \citet{leroy_16}, we adopt 
\begin{equation}
    k = 0.47 r - 0.23 r^2 - 0.16 r^3 +0.43 r^4~.
\end{equation}
This model only accounts for correlation between adjacent channels, but this is sufficient to describe most radio data. For the PHANGS-ALMA \cotwo\ data, we adopt $r=0.05$, which is a good approximation for all the data sets used in this work \citep{sun_20b}.

To measure cloud radii, we correct the extrapolated size measurements to account for the finite resolution of the data. We deconvolve the round Gaussian beam\footnote{The {\sc pycprops} algorithm also supports deconvolution by elliptical Gaussian beams.} from the data, assuming that both the cloud and the beam have a Gaussian profile. We convert from deconvolved major and minor sizes, $\sigma_{\mathrm{maj,d}}$ and $\sigma_{\mathrm{min,d}}$, to a cloud radius measurement using 
\begin{equation}
R = \eta \sqrt{\sigma_{\mathrm{maj,d}} \sigma_{\mathrm{min,d}}}~.
\end{equation}
The factor $\eta$ formally depends on the light or mass distribution within the cloud (e.g., see \citetalias{cprops}). For PHANGS-ALMA, we model the surface brightness of the clouds as a two dimensional Gaussian and use $R$ to denote the half width at half maximum so that $\eta =\sqrt{2\ln 2} = 1.18$.  We also report the position angle of the major axis, P.A., and the aspect ratio of the cloud in terms of the cloud eccentricity, $e = \sqrt{1-\sigma_{\mathrm{min,d}}^2/\sigma_{\mathrm{maj,d}}^2}$, to enable energetics estimates using the approach of \citet{bertoldi-mckee}.

We note that our adopted $\eta = 1.18$ is smaller than the $\eta = 1.91$ used in \citet[][ hereafter \citetalias{srby87}]{srby87}. This factor was determined from an empirical scaling between the measured moment and the cloud boundary defined by a brightness contour in the Massachusetts-Stony Brook CO Galactic Plane Survey \citep{sanders86}.  The survey identified clouds from a wide range of distances in the Galactic plane, so this factor is calculated for clouds that range from marginally to well resolved on a sparsely sampled pixel grid.  This same value of $\eta=1.91$ was adopted in \citetalias{cprops} and has been widely used throughout the GMC literature primarily for consistency. For PHANGS-ALMA, the clouds that we observe are almost always marginally resolved (e.g., see Figure~\ref{fig:zoomfig}) and the map is shaped by the Gaussian beam of our data. We consider our definition of the radius to offer a more realistic representation of the emission distribution. As a result, we expect that size-dependent quantities such as the inferred volume density and surface density will also be closer to physical reality using our definition. While the heterogeneous nature of literature CO data is likely to be a greater source of uncertainty for comparative studies of cloud properties, we nonetheless note that our revised definition of size should be taken into account when comparing the results presented here to previous literature.

Our geometric model approximates the cloud as a spherically symmetric object so that $R$ also characterizes the object in three dimensions. As such, we do not apply any inclination corrections to $R$. This model becomes limited when the size of the cloud approaches the scale height of the molecular medium in a galaxy. As noted by \citet{sun_20}, the 90~pc resolution of our data and our measured cloud sizes can approach or even exceed the ${\sim}100$~pc FWHM height of the molecular gas disc in the Milky Way \citep[e.g.,][]{heyerdame_2015} and nearby galaxies \citep[e.g.,][]{yim14}. This issue will be even more severe considering the 150~pc common resolution of the full PHANGS-ALMA data set. In this case, the conventional assumption that clouds have a spherical geometry with a line of sight depth equal to the projected size on the sky is no longer appropriate.  In these cases we shift our model to a spheroidal geometry for objects where the measured radius would exceed the expected FWHM scale height, $H$, of emission in the galaxy. Concretely, we take the line of sight depth of the cloud to be the lesser of the cloud diameter, $2R$, and the scale height, $H$. This leads to a three dimensional mean radius, $R_{\mathrm{3D}}$, that should be used to calculate volumetric quantities. We take
\begin{equation}
    R_{\mathrm{3D}} = \left\{\begin{array}{rl}
    R & ; \quad R\le H/2\\
    \sqrt[3]{\frac{R^2H}{2}} & ; \quad R > H/2~.
    \end{array}\right.
    \label{eq:r3d}
\end{equation}
In this paper, we adopt a constant $H=100~\mathrm{pc}$ for simplicity in all our targets. More sophisticated models of $H$ that account for the local structure of the galaxy are possible \citep[e.g.,][]{blitzros06} and represent a future direction for development. Over half (54\%) of our clouds are large enough that $R_\mathrm{3D}\ne R$.

We also estimate the virial mass of each cloud, $M_{\mathrm{vir}} = 5\sigma_v^2 R_\mathrm{3D} /G$, and a simple virial parameter, $\alpha_\mathrm{vir}= 2M_{\mathrm{vir}}/{M_{\mathrm{CO}}}$ \citep{bertoldi-mckee}. The factor of 2 arises from our two dimensional Gaussian cloud model, where half the mass is contained inside the FWHM. We thus treat the virial mass as an estimate of the dynamical mass within the FWHM, comparing $M_{\mathrm{vir}}$ to $M_{\mathrm{CO}}/2$.  Practically, we use the virial parameter as a scalar estimate of the relative strength of the gravitational binding energy versus the kinetic energy of a molecular cloud rather than a statement that cloud is virialized, which would require the cloud to be stable and thus long-lived. Following the diction of the field, references to virialization are best interpreted in terms of contributions to the balance of energy terms in complete virial analysis rather than an assertion of dynamical stability.

If we instead were to consider the energy balance in a Gaussian cloud and report the virial parameter $\alpha_\mathrm{vir} = 2K/|U_g|$, we would arrive at a similar result. For a three dimensional Gaussian mass distribution with dispersion $\sigma_r$, \citet{pattle15} show that the gravitational binding energy is 
\begin{equation}
    U = -\frac{1}{2\sqrt{\pi}} \frac{GM^2}{\sigma_r} = -\sqrt\frac{\ln 2}{2 \pi} \frac{GM^2}{R_\mathrm{3D}}~,
\end{equation}
where we have used the result that a 3D Gaussian distribution with dispersion $\sigma_r$ projects to a 2D Gaussian surface density distribution with the same dispersion.  The kinetic energy is $K = \frac{3}{2} M\sigma_v^2$ by construction since the line width is measured as a luminosity-weighted average over the profile and we assume that light traces mass.  Thus, 
\begin{equation}
    \alpha_\mathrm{vir} = 3\sqrt{\frac{2\pi}{\ln 2}} \frac{R_\mathrm{3D} \sigma_v^2}{GM} \approx 9.03  \frac{R_\mathrm{3D} \sigma_v^2}{GM}~,
    \end{equation}
where the numerical prefactor is only 10\% different from the prefactor of 10 used in the simple model we report.

We calculate the average surface density within the FWHM size, $\Sigma_\mathrm{mol} = M_{\mathrm{CO}} / (2{\pi R^2})$. This estimate adopts the two dimensional Gaussian cloud model in which half the mass is contained inside the FWHM.  For each cloud we also calculate the implied turbulent line width on a fixed 1~pc scale: $\sigma_0 = \sigma_v (R_\mathrm{3D}/\mathrm{1~pc})^{-0.5}$. This $\sigma_0$ assumes that the turbulent structure function within all clouds has an index of 0.5 \citep[e.g.,][]{heyer_2004} and scales from the measured $\sigma_v$ and $R_\mathrm{3D}$ to derive $\sigma_0$ at $R_\mathrm{3D}=1$~pc.

\subsection{Completeness Limits}
\label{sec:completeness}

We validate the results of our source identification and property recovery by injecting false sources into signal-free regions (i.e., the complement of $\mathcal{M}$) of the data cube and analyzing them following the methods described above.  This provides a good test of the source identification and characterization algorithms and represents another point of improvement over the algorithm presented in \citetalias{cprops}. We use our real data in this case so that the residual effects of interferometric deconvolution or the influence of faint emission on source recovery are empricially included in the analysis. We emphasize, however, that this analysis does \textit{not} assess the effects of blended emission, e.g., detecting a cloud in a crowded region or separating blended clouds. We empirically characterize the effects of blending in Section~\ref{sec:massdist}.

The false sources have Gaussian profiles in position-position-velocity space. Each cloud has a mass, virial parameter, and surface density drawn randomly from log-uniform distributions for that parameter. This part of the analysis adopts a fixed $\alpha_\mathrm{CO}= 6.7\ M_\odot$ $(\mathrm{pc^2~K~km~s^{-1})}^{-1}$ throughout.

We generate false clouds using uncorrelated sampling of distributions of the cloud mass, surface density, and virial parameter. False cloud masses have a 2.5 dex range centred on $M_0 = 50 \alpha_\mathrm{CO}\Omega_{\mathrm{bm}} d^2 \sigma_T \Delta v$, where $\Omega_\mathrm{bm}$ is the solid angle subtended by the beam, $d$ is the distance to the galaxy, $\sigma_T$ is the median RMS noise level in the data set, and $\Delta v$ is the channel width.  This mass value corresponds to a $\mathrm{S/N}=50$ detection in a single beam and a single channel. False cloud surface densities have a 2.5 dex range centred on $\Sigma_\mathrm{mol} = 150~M_{\odot}~\mathrm{pc}^{-2}$. Virial parameters are drawn in a 2 dex range centred around $\alpha_\mathrm{vir} = 2$.

Once the mass, virial parameter, and surface density are specified, we calculate the implied line width, $\sigma_v = (\alpha_\mathrm{vir} G/5)^{1/2} (\pi M\Sigma_\mathrm{mol}/2)^{1/4}$, and an observed two dimensional radius, $R = (M/2\pi \Sigma_\mathrm{mol})^{1/2}$. The resulting distributions of line width and radius are not exactly uniform, but they span a larger range than we expect for real GMCs. For example, radii range from 3~pc for high surface density, low mass clouds to 800~pc for low surface density, high-mass clouds. The line widths range from 0.5~km~s$^{-1}$ for low virial parameter, low mass, and low surface density clouds to 25~km~s$^{-1}$ for high virial parameter, high mass, and high surface density clouds.

For each data set, we inject $>10^3$ false sources into the signal-free portion of the data cube and process the cubes with the same parameters as used in the main analysis.  For each source, we note whether the source is detected or not and record both the recovered parameters and the injected parameters. 

We use these data to determine the probability of detecting a cloud with a given $M$, $\Sigma_\mathrm{mol}$, and $\alpha_\mathrm{vir}$. To do this, we fit a logistic regression to the detection data, which has the form:
\begin{align}
P(M,\Sigma_\mathrm{mol},\alpha_\mathrm{vir}) = \,& \left\{1+\exp\left[- c_0 - c_1 \log_{10}\left(\frac{M}{10^6 M_{\odot}}\right) \right.\right.\nonumber\\
& - c_2 \log_{10}\left(\frac{\Sigma_\mathrm{mol}}{150~M_{\odot}~\mathrm{pc}^{-2}}\right) \nonumber\\
& - c_3 \left. \left. \log_{10}\left(\frac{\alpha_\mathrm{vir}}{2}\right)
\right]\right\}^{-1}.
\label{eq:complim}
\end{align}
We use the {\sc statsmodels} package \citep{statsmodels} in {\sc python} to perform the regression and obtaining coefficients $c_i$ for the models from fits to the detection statistics. 

In the model described by Equation~\eqref{eq:complim}, the mass limit below which the completeness is $< 50\%$ for a cloud with $\alpha_\mathrm{vir}=2$ and $\Sigma_\mathrm{mol} = 150~M_{\odot}~\mathrm{pc}^{-2}$ is 
\begin{equation}
\label{eq:mcomp}
M_\mathrm{comp} = 10^{-c_0/c_1 + 6}\;M_\odot~.
\end{equation} 
The logistic regression can report the mass scales for different completeness fractions but the 50\% completeness level is the characteristic scale for this model. For the homogenized data ($90$~pc resolution, uniform noise), this completeness limit is \complim\ at $\Sigma_\mathrm{mol}=150~M_\odot\,\mathrm{pc}^{-2}$ and $\alpha_\mathrm{vir} = 2$. 

In Equation~\eqref{eq:complim}, the probability of cloud detection also depends on $\alpha_\mathrm{vir}$ and $\Sigma_\mathrm{mol}$. We find coefficients $c_2$ and $c_3$ that are significantly non-zero, indicating that these other terms are important. On average, we find that $\langle c_0 \rangle = 1.5$, $\langle c_1\rangle = 4.7$, $\langle c_2\rangle = 2.2$ and $\langle c_3 \rangle = -1.4$ across all data sets.

If we shift the fiducial $\Sigma_\mathrm{mol}$ we expect to effectively change $c_0$ in Equation~\eqref{eq:mcomp} by $c_2 $ times the logarithmic change in $\Sigma_{\rm mol}$. Given $\langle c_2\rangle = 2.2$, adjusting to a factor of three (or $0.5$~dex) lower $\Sigma_\mathrm{comp}$, $M_\mathrm{comp}$ will increase by 70\%. The effect of $\Sigma_\mathrm{mol}$ is weaker than $M$, i.e., $c_1 > c_2$ because these clouds are marginally resolved in our data. The convolution with the instrumental effects leads two clouds with the same mass but different $\Sigma_\mathrm{mol}$ to having roughly the same surface brightness as long as both are relatively compact compared to the beam. 

Cloud line widths are typically well resolved in our data and clouds with larger virial parameters have emission distributed over a larger number of channels. For fixed mass, this decreases the peak brightness of the cloud and lowers the signal-to-noise in any given channel, thus lowering the probability of detection. If we raise the fiducial $\alpha_{\rm vir}$ from $2$ to $6$, then the corresponding $M_\mathrm{comp}$ will increase by 40\%.  

\begin{figure}
\includegraphics[width=\columnwidth]{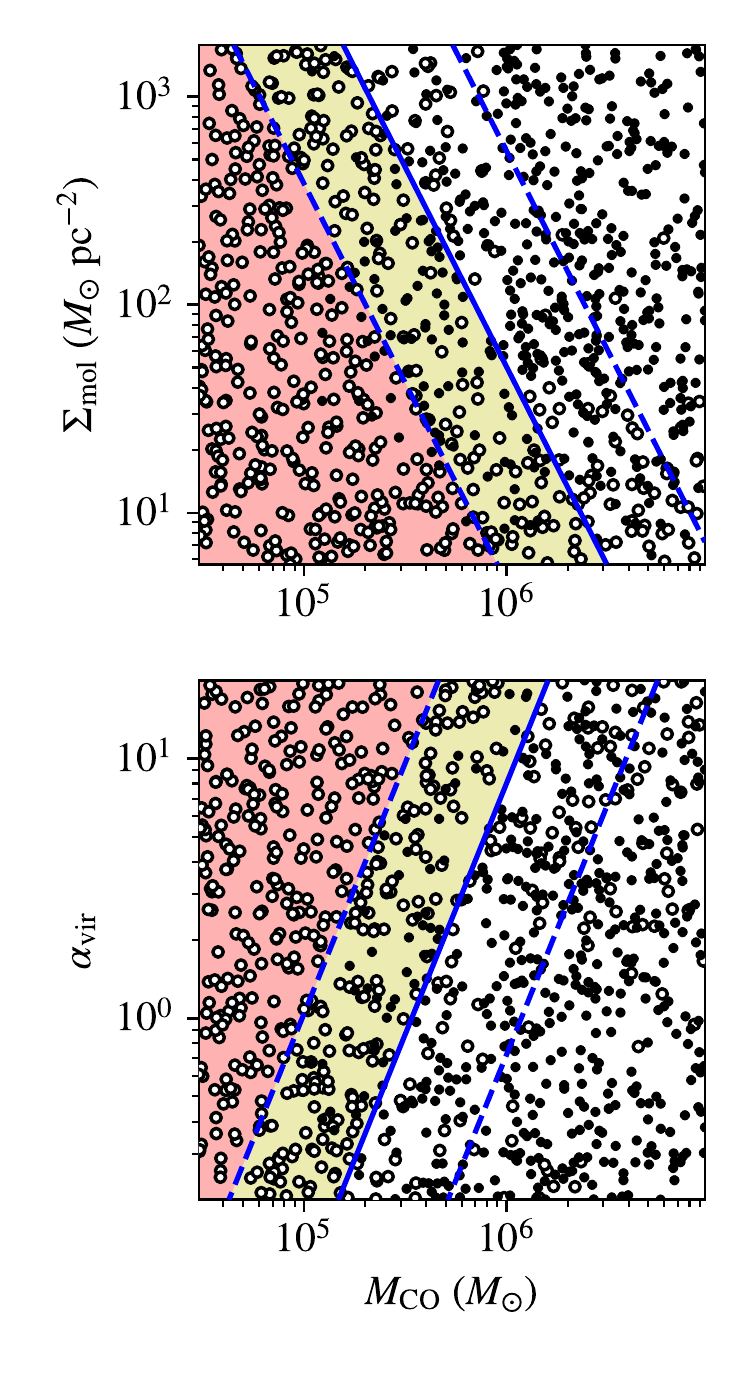}
\vspace*{-25px}
\caption{\label{fig:complim} Recovery of false sources injected into the signal-free region of the homogenized, 90~pc resolution data for NGC~3627. The plots show the injected mass, $M$, against the surface density, $\Sigma_\mathrm{mol}$ (top) and virial parameter, $\alpha_\mathrm{vir}$ (bottom) of the test clouds. Filled points indicate false clouds recovered by \textsc{pycprops}. Empty circles indicate points not recovered. The solid blue line marks  $P(M,\Sigma_\mathrm{mol},\alpha_\mathrm{vir})=0.5$ (Equation~\ref{eq:complim}), taking the fiducial parameters for variables that are not plotted.  This line shows where clouds would be detected with 50\% completeness. The blue dashed lines show the analogous lines of 10\% and 90\% completeness. The figure shows that cloud mass is the primary factor affecting source recovery but surface density and virial parameter have important secondary effects.}
\end{figure}

Figure~\ref{fig:complim} shows the results of our completeness analysis for the homogenized, 90~pc resolution data for NGC~3627. The figure shows that the logistical regression described by Equation~\eqref{eq:complim} is a good approximation to the completeness structure in the data, with roughly half of the clouds near the 50\% completeness limit line being detected.  However, the top panel of the figure indicates that clouds with very low surface densities ($\Sigma_\mathrm{mol} < 10\,M_\odot\,\mathrm{pc}^{-2}$) are poorly recovered irrespective of their total mass because of their low surface brightness.  This can be seen since several clouds in the $>90\%$ regime (bottom right) are still not detected. This behaviour is not captured by the multi-linear logistic regression model (Equation~\ref{eq:complim}), and our algorithm will not detect clouds with low surface brightness.  The tilt of the completeness lines with respect to the coordinate axes illustrate how $M_\mathrm{comp}$ depends on $\Sigma_\mathrm{mol}$ and $\alpha_\mathrm{vir}$, increasing with increasing $\alpha_\mathrm{vir}$ and decreasing with increasing $\Sigma_\mathrm{mol}$.

This analysis gives quantitative completeness limits and qualitatively shows that the algorithm is best at detecting luminous objects that are compact in both space and velocity. For fixed mass, clouds with higher virial parameter or lower surface density have lower signal-to-noise at any individual pixel because the cloud is distributed over a wider region in the spatial and spectral directions. 

Overall, \textsc{pycprops} applied to PHANGS-ALMA reliably extracts ``classical'' molecular clouds with $\Sigma_\mathrm{mol} \gtrsim 10^{2}M_{\odot}~\mathrm{pc}^{-2}$ and virial parameters $\alpha_\mathrm{vir} \lesssim 2$. It is not sensitive to the presence of a diffuse molecular medium, e.g., as inferred from multi-scale analysis of M31, M51, and other nearby galaxies \citep{Pety_2013,calduprimo16,chevance20}. Unbound molecular gas and low surface density clouds are unlikely to be detected as individual objects \citep{romanduval2016}. If they are isolated, they will not be detected and catalogued. If they are in a dense region, they will be assigned to nearby clouds by the decomposition algorithm.

\section{Scaling Relations}
\label{sec:scaling}

The scaling relations between the macroscopic properties of GMCs (mass, radius, line width) illustrate the changing physical conditions of the star-forming ISM across different galactic environments.  These scalings are primary results from the cataloguing processes outlined above. 

Our main results come from an analysis of the homogenized data described in Section~\ref{sec:homogenized}. This homogenization is essential for a robust inter-galaxy comparison of cloud populations \citep{hughes13}. Here, we illustrate the effect and importance of the homogenization in Section~\ref{sec:NeedForHomog}. We report the properties of individual clouds in Table~\ref{tab:gmccat}. We explore correlations among cloud properties in Section~\ref{sec:HomogCat} and summarize distributions of their physical properties in Table~\ref{tab:pardist}.  For comparison, we present the same scaling relationships at the native resolution and noise levels of the data in the Appendix. 

When applied to the homogenized data for our 10 targets, \textsc{pycprops} recovers \numcloudhomogtotal clouds. Of these, \numcloudhomog have a signal-to-noise ratio at the cloud peak of $\mathrm{S/N}>6$, which we consider a threshold for reliable measurements \citepalias{cprops}. 

\begin{table*}
    \begin{tabular}{cccccccccccc}
    \hline
    Galaxy & Number & RA & Dec & $V_\mathrm{LSR}$  & $R$ & PA & $e$ & $\sigma_v$ & $L_\mathrm{CO}$ & $M_\mathrm{CO}$ & $M_\mathrm{vir}$ \\
          &        & (J2000) & (J2000) & (km~s$^{-1}$) & (pc) & ($^\circ$) & & (km~s$^{-1}$) & ($10^5$~K~km~s$^{-1}$~pc$^{2}$) & ($10^6~M_\odot$) & ($10^6~M_\odot$) \\
    \hline
NGC~0628 & 1 & 24.17516879 & 15.76966235 & 631.8 & 115.0 &    3 & 0.41 &   7.6 &   3.3 &   1.9 &   5.8 \\ 
NGC~0628 & 2 & 24.15919146 & 15.76697688 & 625.4 &  98.8 &    3 & 0.55 &   7.2 &   1.0 &   0.6 &   4.8 \\ 
NGC~0628 & 3 & 24.17234447 & 15.76836941 & 626.4 & 118.8 &  179 & 0.90 &   7.1 &   2.4 &   1.4 &   5.3 \\ 
NGC~0628 & 4 & 24.16455515 & 15.76022010 & 623.4 & 119.5 &  153 & 0.94 &   4.4 &   3.7 &   2.6 &   2.0 \\ 
NGC~0628 & 5 & 24.16955384 & 15.76238607 & 626.4 &  99.6 &   84 & 0.86 &   6.1 &   3.5 &   2.4 &   3.4 \\ 
NGC~0628 & 6 & 24.16149093 & 15.76770079 & 626.3 & 107.9 &  100 & 0.40 &   5.1 &   5.4 &   3.5 &   2.5 \\ 
NGC~0628 & 7 & 24.18739848 & 15.74728395 & 628.2 &  57.6 &  138 & 0.95 &   3.1 &   0.8 &   0.7 &   0.6 \\ 
NGC~0628 & 8 & 24.18065112 & 15.75086674 & 631.3 &  58.1 &  118 & 0.67 &   4.9 &   5.2 &   4.3 &   1.5 \\ 
NGC~0628 & 9 & 24.17763321 & 15.75330872 & 632.2 &  93.4 &   28 & 0.76 &   7.5 &   1.5 &   1.2 &   5.0 \\ 
NGC~0628 & 10 & 24.16708744 & 15.75749143 & 624.6 &  64.4 &  115 & 0.65 &   1.1 &   1.0 &   0.8 &   0.1 \\ 
    \hline                
    \end{tabular}
    \caption{\label{tab:gmccat}  Example catalogue of GMC properties determined from PHANGS-ALMA observations of NGC~0628.  The complete table including all targets is available online. Here, $R$ is the deconvolved cloud radius, the position angle of the major axis is given in the PA column, and $e$ is the cloud eccentricity.  The line width is given as a velocity dispersion, $\sigma_v$.  The cloud luminosity in \cotwo\ emission is given as $L_\mathrm{CO}$ and the mass is given as $M_\mathrm{CO}$ after applying a spatially variable CO-to-H$_2$ conversion factor.  Finally the virial mass appears in the $M_\mathrm{vir}$ column.  For more details on how these properties are measured for each cloud, see Section~\ref{sec:derived}.}
    \end{table*}
    \medskip

\subsection{Resolution Effects Bias Cloud Properties}
\label{sec:NeedForHomog}

As described in Section~\ref{sec:properties}, \textsc{pycprops} attempts to correct the measured cloud properties for the finite sensitivity and resolution of the data. These after-the-fact corrections assume that the emission associated with each cloud has been correctly assigned to the cloud by the segmentation algorithm (Section~\ref{sec:segment}). However, the segmentation algorithm itself is affected by limited data sensitivity and resolution.

The segmentation relies on identifying local maxima in the spectral line data cube. Clouds at separations smaller than the beam size will nearly always be blended together into larger structures by this approach. This blending introduces a direct dependence of cloud size and mass on the resolution of the data. The ISM has structure on a wide range of scales \citep[e.g.,][]{snez_01}.  When a seeded watershed algorithm like \textsc{clumpfind} or \textsc{cprops} is applied to such multiscale structure, the algorithm tends to divide the emission into regions comparable to the beam size \citep{pineda_09, hughes13, leroy_16}.

In addition to resolution, noise also plays an important role in segmentation. Local maxima are identified based on a contrast threshold $\delta$, which is set by the noise in the data. In data with insufficient sensitivity, peaks with low contrast may be blended together. The results of our completeness analysis in Table~\ref{tab:data} also show that the mass of clouds that can be recovered is a sensitive function of the physical resolution and the noise of the data.

These effects mean that source identification depends on the resolution and noise properties of the data. Despite attempts to correct for these effects in the property measurements, this dependence of the decomposition algorithm can strongly influence the distribution of measured cloud properties. 

\begin{figure*}
\includegraphics[width=\textwidth]{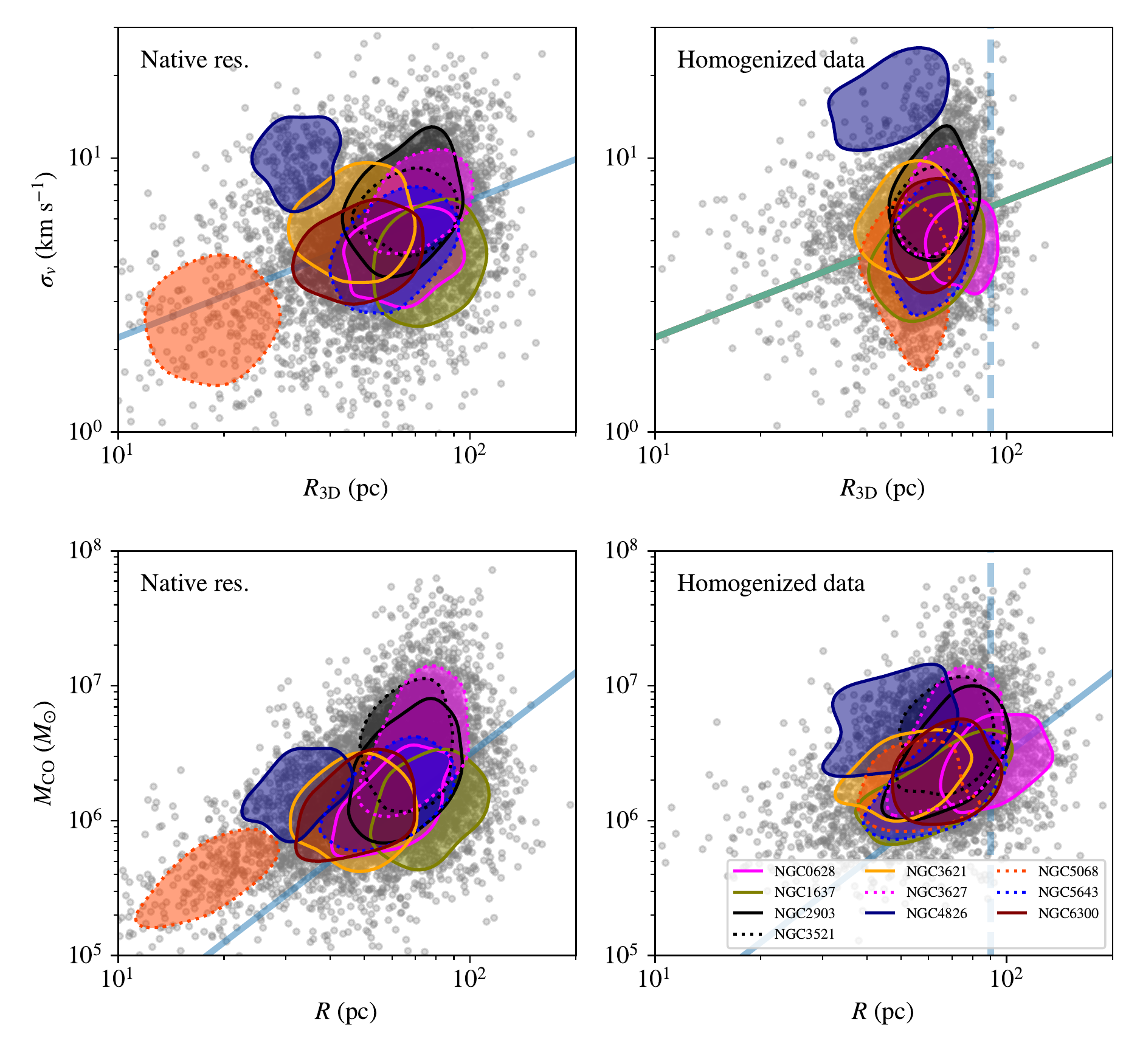}
\vspace*{-25px}
\caption{\label{fig:orig_vs_common} Comparison of the size--line width (top) and mass--radius relationship (bottom) before (left) and after (right) homogenizing the data. The left panels show results for our targets analysed at their native resolutions and noise levels. The right panels show results for the homogenized data. Values for the individual GMCs are shown as grey points. The coloured contours denote the regions containing 50\% of the data for individual galaxies. The solid blue lines show the reference relationships $\sigma_{v}=0.7~\mathrm{km~s}^{-1}(R_\mathrm{3D}/\mathrm{1~pc})^{0.5}$ in the top panel and $\Sigma_\mathrm{mol}=100~M_{\odot}~\mathrm{pc}^{-2}$ in the bottom panel.  The vertical blue dashed lines in the right panels show the beam size of 90~pc data in the homogenized data. Analyzing the native resolution data (see Table~\ref{tab:data} for characteristics) would lead to the conclusion that cloud populations show more variation than what is inferred from a parallel analysis on the homogenized data.}
\end{figure*}

To illustrate this, Figure~\ref{fig:orig_vs_common} compares the size--line width (top) and mass--radius (bottom) relationships for our sample before and after homogenization. The figure illustrates how the segmentation algorithm distinguishes objects near the beam size. Due to the range of angular resolution and distances for our target galaxies (Table~\ref{tab:data}), the native resolution data in the left panels yield a wide range of linear resolutions (projected beam sizes of 26 to $89$~pc) and hence measured cloud sizes. After smoothing all data to have the same resolution in the right panels, the sizes span a much narrower range. Clouds from NGC~4826 and NGC~5068, which begin at linear resolution $\lesssim 30$~pc, show particularly striking contrast between the two panels. Meanwhile, the addition of noise during the homogenization also removes sensitivity to lower mass clouds. With homogenization, clusters of smaller clouds in these two galaxies are blended together and catalogued as a single complex of emission. After homogenization most recovered clouds have $M > 10^6$~M$_\odot$, while before homogenization many galaxies have populations of clouds extending down to $M \lesssim 5\! \times\! 10^5$~M$_\odot$.

Crucially, the measurements without homogenization often suggest variations among cloud populations that diminish or even disappear when the data properties are matched. From the left to right panels of Figure~\ref{fig:orig_vs_common}, our sample shifts from showing a wide range of apparent cloud masses and sizes to showing substantial overlap among the cloud populations in most galaxies.  Higher resolution observations on the distant sources would likely resolve individual clouds into more objects, including substantial numbers of low mass clouds like those seen in our nearest targets like NGC~4826 and NGC~5068.

The homogenization can also highlight differences that would not be apparent without careful matching. For example, we note in Figure~\ref{fig:orig_vs_common} that clouds in NGC~4826 clearly show larger line widths and  surface densities relative to the other targets after homogenization. With homogenization, this becomes a robust measurement that reflects real differences in the cloud properties in this dense starburst.

Homogenizing the data in resolution and noise represents the most direct and robust approach to mitigate these issues. If the data share the same observational properties then the biases introduced by the segmentation will be the same across all data sets \citep{hughes13}. This gives high confidence to comparative analyses, so that variations between the measured cloud populations reflect real differences in the underlying distribution of emission. Other approaches do exist. For example, the original \citetalias{cprops} implementation attempted to address these issues for the specific case of comparing Galactic to extragalactic cloud samples. This approach set prior assumptions for the clouds to be extracted in data sets with radically different resolution and noise levels by setting the algorithm parameters laid out in Section~\ref{sec:algorithm} to fixed physical scales.

\subsection{Homogenized Catalogs}
\label{sec:HomogCat}

\begin{figure*}
\includegraphics[width=\textwidth]{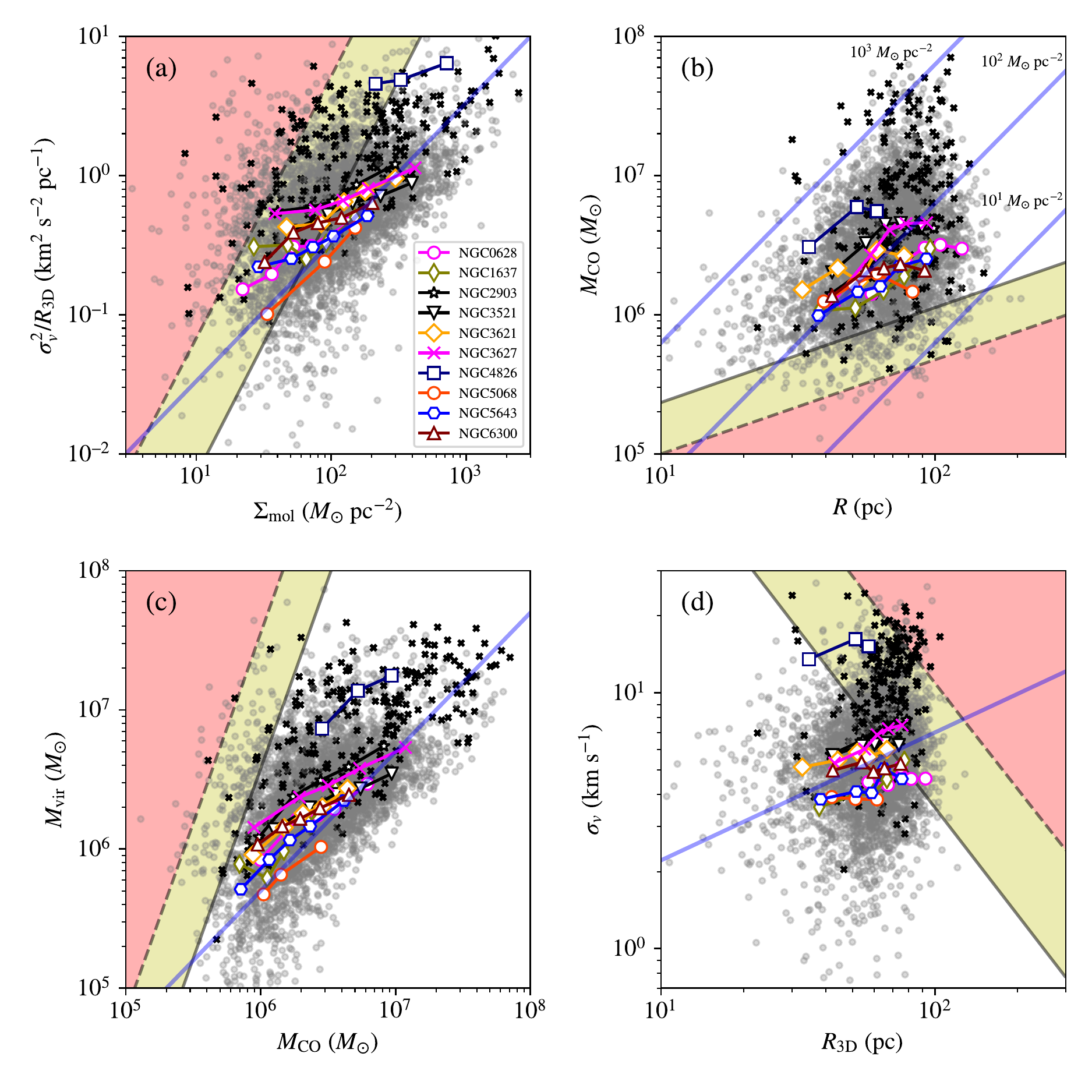}
\vspace*{-25px}
\caption{Scaling relationships among molecular cloud properties in our homogenized cloud catalogue. Individual clouds are presented as grey points except for clouds in galactic centres, which are indicated with a black $\boldsymbol{\times}$ symbol. Colored curves indicate the binned relationships for each individual galaxy, giving the median values in bins set to contain equal numbers of clouds along the $x$-axis. The shaded regions indicate where the false source insertion tests indicate $<50\%$ (yellow) and $<10\%$ (red) recovery rate for clouds in the test sample (Section~\ref{sec:completeness}).  For cases where the completeness region depends on cloud mass, we assume a test mass of $\sqrt{10}\,M_\mathrm{comp} = 1.5\! \times\! 10^6~M_\odot$, where $M_\mathrm{comp}$ is the 50\% completeness mass for the homogenized data: \complim.  solid blue lines indicate reference relationships. Panel (a) shows the relationship between molecular gas surface density, $\Sigma_{\mathrm{mol}}$, and the ratio $\sigma_0^2 = \sigma_v^2 / R_\mathrm{3D}$. The blue line shows the locus of virialization. Panel (b) shows the mass--radius relationship with blue lines showing constant surface densities. Panel (c) shows the relationship between estimated virial mass and the mass estimated from CO luminosity and our location dependent conversion factor. The blue line shows $M_\mathrm{vir} = M_\mathrm{CO}/2$. Panel (d) shows the size--line width relationship with the blue line showing the scaling identified in the Milky Way (S87). In panels (a) and (c), the offset from the reference relationship encodes the virial parameter (see the text).}
\label{fig:scaling_common}
\end{figure*}

We plot the relationships among the properties of molecular clouds catalogued in the homogenized data in Figure~\ref{fig:scaling_common}. In all panels, points show results for individual clouds. Black points indicate clouds in the centres of galaxies, defined as per Section~\ref{sec:envmasks}. Coloured, connected lines show the binned trends for each individual galaxy. These lines trace the median $y$-axis properties measured in up to five bins, each containing an equal number of clouds after sorting the data by the $x$-axis quantity. Blue lines show fiducial relationships based on simple physical expectations or studies of Milky Way molecular clouds (\citetalias{srby87}, \citetalias{heyer_2009}).

In the Appendix, we show the scaling relations for the native resolution data. Enforcing a common physical resolution and sensitivity between galaxies brings the cloud populations into better agreement with one another in all four views into parameter space. It also makes us more confident that any differences we measure among clouds are physical.

The shaded regions in Figure~\ref{fig:scaling_common} highlight regions of parameter space where our survey is not sensitive to cloud detection. These completeness regions are defined by mapping the values fit to Equation~\eqref{eq:complim} to the parameter values plotted. In cases where the mapping applies for a specific cloud mass (such as for the size--line width relationship in panel~d), we plot the region for $M=\sqrt{10}\,M_{50}$ where $M_{50}$ is the 50\% completeness limit for the homogenized data: \complim. Since the most common clouds detected in our study are found near the completeness limit, this represents the censored regions of parameter space for the most common clouds. Higher mass clouds can still be found outside these regions. The yellow shaded region indicates $10{-}50$\% recovery of these sources over the range of virial parameters and surface densities considered. The red shaded regions indicate $<10\%$ recovery.

{\bf The \boldmath$\Sigma_\mathrm{mol}{-}\sigma_0$ relationship} -- Figure~\ref{fig:scaling_common}(a) shows the relationship between the mean molecular gas surface density, $\Sigma_{\mathrm{mol}}$, and the normalized turbulent line width $\sigma_v^2/R_\mathrm{3D}\equiv \sigma_0^2$. We compare the data to the relationship
\begin{equation}
    \frac{\sigma_v^2}{R_\mathrm{3D}} = \frac{\pi G \Sigma_\mathrm{mol}}{5} \left(\frac{R}{R_\mathrm{3D}}\right)^2,
\end{equation}
which is shown as a solid blue line and expected for clouds with $\alpha_\mathrm{vir}=1$ and $R=R_\mathrm{3D}$.

Overall the population of giant molecular clouds appears consistent with the locus of virialization. This result is broadly consistent with \citet{heyer_2009} for the Milky Way and a non-parametric analysis of the first PHANGS-ALMA targets by \citetalias{sun_18}. Both of these studies find that the molecular ISM tends to be found in a state consistent with self-gravitation ($\alpha_\mathrm{vir} \approx 2$) across a wide range of physical scales. Here we investigate a new sample of galaxies, add information on cloud sizes, and find that the molecular ISM across our sample shows $\alpha_\mathrm{vir}\approx 1.5$ with a broad distribution of virial parameters ($\pm 1\sigma$ ranges of 0.7 to 3.6).  Given our uncertainties are expected to be ${\sim}50\%$ from the analysis of false source injection, the range of virial parameters could be explained by observational errors.

Clouds in galaxy centres (black markers) often appear to deviate systematically higher than the $\alpha_\mathrm{vir} = 1$ relationship. They show consistently larger line width at fixed size scale and fixed $\Sigma_\mathrm{mol}$ than clouds in the outer parts of galaxies. This likely reflects a combination of source blending, the presence of a diffuse molecular phase, the influence of external forces, and unresolved motions contributing to the line width in galaxy centres \citep{kruijssen19orb, dale19, sun_20}. 

Most galaxies show general agreement in this plot, although NGC~4826 (M64) is an outlier here and below.  Previous work \citep{m64-gmcs} showed that GMCs in this system have high surface densities and line widths compared to Local Group GMCs but had energetics consistent with being virialized.  The starburst in this system appears to be driven by the unique evolutionary state of the gas in that galaxy relative to the rest of our sample.  The high gas surface density is conjectured to come from a retrograde galaxy collision that drove most of the neutral ISM into the central $1.5$~kpc of the galaxy \citep{counterrot_m64}.  

{\bf Mass--radius relationship} -- The mass--radius relationship is shown in Figure~\ref{fig:scaling_common}(b). Here we compare our measurements to lines of constant surface density, $\Sigma_\mathrm{mol}$ (solid blue lines). As in previous GMC studies, our measurements show that most GMCs in this sample have average surface densities of $\Sigma_{\mathrm{mol}}\approx 10^2~M_{\odot}\,\mathrm{pc}^{-2}$. The completeness regions show that the apparent increase in surface density at small radii is a censoring effect because only high surface brightness clouds with such small sizes are detected. Based on Milky Way studies (\citetalias{srby87}, \citealt{hc01}, \citetalias{heyer_2009}) there are likely to be many clouds with smaller masses and radii that we simply cannot detect in the (homogenized) PHANGS-ALMA data.

Nevertheless, there is real variation in the surface densities of our identified clouds. As seen in panel (a), the high $\Sigma_\mathrm{mol}$ clouds also tend to exhibit high $\sigma_v^2 / R_\mathrm{3D}$, thus, maintaining $\alpha_\mathrm{vir} \approx 1.5$ even as $\Sigma_{\rm mol}$ varies. Once again, the cloud population in NGC~4826 stands out as having relatively high surface densities.

Figure~\ref{fig:scaling_common}(b) also shows that clouds located in the central regions of galaxies are associated with larger values of $\Sigma_\mathrm{mol}$ but these targets are found throughout this parameter space. We caution that in the dense central regions of galaxies, the molecular medium is more luminous per unit mass compared to galaxy discs \citep[e.g.,][see also Section~\ref{sec:discussion}]{Solomon_1997, Bolatto_2013, Sandstrom_2013}.

Interestingly, we do not recover large clouds in our data set. We find essentially no clouds with $R>200$~pc and very few clouds with $R>100$~pc. The cloud radii are limited by intracloud spacing which is typically only $100{-}250$~pc \citep[e.g.,][]{chevance20}, which is consistent with this scale.  However, we are wary of ascribing physical meaning to the upper limit in cloud sizes because of the behaviour of the {\sc pycprops} segmentation algorithm. CO emission appears clumpy on the scale of the beam even when the emission forms larger structures, e.g., see the maps in Figure~\ref{fig:ngc3627atlas} and the online supplement. The algorithm selects local maxima, but this selection is limited by the resolution. Thus, the sizes tend to cluster around the minimum values which is determined by the beam scale \citep{pineda_09, leroy_16}, which because of the design of the PHANGS-ALMA observations, is comparable to the molecular disk scale height \citep{PHANGSSurvey}. This effect is so sharp as it acts as another form of censoring that is imposed by the segmentation algorithm rather than the sensitivity. In principle, we can detect larger, smooth structures, but the multiscale structure of the ISM combined with our selection based on local maxima precludes us identifying such objects.


{\bf Dynamical versus Luminous Mass} -- Figure~\ref{fig:scaling_common}(c) shows the correlation between dynamical mass estimated from the virial theorem, $M_{\mathrm{vir}}$, and the mass estimated from CO luminosity, $M_{\mathrm{CO}}$. We plot the line $M_\mathrm{vir} = M_\mathrm{CO}/2$, expected for $\alpha_\mathrm{vir}=1$. The tight correlation is expected given the results in panel (a), which shows that the data cluster around the relationship expected for $\alpha_\mathrm{vir} = 1.5$. The consistency of the two mass estimates provides additional, independent support for our adopted CO-to-H$_2$ conversion factor prescription.
                                                                        
Clouds found in the centres of galaxies appear systematically offset to high values above the reference line \citep[see also][]{sun_20b}. While galactic centres can have systematically higher luminosity per unit molecular gas mass \citep{Sandstrom_2013}, such effects would only exacerbate the discrepancy. Instead, the offset seen in both panel (a) and panel (c) suggests that dynamical effects such as streaming motions could be significant for these clouds \citep{meidt18}. The objects that we identify in galaxy centres therefore seem to be less gravitationally bound than the clouds identified in the main discs of our targets. There may be smaller bound structures within the clouds that we identify in galaxy centres, or these structures may be gravitationally bound when the broader environment is considered.  However, at our working resolution while treating the molecular gas in isolation, these central clouds appear to include at least some gravitationally unbound, high line width material.

We also examined the set of points at $M_\mathrm{CO} \approx 10^6~M_\odot$ with small virial masses. These clouds have highly uncertain deconvolved radii and line widths that likely result in underestimates of the virial mass.  Since the virial mass depends on the second moment of the intensity distributions whereas the luminous mass estimate only depends on the zeroth-moment, the luminous masses are more stable in these cases.

{\bf Size--line width relationship} -- Figure~\ref{fig:scaling_common}(d) shows the size--line width relationship. The blue line shows the fiducial Milky Way relationship with $\sigma_v = \sigma_0 R_\mathrm{3D}^{0.5}$. We adopt the normalization $\sigma_0 = 0.7~\mathrm{km~s}^{-1}$ \citepalias{srby87} as the reference line.

The median trends show significant galaxy-to-galaxy variation between the different cloud populations. Based on the previous three panels, these can be at least partially attributed to variations in cloud surface density, $\Sigma_\mathrm{mol}$, from galaxy to galaxy. The data cluster around the fiducial Milky Way line. The different galaxies all appear roughly consistent with a slope of $\beta = 0.5$, though they would have systematically different values of $\sigma_0$. However, a combination of censoring effects (shaded regions) and coarse resolution create limits on the dynamic range of our study, so we are not in a good position to measure an independent size--line width relationship based only on this analysis. Like the other panels, clouds in the central regions of galaxies have higher velocity dispersions for a given radius compared to those in the outer parts of discs, and again NGC~4826 appears as a distinct system.

{\bf Summary} -- We find general consistency between our catalogued clouds and the basic physical picture formed from studies of the Local Group. In particular, outside the centres of galaxies we find that GMCs show a virial parameter slightly larger than unity. They show a range of surface densities that is broadly consistent with commonly adopted characteristic values for GMCs. We find good agreement between our CO luminosity-based masses and virial mass estimates. Finally, our clouds show the size--line width relations estimated with the binned medians consistent with the Galactic size--line width relation given their sizes.

The nuclear regions of galaxies show significant deviations from all of these trends. In these regions, many of our targets show high surface densities, high line widths, and evidence for less strongly bound gas when considered in isolation. Some of these trends may be explained if CO is overluminous in these regions compared to the discs of galaxies. In contrast, accounting for variations in the CO-to-H$_2$ conversion factor would only make the observed differences between the virial and luminous masses starker.

The plots in Figure~\ref{fig:scaling_common} also show how censoring effects shape the trends exhibited by our data. We only detect clouds with reasonably high $\Sigma_\mathrm{mol}\gtrsim 30~M_\odot~\mathrm{pc}^{-2}$, high $M_\mathrm{CO} \gtrsim 10^{5.5}$~M$_\odot$, and line width $\sigma_v \gtrsim 2.5$~km~s$^{-1}$. The segmentation algorithm also imposes a bias against finding large structures due to its focus on finding compact sources associated with local maxima.

\subsection{Cloud Population Characteristics}
\label{sec:clouddist}

\begin{figure*}
\includegraphics[width=\textwidth]{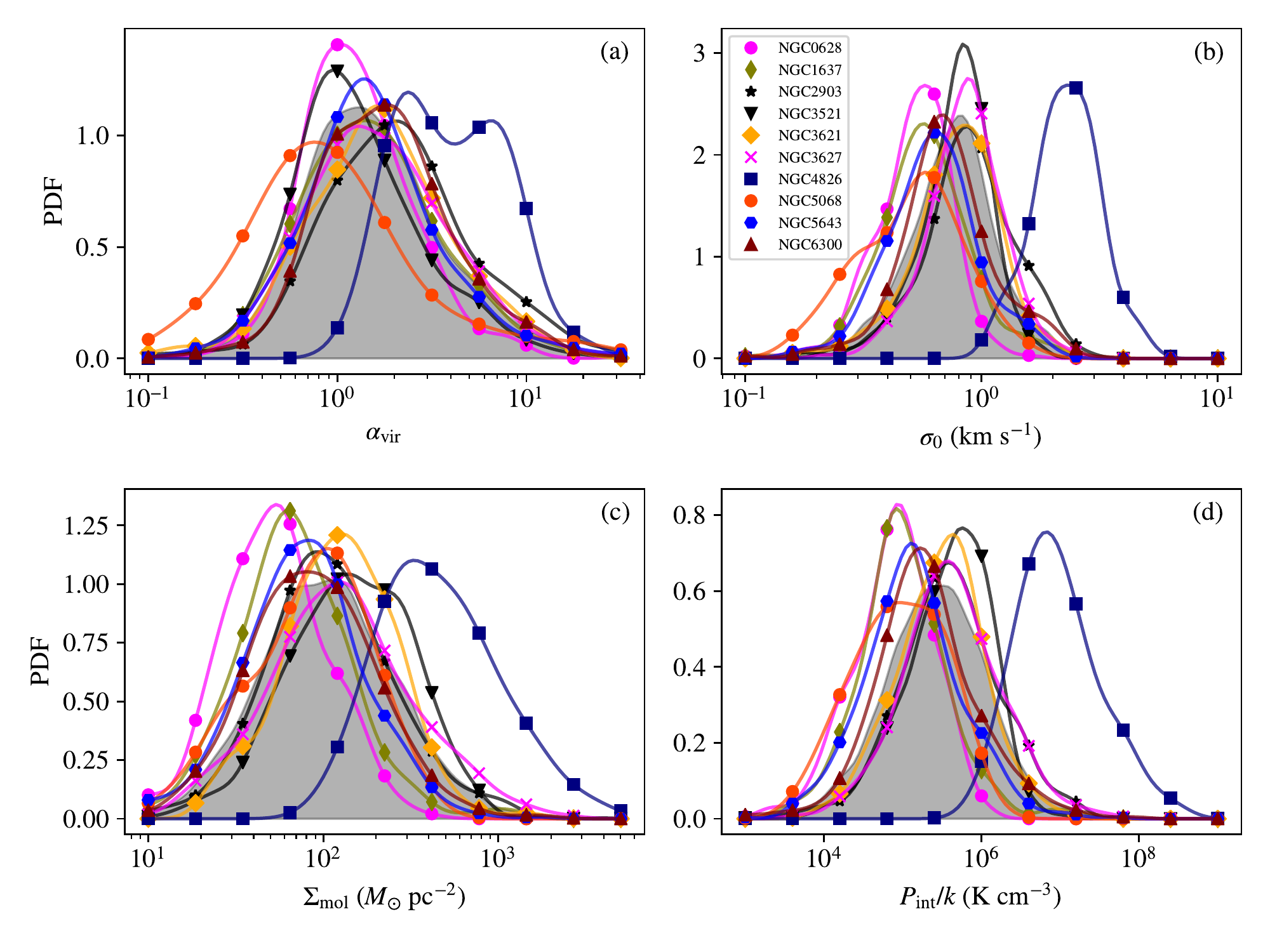}
\vspace*{-25px}
\caption{\label{fig:parhist_common} Probability density functions (PDFs) of parameters describing the GMC population conditioned by galaxy for the homogenized data sets. These PDFs show the distribution of (a) the virial parameter $\alpha_\mathrm{vir}$, (b) the line width on 1~pc scales $\sigma_0$, (c) the mean surface density $\Sigma_{\mathrm{mol}}$, and (d) the internal pressure $P_\mathrm{int}/k$. The PDFs are calculated for the $\log_{10}$ of the quantity in question and represent probability per decade. The grey distribution in the background shows the PDF of all GMCs in the sample.}
\end{figure*}

Our data show overall consistency with the picture of GMCs as approximately self-gravitating objects, though there is a wide range of observed virial parameters spanning nominally bound and unbound systems. In Figure~\ref{fig:parhist_common} and Table~\ref{tab:pardist} we directly quantify the distributions of virial parameter, $\alpha_\mathrm{vir}$, surface density within the FWHM, $\Sigma_\mathrm{mol}$, and normalized line width, $\sigma_0 \equiv \sigma_v/\sqrt{R_\mathrm{3D}}$. These can be viewed as the basic physical properties of the clouds in our catalogues. They also correspond to the normalizations of many commonly adopted GMC scaling relationships, e.g., $M_{\mathrm{vir}} \propto M_{\mathrm{CO}}$, $M_{\mathrm{CO}} \propto R^{2}$, and $\sigma_v \propto R_\mathrm{3D}^{0.5}$ (\citetalias{srby87}, \citetalias{heyer_2009}, \citealt{bolatto08}, \citealt{fukui-kawamura}).

We also characterize the distribution of the mean internal pressure $\langle P_{\mathrm{int}}\rangle = \rho \sigma_v^2$, where $\rho$ is the mass density. Here, we use $R_\mathrm{3D}$ from Equation~\ref{eq:r3d} to calculate the density. Expressing this in terms of the other properties of molecular clouds: 
\begin{equation}
P_{\mathrm{int}} = \frac{3M\sigma_v^2}{8\pi R_\mathrm{3D}^3}~.
\end{equation}
Using the measured density, we also calculate the implied free-fall collapse time,
\begin{equation}
    t_\mathrm{ff} = \sqrt{\frac{3\pi}{32 G \rho}} = \sqrt{\frac{\pi^2 R_\mathrm{3D}^3}{4GM}}~.
\end{equation}

\begin{table*}
\begin{tabular}{l c c c c c}
\hline
Galaxy & $\alpha_{\mathrm{vir}}$ &  $\sigma_0$ & $\Sigma_{\mathrm{mol}}$ & $P_{\mathrm{int}}/k_{\mathrm{B}}$ & $t_\mathrm{ff}$ \\ 
& & (km s$^{-1}$) & ($M_{\odot}~\mathrm{pc}^{-2}$) & ($10^5~\mathrm{cm^{-3}~K}$) 
& (Myr) \\
\hline
NGC~0628 & $1.2^{+1.3}_{-0.5}$ & $0.54^{+0.18}_{-0.18}$ & $ 53^{+ 57}_{- 26}$ & $  1^{+  2}_{-  1}$ & $9.1^{+3.6}_{-2.9}$  \\ 

NGC~1637 & $1.4^{+1.9}_{-0.8}$ & $0.57^{+0.28}_{-0.18}$ & $ 66^{+ 65}_{- 33}$ & $  1^{+  2}_{-  1}$ & $8.0^{+3.3}_{-2.7}$  \\ 

NGC~2903 & $2.0^{+3.3}_{-1.1}$ & $0.90^{+0.52}_{-0.30}$ & $100^{+140}_{- 54}$ & $  4^{+ 13}_{-  3}$ & $6.3^{+3.0}_{-2.3}$  \\ 

NGC~3521 & $1.2^{+1.6}_{-0.5}$ & $0.81^{+0.25}_{-0.25}$ & $140^{+180}_{- 82}$ & $  4^{+  8}_{-  3}$ & $5.4^{+3.0}_{-1.8}$  \\ 

NGC~3621 & $1.6^{+2.1}_{-0.9}$ & $0.81^{+0.32}_{-0.28}$ & $120^{+130}_{- 61}$ & $  3^{+  6}_{-  3}$ & $5.8^{+2.5}_{-2.0}$  \\ 

NGC~3627 & $1.6^{+2.5}_{-0.9}$ & $0.87^{+0.35}_{-0.26}$ & $120^{+210}_{- 74}$ & $  4^{+ 14}_{-  3}$ & $5.9^{+3.4}_{-2.4}$  \\ 

NGC~4826 & $4.1^{+4.0}_{-2.0}$ & $2.30^{+0.73}_{-0.53}$ & $430^{+470}_{-200}$ & $ 84^{+190}_{- 56}$ & $2.9^{+1.4}_{-1.2}$  \\ 

NGC~5068 & $0.8^{+1.0}_{-0.4}$ & $0.54^{+0.24}_{-0.24}$ & $ 86^{+ 64}_{- 53}$ & $  1^{+  2}_{-  1}$ & $6.8^{+4.7}_{-2.0}$  \\ 

NGC~5643 & $1.5^{+1.8}_{-0.8}$ & $0.62^{+0.30}_{-0.22}$ & $ 78^{+ 88}_{- 41}$ & $  1^{+  4}_{-  1}$ & $7.4^{+3.5}_{-2.5}$  \\ 

NGC~6300 & $1.7^{+2.1}_{-0.9}$ & $0.69^{+0.37}_{-0.21}$ & $ 85^{+110}_{- 45}$ & $  2^{+  6}_{-  1}$ & $7.2^{+3.4}_{-2.6}$  \\ 
\hline
Average & $1.5^{+2.1}_{-0.8}$ & $0.77^{+0.35}_{-0.27}$ & $100^{+160}_{- 60}$ & $  3^{+  9}_{-  2}$ & $6.4^{+3.5}_{-2.5}$  \\ 
\hline
Bar & $2.6^{+4.3}_{-1.5}$ & $1.19^{+0.54}_{-0.41}$ & $130^{+230}_{- 82}$ & $  9^{+ 29}_{-  7}$ & $5.8^{+3.9}_{-2.4}$  \\ 
Disc & $1.4^{+1.6}_{-0.7}$ & $0.71^{+0.29}_{-0.25}$ & $ 96^{+120}_{- 53}$ & $  2^{+  6}_{-  2}$ & $6.6^{+3.4}_{-2.3}$  \\ 
\hline
Arm & $1.2^{+1.2}_{-0.6}$ & $0.64^{+0.29}_{-0.21}$ & $ 87^{+130}_{- 49}$ & $  2^{+  6}_{-  1}$ & $6.9^{+3.7}_{-2.6}$  \\ 

Interarm & $1.4^{+1.7}_{-0.7}$ & $0.65^{+0.29}_{-0.24}$ & $ 75^{+ 95}_{- 42}$ & $  1^{+  4}_{-  1}$ & $7.5^{+4.0}_{-2.6}$  \\ 
\hline
\end{tabular}
\caption{\label{tab:pardist} Parameter distributions for the GMC populations in our target galaxies. The Bar, Disc, Arm and Interarm populations show the average parameters for GMCs in these environments (defined in Section~\ref{sec:envmasks} and discussed in Section \ref{sec:environment}).}
\end{table*}

Figure~\ref{fig:parhist_common} shows the probability density functions (PDFs) for (the $\log_{10}$ of) four characteristic properties obtained from the homogenized catalogue. Our targets have relatively uniform distribution of virial parameters with $\langle \alpha_\mathrm{vir} \rangle \approx 1.5$ and a standard deviation of $0.3$~dex. 

Panel (a) reveals small differences between targets. Clouds in NGC~5068 have a lower mean virial parameter than other galaxies whereas clouds in NGC~4826 and NGC~2903 have a higher mean virial parameter.  Since the virial parameter scales inversely with the CO-to-H$_2$ conversion factor, some of these variations may stem from limitations of our prescription for $\alpha_\mathrm{CO}$.  NGC~4826 represents a unique starburst case and shows high $\alpha_\mathrm{vir}$, NGC~2903 and to a lesser extent NGC~3621, NGC~3627, NGC~6300 also show regions with high $\alpha_\mathrm{vir}$.  These galaxies all show prominent stellar bars and the clouds with high virial parameter are preferentially found in those environments.  We revisit the effect of dynamical environment on GMC properties in Section~\ref{sec:environment}.

We note that the outlier systems, NGC~4826 (high $\alpha_\mathrm{vir}$) and NGC~5068 (low $\alpha_\mathrm{vir}$) also exhibit extreme values for the amount of emission identified in the bright emission mask ($f_\mathrm{mask}=0.86$ for NGC~4826 and 0.33 for NGC~5068; see Table~\ref{tab:data}).  This correlation indicates that the cloud properties are reflecting the spatial distribution of bright emission since the mask is defined from the spatial and spectral association of high significance emission. The CO emission morphologies of these two systems are distinct, with NGC~4826 showing a nearly-continuous molecular disk where the catalogued clouds may blend together low-mass neighbouring clouds and contributions from a diffuse molecular medium.  In contrast, the emission from NGC~5068 appears to be isolated into individual clouds.

Figures \ref{fig:parhist_common}(b) and \ref{fig:parhist_common}(c) show the distributions of $\sigma_0$ and $\Sigma_{\mathrm{mol}}$.  Compared to the results for the virial parameter, there is substantially more variation between our targets for these cloud properties. The homogenized data exhibit real galaxy-to-galaxy variations in the cloud surface densities, $\Sigma_\mathrm{mol}$, and scaled line widths, $\sigma_0$. As is illustrated in Figure~\ref{fig:scaling_common}, these two properties tend to scale with one another (see also \citetalias{sun_18}) such that $\alpha_{\mathrm{vir}}$ is of order unity for the GMC populations of all our targets. 

Similar to $\Sigma_\mathrm{mol}$, Figure~\ref{fig:parhist_common}(d) shows a wide range of internal pressures. We find a range of at least 1~dex in $P_\mathrm{int}$ within most galaxies and 4~dex across our whole sample.  \citet{sun_20} show that the internal pressure measurements on cloud scales are closely coupled to (but generally exceed) estimates for the pressure in the diffuse ISM on kpc scales.  The pressure in the diffuse ISM varies by a similar range and can qualitatively explain the wide range of cloud internal pressures observed. \citet{meidt18} further argue that these two observations are necessarily linked since the GMCs we are cataloguing are a continuous part of a self-gravitating ISM in a disc geometry and not dynamically distinct.

In Table~\ref{tab:pardist}, we also report the typical free-fall times for the GMCs in each galaxy. For the whole sample, we find a characteristic value of $6.4$~Myr with a range of $0.2$~dex and relatively little galaxy-to-galaxy variation. Again, NGC~4826 is an outlier, showing smaller typical free-fall time ($2.9$~Myr) as a consequence of the higher density clouds in the starburst environment.

\section{Mass Distributions}
\label{sec:massdist}

The cloud mass distribution describes the structure of the molecular ISM and is theoretically linked to the stability and formation of the molecular medium \citep{reinacampos17}. Unlike the distributions of, e.g., pressure or surface density, we have a prior expectation for the shape of the mass function \citep[e.g.,][]{mspec,Mok19,Mok20} as a distribution that declines strongly with increasing mass, where there are a few high-mass clouds and many low mass clouds.  The function is frequently described as a power-law or exponential distribution. Establishing an analytic form of the mass distribution allows us to fit the parameters of the function using the maximum likelihood approach advocated by \citet{Mok19, Mok20}, which is based on the approach used in galaxy population studies \citep[e.g.,][]{MvdBW}.  These parameters can then be directly compared to theoretical expectations.  In this section, we amend the maximum likelihood formalism to account for the effects of completeness and to compare different models using the Bayes Information Criterion.

In this work, we adopt a Schechter-like, truncated power-law function to describe the probability density function (PDF) for finding a molecular cloud of a given mass $M$:
\begin{equation}
p(M|M_c, \beta) =  C \left(\frac{M}{M_c}\right)^\beta \exp\left(-\frac{M}{M_c}\right)~.
\label{eq:powerlaw}
\end{equation}
Here, $C$ is a normalization coefficient, $M_c$ is the cutoff mass, and $\beta$ is the power-law index.

Incompleteness complicates direct comparison between Equation~\eqref{eq:powerlaw} and the data.  To account for this, we combine Equation~\eqref{eq:powerlaw} with a logistic selection function following Section~\ref{sec:completeness}. The amended PDF is:
\begin{eqnarray}
    p(M | M_c, \beta, c_0, c_1) & = &C \left[1+\exp\left(-c_0 - c_1 \frac{M}{M_c} \right)\right]^{-1} \nonumber\\ && \times \left(\frac{M}{M_c}\right)^\beta \exp\left(-\frac{M}{M_c}\right)~.
    \label{eq:amendedpdf}
\end{eqnarray}
Unlike in Section~\ref{sec:completeness}, we do not consider the effects of the virial parameter or surface density on detection. We leave the parameters of the selection function, $c_0$ and $c_1$, free for optimization along with $M_c$ and $\beta$. This functional form equals a power-law mass distribution between a cutoff at the low-mass end from the completeness term and a cutoff at the high-mass end by the exponential term. Directly fitting for the completeness terms allows for the possibility that the true completeness is higher than we estimate from the false cloud injection tests. Such an effect is expected because our false cloud tests do not account for blending or source extraction in crowded regions. However, in cases where the completeness mass approaches or exceeds the cutoff mass, the power-law slope of the completeness function will be unconstrained.

We also account for uncertainty in the cloud mass determination. Based on the false source injection (Section~\ref{sec:completeness}), we estimated a characteristic $0.2$~dex uncertainty in cloud masses. To reflect this, we adopt the approach from \citet{Mok20} and convolve Equation~\eqref{eq:amendedpdf} with a log-normal distribution with width $\sigma = 0.2$~dex before fitting, to transform the true masses of the clouds $M$ into the observed masses~$M'$.

We find the parameters that maximize the likelihood:
\begin{equation}
\label{eq:likelihood}
\underset{M_c, \beta, c_0, c_1}{\mathrm{argmax}} L = \underset{M_c, \beta, c_0, c_1}{\mathrm{argmax}} \prod_i \frac{p(M_i | M_c, \beta,c_0,c_1)}{\int_{M_\mathrm{min}}^{M_\mathrm{max}}p(M' | M_c,c_0,c_1, \beta) dM'}
\end{equation}
where $L$ is the likelihood. We set $M_\mathrm{min}$ and $M_\mathrm{max}$ to be two orders of magnitude lower and higher than the observed minimum and maximum cloud mass.  We determine the credible ranges of these parameters using the {\sc emcee} sampler \citep{emcee} to sample the likelihood function. For the optimization and sampling, we assume that the prior distribution of $\beta$ is uniformly distributed on the interval $[-5, 1]$ and adopt improper, uninformed priors for $c_0$ and $c_1$, and that $\log_{10}(M_c/M_\odot)$ is uniformly distributed on $[4, 9]$.

\begin{figure*}
\includegraphics[width=0.7\textwidth]{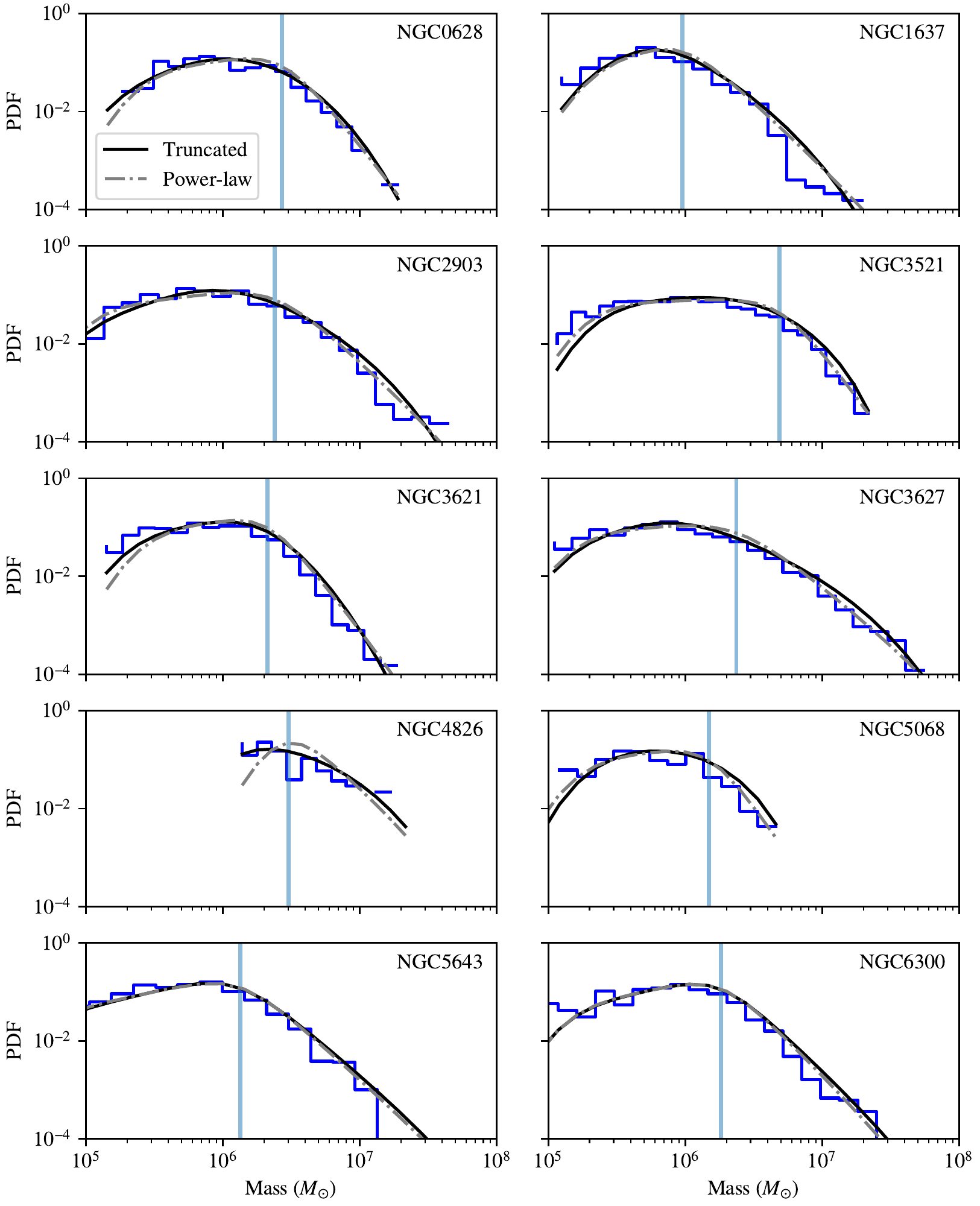}
\caption{\label{fig:massdist} Empirical and best-fitting estimates of the mass distributions for the GMCs in ten galaxies.  An empirical estimate for the mass PDF is shown as a stepped histogram.  The two curves illustrate the best-fitting relationships for a power-law mass distribution.  The curves include the effects of completeness and blending at the low mass end. The Truncated model includes a high-mass exponential truncation to the distribution that is not included in the Power-law distribution.  The vertical line indicates the maximum likelihood estimate of the completeness limit for the pure power-law model. For most galaxies, both functional forms provide good models for the PDF shape.}
\end{figure*}

We also consider a model in which $M_c \to \infty$, i.e., there is no truncation in the mass distribution and the distribution is represented as a pure power-law above the completeness limit. We maximize a likelihood analogous to Equation~\eqref{eq:likelihood} but without $M_c$ and so only using three free parameters. 

For these two models, we report the best-fit power-law index, $\beta$, and cutoff mass, $M_c$, in Table~\ref{tab:mspec}. We show these fits in comparison to an empirical estimate of the PDF determined from binned mass distributions in Figure~\ref{fig:massdist}.

The empirical distributions in Figure~\ref{fig:massdist} show substantial variation in the mass distributions between galaxies.  While the maximum likelihood approach does not fit these binned empirical estimates, they trace the functional form that our fitting approach is trying to match.  In the homogenized data, the maximum mass of clouds reaches much higher scales in some galaxies (NGC~3627) than others (NGC~5068). Some systems show a steep truncation and then a few high-mass clouds (NGC~1637 and NGC~2903).  These high-mass clouds are found in the centres or bars of the galaxies, and this likely reflects the combination of multiple different galactic environments in a single population.  Only a few galaxies show a clear power-law tail at the high-mass end (NGC~5643, NGC~6300) above the completeness limit.

The mass distributions do not show a declining tail over a large dynamic range in mass. There is usually only a factor of 10 between the peak of the PDF and the highest mass cloud.  Thus, our fitting to the mass PDFs shows that both functional forms (i.e., a power-law model with and without a high-mass truncation) are reasonable representations of the mass distribution. The empirical estimates of the censoring effects that arise from crowding in our source extraction provide a good model for the turnover in mass distribution at the low-mass end.  By explicitly including this censoring effect, we can include all of the clouds in the catalogue in our estimates of the mass distributions rather than rejecting the low mass clouds below a limit defined by-eye.  This has the advantage that we can statistically assess the effects of blending but the approach has the drawback that it relies on Equation~\eqref{eq:amendedpdf} being a good model for the observed data.

To assess whether there is a preferred functional form for the mass distribution, we use model selection with the Bayes Information Criterion \citep[BIC,][]{BIC}.  We use 
\begin{equation}
\mathrm{BIC} = k \ln N - 2 \ln (\hat{L})~,
\end{equation}
where $k$ is the number of parameters in the model (3 for a pure power-law, 4 for the truncated power-law), $N$ is the number of clouds in the fit, and $\hat{L}$ is the maximum of the likelihood function. We report the difference in BIC between a pure and power-law model in Table~\ref{tab:mspec} ($\Delta \mathrm{BIC} = \mathrm{BIC_{PL} - BIC_{trun}}$).

\begin{table*}
\begin{tabular}{@{\extracolsep{4pt}}l r r r r r r r r@{}}
\hline
& & \multicolumn{3}{|c|}{Truncated Power-Law} & \multicolumn{2}{|c|}{Pure Power-Law}\\
\cline{3-5} \cline{6-7}\\
NGC~ & $N$ & $\beta_\mathrm{trun}$ & $M_{c}$ & $M_\mathrm{comp}$ & $\beta_\mathrm{PL}$ & $M_\mathrm{comp}$ & $\Delta \mathrm{BIC}$ \\
& & & ($10^6~M_{\odot}$) & ($10^6~M_{\odot}$) & & ($10^6~M_{\odot}$) \\
\hline
0628 & 472 & $ 0.6^{+ 0.7}_{- 0.3}$ & $ 1.1^{+ 0.2}_{- 0.1}$ &  1.0 & $-3.2^{+ 0.2}_{- 0.2}$ &  2.7 & $10.9$\\
1637 & 275 & $-1.4^{+ 0.6}_{- 0.7}$ & $ 2.6^{+ 3.9}_{- 1.2}$ &  0.6 & $-2.5^{+ 0.2}_{- 0.2}$ &  1.0 & $-0.1$\\
2903 & 810 & $-0.9^{+ 0.4}_{- 0.4}$ & $ 4.6^{+ 2.2}_{- 1.2}$ &  0.9 & $-2.5^{+ 0.1}_{- 0.1}$ &  2.4 & $19.4$\\
3521 & 1432 & $-3.4^{+ 1.3}_{- 1.5}$ & $ 0.8^{+ 0.2}_{- 0.1}$ & 26.6 & $-3.4^{+ 0.2}_{- 0.2}$ &  4.9 & $72.5$\\
3621 & 394 & $-1.3^{+ 1.2}_{- 1.7}$ & $ 1.2^{+ 1.8}_{- 0.5}$ &  1.9 & $-3.4^{+ 0.3}_{- 0.2}$ &  2.1 & $-2.5$\\
3627 & 1048 & $-1.0^{+ 0.2}_{- 0.2}$ & $ 9.9^{+ 2.5}_{- 1.9}$ &  0.8 & $-2.2^{+ 0.1}_{- 0.1}$ &  2.4 & $43.9$\\
4826 & 48 & $ 0.1^{+ 0.7}_{- 0.6}$ & $ 3.4^{+ 2.6}_{- 1.1}$ &  1.2 & $-2.3^{+ 0.5}_{- 0.3}$ &  3.0 & $ 6.2$\\
5068 & 74 & $-2.2^{+ 2.6}_{- 2.3}$ & $ 0.2^{+ 0.3}_{- 0.1}$ &  6.9 & $-3.7^{+ 0.9}_{- 0.6}$ &  1.5 & $-0.7$\\
5643 & 695 & $-2.3^{+ 0.1}_{- 0.2}$ & $50^{+200}_{-31}$ &  1.2 & $-2.5^{+ 0.1}_{- 0.1}$ &  1.3 & $-4.4$\\
6300 & 510 & $-2.6^{+ 0.2}_{- 0.3}$ & $55^{+300}_{-40}$ &  1.7 & $-2.7^{+ 0.1}_{- 0.1}$ &  1.8 & $-5.2$\\
\hline
Bar & 552 & $-0.9^{+ 0.2}_{- 0.2}$ & $12.5^{+ 4.7}_{- 2.8}$ &  0.9 & $-2.2^{+ 0.1}_{- 0.1}$ &  3.7 & $22.2$\\
Disc & 2454 & $-1.2^{+ 0.2}_{- 0.2}$ & $ 4.7^{+ 1.0}_{- 0.8}$ &  0.9 & $-2.4^{+ 0.1}_{- 0.1}$ &  1.7 & $66.0$\\
\hline
Arm & 931 & $-1.0^{+ 0.2}_{- 0.2}$ & $ 5.8^{+ 1.7}_{- 1.1}$ &  0.7 & $-2.3^{+ 0.1}_{- 0.1}$ &  1.9 & $39.6$\\
Interarm & 903 & $-0.7^{+ 0.3}_{- 0.3}$ & $ 2.3^{+ 0.7}_{- 0.5}$ &  0.5 & $-2.5^{+ 0.1}_{- 0.1}$ &  1.5 & $31.8$\\
\hline
\end{tabular}
\caption{\label{tab:mspec} Parameters of mass distribution fits including the number of clouds catalogued ($N$), the index of the mass distribution with and without a truncation ($\beta_\mathrm{trun}, \beta_\mathrm{PL}$), the truncation mass ($M_c$), the derived completeness limits for the two fits ($M_\mathrm{comp}$) for the truncated and pure power-law cases), and the difference in the BIC between the two models, $\Delta \mathrm{BIC} = \mathrm{BIC_{PL} - BIC_{trun}}$.}
\end{table*}

Table~\ref{tab:mspec} shows that the truncated and pure power-law fits distribution are usually both good representations of the molecular cloud mass distribution. The BIC indicates a very strong preference \citep[$\Delta \mathrm{BIC}>10$,][]{bayes-factor} for the truncated power-law mass distribution in NGC~0628, NGC~2903, NGC~3521, and NGC~3627. In these cases, the truncation plus relatively shallow power-law mass distributions are reflecting the behaviour of the mass distribution below the cutoff where it is affected by blending.  

The completeness limits shown in Figure~\ref{fig:massdist} come from the empirical fit using Equation~\eqref{eq:mcomp}. These values exceed the limit established by the completeness tests, \complim, by $0.5$ to $1.0$ dex. In the mass distribution fits, our completeness limit is generally near the peak of the PDF. As mentioned above, we expect that this higher fit completeness reflects crowding and blending of GMCs. However, the fit values will only be meaningful as long as our adopted functions are actually good descriptions of the underlying mass distribution.

In NGC~3521, the truncated power-law shows a completeness limit $30$ times larger than the truncation mass $M_c$, a byproduct of the model finding a solution to explain the large mass range where the PDF is flat. The optimization provides a better fit by having the power-law component of the mass distribution represent this flat section with a shallow index.  Thus, we are wary of interpreting this as a physical result and believe that this is reflecting the limitations of our algorithm applied to coarse resolution data.

The pure and truncated power-law models are equivalently good representations of the mass distributions for most systems ($|\Delta \mathrm{BIC}| < 10$).  However, these show steep indices compared to Local Group studies \citep{mspec}, sometimes showing $\beta < -3$.  In the cases with steep indices, the completeness limit is relatively high so there is less than an order of magnitude of dynamic range in the well-sampled portion of the mass spectrum. Observations with higher linear resolution  would allow for a wider dynamic range in mass to be observed without blending effects, providing an improved measurement of the mass distribution.

For NGC~5643 and NGC~6300, the cutoff mass in the truncated mass distribution happens at high mass ($>10^7~M_\odot$), leading to a slightly shallower slope in the power-law component of the mass distribution.  In both of these cases, the BIC does not show strong evidence for preferring a high-mass truncation.  Similar to the results of \citet{Mok19} and \citet{Mok20}, we see no evidence for galaxies that show a clear power-law mass distribution with or without high-mass truncations.  Our derived truncation masses are typically between $10^6$ $M_\odot$ and $10^7~M_\odot$ so that the exponential cutoff in Equation~\eqref{eq:amendedpdf} is describing the entire high-mass tail of the distribution rather than just providing a cutoff at the highest masses above a power-law section. The marginal differences between models seen in most of the BIC results are reflecting that there is not a clear evidence that prefers for exponential versus power-law tails in the mass PDF.

Overall, we see that the homogenized mass distributions change between galaxies, reflecting real differences in the emission distributions among systems.  At the $90$~pc resolution of the homogenized data, blending effects are severe and the resulting data only have a limited dynamic range over which we can fit the mass distribution.  These blending effects will be mitigated by observations with higher physical resolution or in media where the atomic-to-molecular gas transition is defining the structure we observe.  While resolution plays a major role, we are also combining cloud populations over many different dynamical environments within these galaxies and separating the analysis of these regions will produce clearer results (Section~\ref{sec:environment}).

\section{Spatial Variations in GMC Properties}
\label{sec:variations}

Once homogenized, GMC properties are broadly consistent across the different galaxies in this sample.  While we see the same basic trends across systems, the central regions are clearly outliers compared to the remainder of the galaxy.  This distinction motivates for a more careful analysis of whether there are variations with galactocentric radius in the characteristic cloud properties.

\subsection{Variations with Galactocentric Radius}

\begin{figure}
\includegraphics[width=\columnwidth]{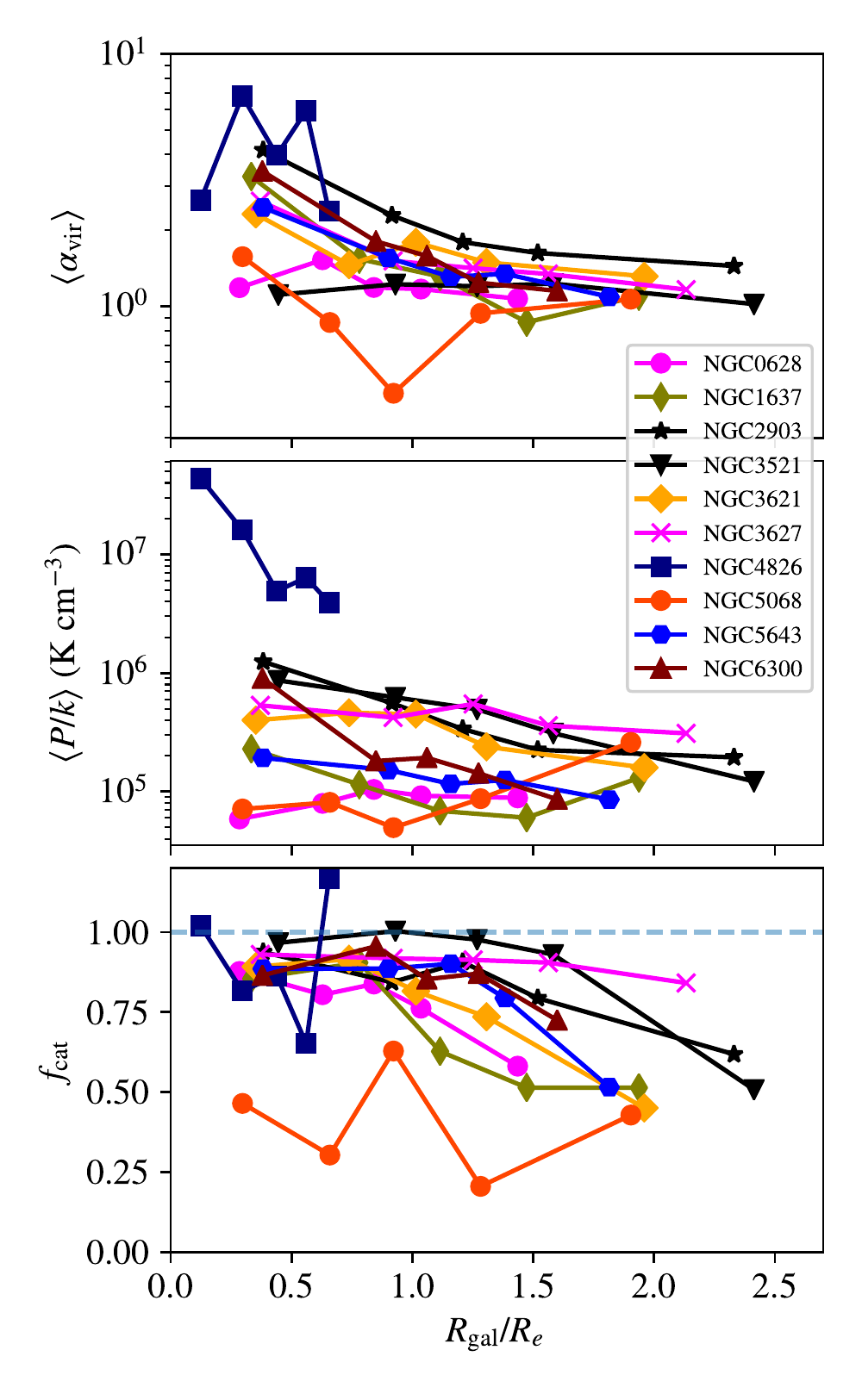}
\vspace*{-25px}
\caption{\label{fig:param_vs_rad} Characteristic GMC parameters plotted as a function of galactocentric radius in 5 even quantiles.  For a given galaxy, the same number of GMCs are found in each bin. The top panel illustrates the median virial parameter for clouds.  The second panel shows the median internal pressure estimated for the molecular clouds.   The bottom panel shows the fraction of flux in the GMC catalogues compared to the total flux calculated by summing all the CO emission in that radial bin. The virial parameter of the molecular clouds is remarkably constant in most galaxies, except for the high values observed in NGC~4826 and the inner parts of barred galaxies. On the contrary, we observe significant galaxy-to-galaxy and radial variations in the other parameters, which trace the organization and brightness of the emission.}
\end{figure}

In Figure~\ref{fig:param_vs_rad}, we show the variations in the cloud populations as a function of galactocentric radius normalized to the effective radius quoted in Table~\ref{tab:sample}. The top two panels show the parameters discussed in \citetalias{sun_18}: the mean virial parameter and the internal pressure. The bottom panel shows the fraction of mass in the surveyed area that is included in the GMC catalogue.

For many galaxies, $\alpha_\mathrm{vir}$ remains relatively flat with radius. In NGC~4826, the virial parameter peaks toward the centre of the galaxy. NGC~1637, NGC~2903, NGC~3627, NGC~3621, and NGC~5643 show milder but still significant increased $\alpha_\mathrm{vir}$ in the inner $0.5{-}1\,R_e$ and all of these are barred systems except NGC~5643. In good agreement with \citet{sun_20b}, our cloud-based analysis suggests that molecular gas with high line widths and apparently high $\alpha_\mathrm{vir}$ values are preferentially associated with the inner parts of strongly barred galaxies. Stellar bars can induce streaming motions and concentrate gas in the inner parts of galaxies.  These streaming motions likely contribute to the observed line widths of the identified GMCs.  At $90$~pc resolution, it is unclear as to whether the high line widths indicate a significantly different dynamical state in the molecular medium (i.e., GMCs are stretched or sheared apart, contributing to an unbound or weakly self-gravitating diffuse molecular medium; e.g., \citealt{meidt18, dale19, kruijssen19orb}) or whether GMCs remain near-virialized entities moving along the local galactic potential with the large line widths arising from clouds on shared orbits overlapping in PPV space or from intersections of gas streamlines (along dust lanes or at bar ends; e.g., \citealt{beuther17,sormani19}).



NGC~4826 is a high density, centrally concentrated molecular gas disc that may be strongly affected by (or even result from) a recent interaction. This high concentration stands out in Figure~\ref{fig:param_vs_rad}, where all of the clouds lie within $R_e$. The high ambient ISM pressure, and large gas and stellar surface densities also make this galaxy a good candidate to host a widespread diffuse molecular gas phase \citep[e.g.,][]{m64-gmcs}. This could also explain the high line widths, internal cloud pressures, and $\alpha_\mathrm{vir}$ values observed in this galaxy.  This is also consistent with the high value of $f_\mathrm{mask}$ in the galaxy, indicating most of the emission is bright and blended together.

As already seen in Section~\ref{sec:clouddist}, our targets show a much wider range in internal pressure than in $\alpha_\mathrm{vir}$. This agrees with the results of \citetalias{sun_18} and \citet{sun_20b} in which pressure represents a main axis of variation of the molecular cloud population in galaxies. 

The internal pressure of the molecular medium typically declines with galactocentric radius, with variations of order a factor of $2$ from the centre to $2R_e$. Not all systems show this trend. Some appear relatively flat (NGC~5068, NGC~5643) or even increase with radius (NGC~0628). Some decline in internal pressure with radius is expected, given that internal pressures in the molecular clouds appear to be coupled to the external pressure in the ISM \citep{hughes13,schruba19,sun_20}. Because ISM pressure declines with galactocentric radius as stellar and gas densities decrease \citep[e.g.,][]{blitzros06}, we expect internal pressure to also decrease with radius. In our sample, there are two galaxies where the internal pressure increases by a factor of $1.5$ with radius NGC~0628 and NGC~~5068.  Both of these systems also have the pressures in their diffuse ISM peaking outside of their centres \citep[e.g.,][and J.~Sun et al.\ in preparation]{herrera20}, which is consistent with the internal pressure being closely linked to the diffuse ISM pressure.


The bottom panel in Figure~\ref{fig:param_vs_rad} shows the fraction of flux catalogued in clouds, $f_\mathrm{cat}$, compared to the total mass implied by the emission in the data cube.  Values of $f_\mathrm{cat}>1$ result from inaccuracies in the extrapolation process. Low values suggest a large fraction of mass is missed from the catalogue, implying such emission is distributed at low surface brightness at 90~pc resolution.  Thus, we can interpret $f_\mathrm{cat}$ as a coarse proxy for the CO emission distribution.

Some galaxies show relatively flat values of $f_\mathrm{cat}$ with $R_\mathrm{gal}$.  For example, NGC~3627 and NGC~5068 consistently show $f_\mathrm{cat}\approx 0.9$ and $0.4$, respectively, implying the ISM emission is not changing in structure radially.  NGC~3521 shows $f_\mathrm{cat}\approx 1$ except in the last bin, which indicates that the emission structure changes in the outer disc. Some systems like NGC~1637 and NGC~5643 show a steadier decline with radius. The low values in NGC~5068, and to a lesser extent NGC~0628, reflect the low overall surface brightness in these galaxies. As discussed in Section~\ref{sec:massdist}, this likely reflects a mixture of CO-to-H$_2$ conversion factor effects and the brightness distribution of the ISM in these galaxies. Most systems show $f_\mathrm{cat}>0.75$ over most of the surveyed regions, showing we are not missing substantial fractions of gas from our analysis provided the extrapolation prescription adopted here is accurate. We note that, at higher resolution, the higher mass clouds will resolve into smaller clouds changing the apparent mass distribution.

Overall, the radial trends show mostly modest variations in cloud internal conditions and emission recovery fraction. In the case of internal pressure, the variations between galaxies are strong and more significant than the variations within galaxies. Signatures of changing dynamical environment and galaxy structure are nevertheless present in our data. In general, we observe high $\alpha_\mathrm{vir}$ values in galaxy centres and declining internal cloud pressure with galactocentric radius.

\subsection{Variations with Dynamical Environment}
\label{sec:environment}

\begin{figure}
\includegraphics[width=0.45\textwidth]{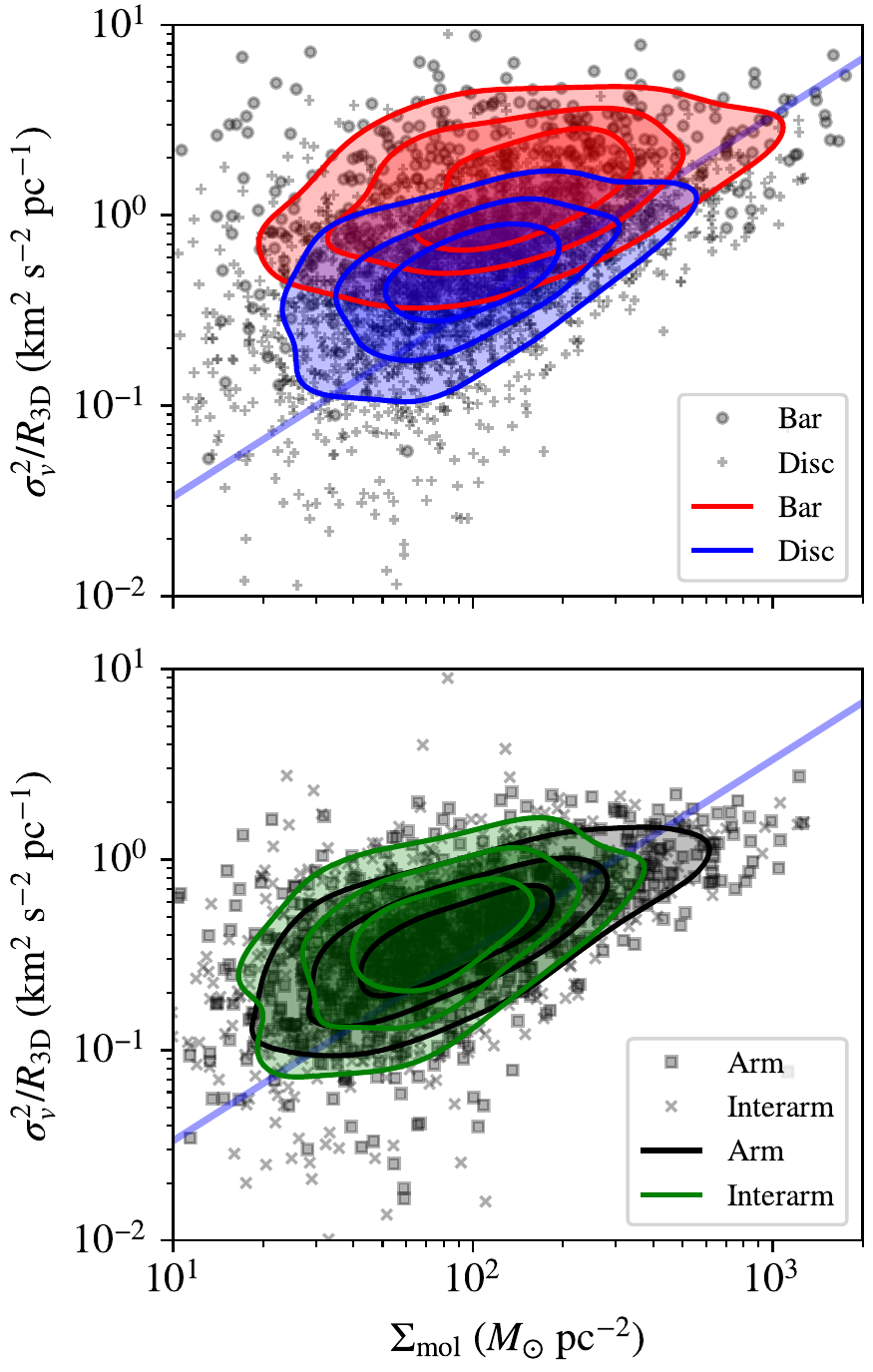}
\caption{\label{fig:environment} The $\Sigma{-}\sigma_0$ relationship showing GMCs grouped by environment. The grey points show the data from the full survey indicating the general distribution of clouds.  The contour levels show the bounds containing 25\%, 50\% and 75\% of the population for clouds found in regions labelled as Spiral Arms, Bars, and Interarm regions (not Bar or Spiral Arm).  The relationship shows the variation between the surface density and turbulent line widths on 1~pc scales.  The solid blue line shows the relationship expected for cloud with virial parameter $\alpha_\mathrm{vir}=1$.}
\end{figure}

\begin{figure*}
    \centering
    \includegraphics[width=0.8\textwidth]{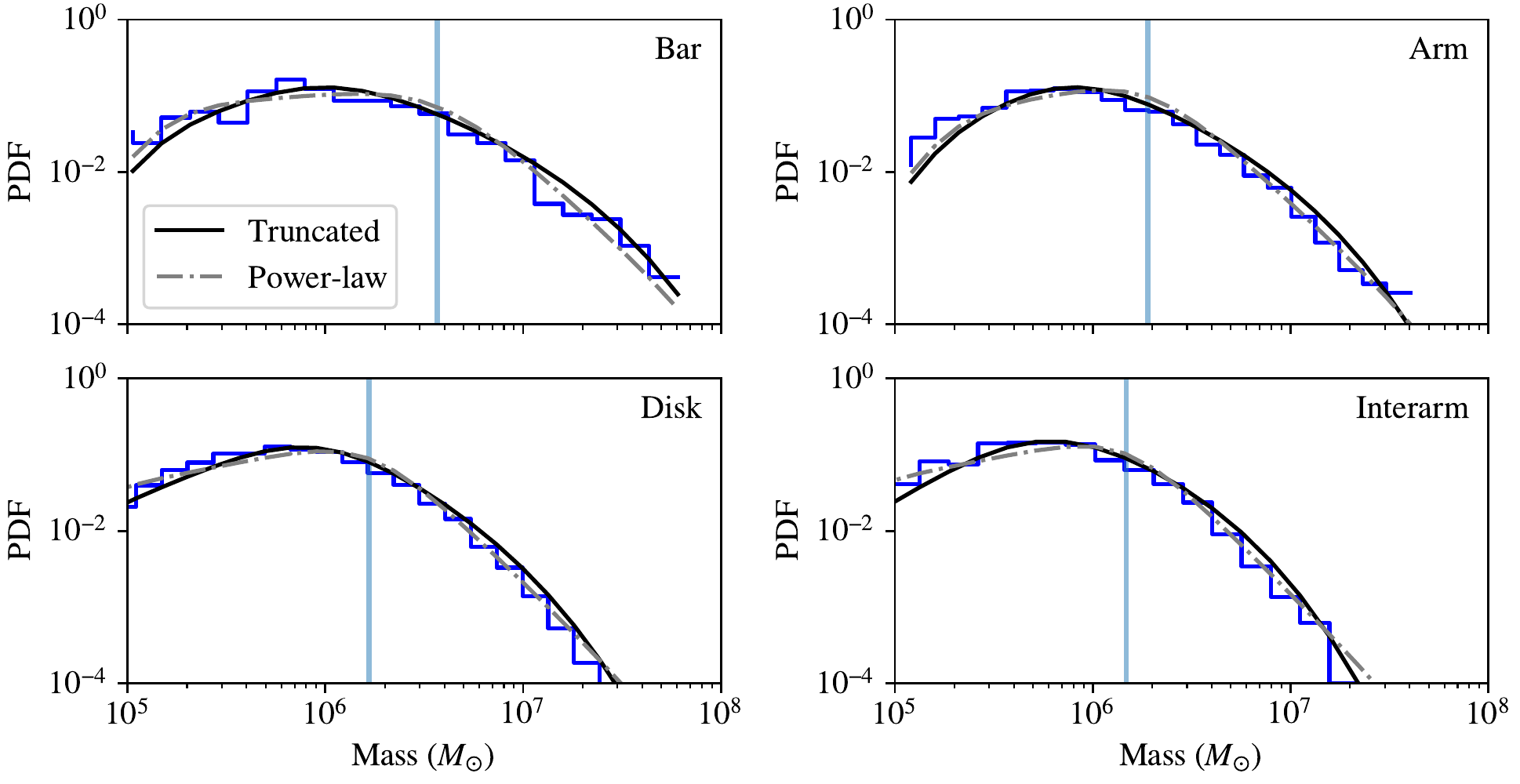}
    \caption{Mass distributions of clouds in different environments for homogenized data.  The stepped measurement of the PDF shows the empirical estimate and the curves show the two different functional models for the mass distribution.  The vertical line indicates the empirically estimated completeness level for our galaxy sample using a pure power-law model. Clouds in spiral arms and bars are well fit by power-law distributions, but interarm clouds tend to have lower masses.}
    \label{fig:massdist_by_label}
\end{figure*}

While there are clear trends in the internal conditions of clouds with $R_\mathrm{gal}$, the visual inspection of the CO maps shows that the changes in molecular emission appear more closely tied to the local dynamical environment. Here we examine how cloud properties and mass distributions depend on dynamical environment in our sample. The resolved maps from PHANGS-ALMA show that the molecular medium is highly organized within the galaxy, and that the location of clouds is strongly influenced by the presence of stellar bars and spiral arms (e.g., see Appendix and \citetalias{sun_18}). In M51, the molecular gas structures in different dynamical environments were also observed to exhibit distinct properties \citep[e.g.,][]{koda09,colombo14}. 

Using the environment labels for individual GMCs (Section~\ref{sec:envmasks}), we establish two comparisons: Bar versus Disc and Arm versus Interarm where the subpopulations of GMCs in those comparisons are mutually exclusive. Figure~\ref{fig:environment} plots the distributions of clouds grouped by environment in $\sigma^2/R_\mathrm{3D}$ versus $\Sigma_\mathrm{mol}$ space and summarizes the cloud properties in each environment in the last four rows of Table~\ref{tab:pardist}. We also show the mass distributions for clouds grouped by environment in Figure~\ref{fig:massdist_by_label}.

\subsubsection{Bar versus Disc Clouds}

In Figure~\ref{fig:environment}, we explore these two comparisons in the context of the  $\sigma_0^2{-}\Sigma_\mathrm{mol}$ plane. Like Figure~\ref{fig:scaling_common}, the solid blue line indicates the relationship expected for clouds in virial equilibrium, i.e., $\alpha_\mathrm{vir}=1$ and $R=R_\mathrm{3D}$. In this parameter space, Bar clouds appear displaced upward relative to the Disc population, showing that they are typically less gravitationally bound with higher line widths at a given surface density. This is reflected in Table~\ref{tab:pardist} where the typical virial parameter for the Bar clouds ($\alpha_\mathrm{vir}=2.6$) is significantly higher than for the Disc clouds ($\alpha_\mathrm{vir}=1.4$). We also see marginally higher surface densities and internal pressure in Bar clouds versus the Disc clouds.  This result shows clear evidence for the influence of dynamical environment on GMC properties \citep[see also the PHANGS-ALMA results in][]{sun_18,sun_20b}.

Considering the Bar versus Disc comparison in Figure~\ref{fig:massdist_by_label}, we see that the mass distribution of Bar clouds extends to higher values and that the completeness mass is higher than for the Disc clouds, reflecting more blended emission in the Bars.  Referring to Table~\ref{tab:mspec}, we see that cutoff mass scale, $M_c$, is a factor of $2.5$ higher, though the power-law components of the truncated mass distributions are indistinguishable ($\beta \approx 2.3$).

Overall, clouds in bars appear less dominated by self-gravity than in galaxy discs. The high masses of Bar clouds with comparable surface densities to the other regions also suggest larger sizes. This in turn suggests that the line widths may be elevated by the strong non-circular orbital motions associated with dynamical features such as stellar bars \citep{meidt18}.  The higher degree of blending and the higher characteristic mass scales implied by the analysis of the mass distribution indicate a molecular ISM with different structure compared to the disc regions of the same galaxies.  These elevated values may also reflect changing mass-to-light ratio in these regions arising from more diffuse emission. 

\subsubsection{Arm versus Interarm Clouds}

Spiral Arm and Interarm GMCs overlap in the $\sigma_0^2{-}\Sigma_\mathrm{mol}$ space shown in Figure~\ref{fig:environment}, with only slight differences in the populations.  Spiral Arm clouds show slightly higher surface densities, comparable turbulent line widths, and slightly lower virial parameters compared to Interarm clouds.  Compared to the Interarm clouds, the Spiral Arm population contains more clouds with $\Sigma_\mathrm{mol} \gtrsim 10^{2.5}$ M$_\odot$~pc$^{-2}$ and low $\sigma^2/R_\mathrm{3D}$, implying low $\alpha_\mathrm{vir}$.  This sub-population comes from the clouds in the spiral arms of NGC~3627, extending outward from the bar ends (Figure~\ref{fig:ngc3627atlas}).

Comparing the Arm and Interarm mass distributions in the bottom panels of Figure~\ref{fig:massdist_by_label}, there are only small differences between the two populations. The truncated power-law model is again preferred over the pure power-law model, but the truncation is at lower mass scales so that it shapes the tail of the distribution. Notably, the truncation mass in the Spiral Arms is a factor of $2.5$ higher than the truncation mass in the Interarm regions, showing that more massive clouds are found in Spiral Arms.  The comparison is meaningful: this population only includes clouds that are outside of the region of influence of the bar and only compares the distributions for galaxies where there are clearly defined arms.   This finding is broadly consistent with the idea that Interarm regions will limit cloud growth as proposed by \citet{meidt15} and \citet{jeffreson18}.  While the pure power-law form is not preferred by the BIC, examining the indices of the distribution shows that the mass distribution of Interarm clouds is marginally steeper than the Spiral Arm clouds. 

The differences between Spiral Arm and Interarm clouds would likely become enhanced if we used a narrower definition of the Spiral Arm region.  The Spiral Arm regions defined in environment masks from M.~Querejeta et al.\ (in preparation) set Spiral Arm regions to have a minimum width of 1~kpc.  If we defined the Spiral Arm region with a more restricted area around the molecular gas ridge line, the Arm--Interarm contrast would likely become stronger.

Our observational results are broadly consistent with \citet{hirota-m83}, who used 45~pc resolution observations of M83 to create a cloud catalogue and examine how cloud properties vary in different environmental regions.  Those authors also found that clouds in the spiral arms tend to have higher masses, lower virial parameters, and shorter free-fall times compared to clouds in the interarm and bar regions of that galaxy.  However, the differences we observe are modest and smaller compared to the differences between the Bar and Disc populations. 

\section{Discussion}
\label{sec:discussion}

By applying a cloud cataloguing approach to the CO emission from our ten PHANGS-ALMA galaxies, we find that the molecular gas in our targets has relatively uniform physical properties, in the sense that the clouds in all targets tend to achieve an approximate balance between their gravitational binding and kinetic energies.  There are modest departures from this balance in certain systems (NGC~4826) and in different environments (e.g., bars).  Here, we compare our conclusions to other structural analysis approaches for the ISM and examine some of the assumptions that underpin our results.

\subsection{Cloud Decomposition and Line of Sight Approaches}

Cloud catalogues have been the historical approach to characterizing the internal conditions of the molecular ISM.  With ALMA, these methods are being extended to CO observations beyond the Local Group. Fundamentally, the cataloguing approach is motivated by the clumpy nature of molecular emission, where there are clearly defined local maxima in the emission.  When applied to high resolution observations of relatively molecule-poor galaxies \citep{schruba17,faesi-ngc300,imara2020} these methods still yield good results that can be directly compared to Galactic and Local Group studies after the careful treatment of systematic effects discussed above (Section~\ref{sec:homogenized}). 

In the ideal case, cloud identification schemes can be trained to identify objects based on physically-motivated criteria, e.g., by tuning the algorithm with prior expectations (see more discussion in \citetalias{cprops}). New approaches are currently being developed to do exactly this \citep{scimes, riener20}, taking advantage of the explosure of high-detail, wide area observations of molecular gas in the Milky Way and the nearest galaxies. When applied to data with sufficient spatial and spectral resolution, these methods offer a powerful, physically-motivated tool to study the nature of the cold ISM.

The cloud identification based approach encounters limitations when the beam reaches scales that are comparable to the extent of emission or when applied to molecule-bright regions where the emission becomes blended together. In these cases, the beam scale becomes the fundamental spatial scale for the decomposition, and many popular cloud identification algorithms return beam-sized objects \citep{pineda_09, leroy_16}. 

These considerations motivated our homogenization of the data as discussed in Section~\ref{sec:homogenized} and shown Section~\ref{sec:NeedForHomog}. With these procedures in place, \textsc{pycrprops} can act as a structure characterization algorithm that allows highly rigorous comparison between different data sets. The objects identified by \textsc{pycrprops} at ${\sim}100$~pc resolution may not be exactly the GMCs that one would pick out with arbitrary spatial resolution, spectral resolution, and sensitivity. However, comparison between cloud catalogues derived from homogenized data will reveal differences in the structure, density, and dynamical state of the molecular ISM between systems.


Faced with the same limitations that motivate our careful homogenization, \citet{leroy_16} advocate for using a non-parametric analysis to characterize the properties of molecular line emission and compare different systems \citep[see also][and N. Brunetti et al.\ in preparation]{hughes13,egusa18,sun_18,sun_20b}. In this approach, one measures the properties of molecular gas along lines of sight (LoS) at one or more fixed spatial scales. The LoS approach has the advantage of simplicity and reproducibility, making it a good tool for comparing to simulations \citep[e.g.,][]{jeffreson20}. It does lack the natural interpretation framework offered by treating each {\sc pycprops} identified object as an individual GMC. Instead, to calculate higher order physical parameters like density and dynamical state, the approach requires an estimate of the spatial extent of emission along the line of sight and a sub-beam structure. Following the framework developed in \citet{sun_18,sun_20,sun_20b}, the simplest such framework is to treat each beam similar to an individual GMC and adopt a model for the thickness of the molecular disc.


The LoS approach has been applied to a subset of PHANGS galaxies in \citetalias{sun_18} and the full PHANGS sample in \citet{sun_20b}.  The cloud cataloguing approach presented here reaches the same general conclusions that were derived from the LoS approach in \citetalias{sun_18}: the molecular ISM shows equipartition between the gravitational binding energy and kinetic energy across a vast range of galactic environment with notable second-order variations due to galaxy centres and linked to stellar bars.  In this current study and both of the LoS studies, the conditions in the molecular ISM show significant variations in the surface density and turbulent gas motions.  These variations combine to manifest as the approximate energy balance, which then implies several orders of magnitude variation in the internal pressure of the clouds. 

This good match could be expected given that the two approaches will have similar behavior when applied to data with marginally resolved clouds and bright, extended emission. In \textsc{pycprops}, the beam size or algorithm places a minimum distance on separations between local maxima. Combined with the sensitivity thresholds for local maxima, the local maxima in blended regions are always a few beam widths apart. Because the spacing and measured sizes of GMCs are tied to the beam size in these regions, the analysis is similar to measurements on a fixed-resolution grid.

Despite these similarities, there are some important differences between the two approaches. The most fundamental difference is that we attempt to deconvolve the effects of the beam from our measurements. The estimated surface density, volume density, and dynamical state thus all represent estimates of the true properties of the identified object. By contrast, the LoS approach treats the spatial scale of the measurements as the averaging scale. In \citet{sun_20,sun_20b}, interpretation of these measurements may consider some sub-beam density distribution but the measurements themselves occur at a fixed spatial scale.

The cloud decomposition approach also captures the spatial structure of emission. The measured cloud sizes and the decomposition into clouds are set by a mixture of the beam size, cloud spacing, and cloud structure. The cloud cataloguing approach captures this through both the relative position of clouds and their measured sizes. When the clustering of local maxima changes, the new spacing will alter the measured sizes of clouds. 

The combination of deconvolution and focusing on local maxima leads to a typical virial parameter of $\alpha_\mathrm{vir}\approx 1.5$ in the cloud case compared to $\alpha_\mathrm{vir}\approx 2$ in \cite{sun_18, sun_20b}, mostly attributable to the smaller spatial sizes of the clouds. The mass distributions (Section~\ref{sec:massdist}) are also not directly comparable. The cloud mass functions result from summing mass over identified objects. The analogous quantity in the LoS case is the brightness or mass surface density distributions at fixed spatial scale. Comparing our results to \citet{sun_18}, we see that these distributions change within galaxies, with the nuclear and bar regions of galaxies having brighter lines of sight and more massive clouds: e.g., compare figure~2 in \citetalias{sun_18} to Figure~\ref{fig:massdist_by_label} in this work. 

Improving our measurement and interpretation of the spatial structure of molecular gas in galaxies represents a clear next step for both approaches. A convolution of the beam, the structure of individual clouds, and the spacing of clouds sets the spacing of local maxima and the resulting properties of clouds in \textsc{pycprops}. These characteristics trace real features of the ISM emission that are not recovered by an LoS-based approach focused on a single scale and one point statistics. However, the interpretation of these measurements is complex, and this algorithmic approach means that its primary utility must come from differential measurements of cloud catalogues carried out under homogeneous analysis as presented here and in A.~Hughes et al.\ (in preparation). Meanwhile, the LoS-based approach will be strengthened by pairing it with two-point statistical descriptions of the molecular ISM. These two-point statistics can be compared to results from cloud cataloguing, to predictions from theoretical models \citep[e.g.,][]{hopkins12}, and to results from simulations.

\subsection{The CO-to-\texorpdfstring{H\boldmath$_2$}{H2} Conversion Factor}

Our estimates of surface density, mass, free-fall time, and pressure all rely on translating the observed CO luminosities into an estimate of the molecular gas mass. The details of this scaling can affect our measured values and impact our interpretation regarding the dynamical state of the ISM.

There is abundant evidence that the CO-to-H$_2$ conversion factor varies as a function of metallicity and dynamical state of the gas and other conditions, like gas temperature, may also play a role \citep[for a review see][]{Bolatto_2013}. We follow \citet{sun_20} in adopting a local prescription that depends on local metallicity, which we estimated via scaling relations adopted from \citet{sanchez14} and \citet{sanchez19}. The metallicity dependence itself is close to the suggested scaling by \citet{accurso2017} and intermediate among the range of scalings that have been adopted in the literature. Nevertheless, we note that \citet{accurso2017} do not derive any resolved (i.e., within galaxies) calibration, and that they include a term related to the galaxy-integrated offset from the star-forming main sequence that \citet{sun_20} do not consider. The comparisons between predictions in \citet{sun_20} suggest that our inferences about the dynamical state of the gas are relatively robust to our choice of prescription. However, it is evident that a robust prescription for $\alpha_\mathrm{CO}$ within galaxies is lacking and that imperfect knowledge of $\alpha_\mathrm{CO}$ represents the limiting factor for many aspects of our analysis.

\citet{sun_20} only attempt to account for metallicity-related variations in $\alpha_\mathrm{CO}$ and thus do not incorporate variations due to changes in the physical properties of the molecular ISM itself. Broader line widths, increased gas temperatures, and variations in opacity can all influence $\alpha_\mathrm{CO}$ and are often cited as the physical drivers for the low ``starburst'' $\alpha_\mathrm{CO}$ seen in merging galaxies and galaxy centres \citep[see][]{Bolatto_2013}. The main sense of these effects is that the true $\alpha_\mathrm{CO}$ will be lower than the one that we adopt in the central regions of galaxies and in stellar bars. In those regions, the wide line widths of the molecular gas at fixed surface density lead to a lower opacity and lower $\alpha_\mathrm{CO}$. 

Accounting for these effects and adjusting to a lower $\alpha_\mathrm{CO}$ in bars and nuclear regions would amplify our conclusion that the gas in these regions has larger virial parameters compared to spiral arms and interarm regions. On the other hand, another result from our analysis is that the mass distribution of clouds in bar regions extends to higher masses compared to the populations in other regions. This difference would be less pronounced -- or perhaps even erased -- using a lower $\alpha_\mathrm{CO}$ for the Bar environment.

Finally, we did not account for a variable $\alpha_{\rm CO}$ in our data homogenization. In principle, the homogenization could be implemented such that the input data cubes have matching sensitivity to molecular mass as opposed to uniform sensitivity to CO intensity, as is in the current implementation. For comparisons where mass or a mass-related quantity is fundamental, this may be an issue for our lower mass galaxies, including NGC~1637, NGC~3511, and especially NGC~5068, which also represents our case with the lowest fraction of catalogued CO emission. In practice, homogenization in terms of mass sensitivity would enforce a much worse common noise floor, significantly degrading the quality of the input data. We consider that existing $\alpha_\mathrm{CO}$ prescriptions are too unreliable relative to the cost in signal-to-noise to justify this approach at present for the PHANGS-ALMA data, but future work with larger samples and more sensitive data may find that uniform mass sensitivity is a preferable basis for comparative analysis. 

\subsection{Evaluating GMCs in Simulations}

The properties of molecular cloud populations often serve as a point of comparison with numerical simulations.  Just as observations have progressively improved to allow cloud-scale imaging of many galaxies, so too have numerical simulations advanced, allowing full-galaxy simulations with resolution that can identify molecular clouds and measure their resolved properties \citep[e.g., ][]{tasker09, renaud13, dobbs13,benincasa13}.  These and subsequent works find concordance between the simulated and observed cloud properties and interpret the agreement as evidence for the validity of the simulation physics.  The comparisons include both the Larson scaling relationships and the mass distributions of the molecular clouds.  \citet{guszejnov20} further showed that, in their simulation set, these GMC scaling relationships and distributions of properties exhibited little variation for an isolated galaxy over cosmic time. More recent work has also proposed using variations in the scaling relationships and mass distributions to critically evaluate the star formation models in the star-forming clouds \citep{fujimoto19, li20, grisdale20}.

In this work, as well as \citetalias{sun_18} and \citet{sun_20b}, we have identified variations in cloud properties with environment that can be used as additional points of comparison with simulations.  In uniformly characterizing GMCs across several galaxies, we find a range of cloud properties and modest changes in the scaling relationships that can give context to any discrepancies between simulations and the canonical cloud population. Several simulation studies have already explored the role of dynamical environment on the properties of molecular clouds finding similar variations to what we observe in bars and spiral arms (viz., increased mass scales and changing virial parameters in bars and spiral arms; \citealt{fujimoto14, pettitt18, pettitt20}). Other studies find that simulated clouds outside of bars show little sensitivity to the dynamical environment of the disc \citep{duartecabral16, jeffreson20, tress20b}.

Such comparisons between simulations and observations must recognize the limited information available in the observational domain and the critical role that observational effects play at shaping the property and mass distributions that are extracted from observations (see Section~\ref{sec:NeedForHomog}).  While our observational work has made our best effort to measure the ``true'' properties of GMCs in other galaxies, the most rigorous comparisons between simulations and observations are carried out by generating mock observations \citep{smith14, pan15, khoperskov16, duartecabral16,izquierdo20} and analyzing those mock observations with analysis algorithms from the observers' toolkit such as {\sc pycprops}, {\sc scimes} \citep{scimes}, or the line-of-sight approach \citep{leroy_16}. \citet{dobbs19} adopt this full analysis arc to identify which star formation and feedback approaches in their simulations yielded best matches to CO observations of M33, using the Larson scaling relations and mass distributions as points of comparison. For comparison to the PHANGS observations, we recommend generating mock observations of \cotwo\ as well as homogenizing the data to a common resolution and noise level (in this work, 75~mK in a 2.5 km~s$^{-1}$ channel and 90 pc resolution element). Mock GMCs extracted with {\sc pycprops} can then be directly compared to this work (e.g., Tables \ref{tab:pardist} and \ref{tab:mspec}).

\section{Summary}

We present improved methods for identifying and characterizing molecular clouds based on high resolution CO observations of external galaxies. These methods update the \textsc{cprops} algorithm \citep{cprops} to be better optimized for the ALMA era. We implement them in a publicly available code, \textsc{pycprops} (URL given in footnote~\ref{footnote:pycprops}). The methods largely follow \citetalias{cprops} with the following notable improvements: 

\begin{enumerate}
\item Before any processing, we convolve all data sets to a common physical resolution and add artificial noise to bring all data to a common noise level. Following \citet{hughes13}, this step is essential for rigorous comparative analysis.

\item We carry out rigorous completeness tests by placing mock clouds in the signal-free regions of the data cubes.  We find that our application of \textsc{pycprops} is sensitive to ``classical'' GMCs with $\Sigma_\mathrm{mol} \approx 10^2~M_\odot~\mathrm{pc}^{-2}$ and $\alpha_\mathrm{vir} \approx 2$.  We find a 50\% completeness limit of \complim\ in our sample.  Low surface density or unbound gas will not be readily detected by our approach.

\item We shift the default segmentation algorithm to a seeded watershed with a preference for compact structures. The key change from \textsc{cprops} is that we decompose all significant emission into clouds. This makes the final cloud properties less sensitive to clipping corrections (``extrapolation''), but risks artifically assigning diffuse CO emission into clouds.

\item We make minor improvements to the treatments of cloud size, the line spread function, and beam deconvolution compared to \citetalias{cprops}.
\end{enumerate}

The homogenization and completeness tests are by far the most important changes, and are prerequisities for a rigorous comparative analysis of the large sets of new CO maps of nearby galaxies now being produced by ALMA. 

We apply our methods to catalogue molecular clouds based on the $^{12}$\cotwo\ emission for ten galaxies from PHANGS-ALMA. These targets were selected to capture a diverse set of morphology, and are close enough such that the PHANGS-ALMA data access a common linear scale of 90~pc. After homogenizing the data, we characterize \numcloudhomog\ individual clouds with $\mathrm{S/N}>6$ and robustly determine their basic physical properties including the effects of a spatially-varying CO-to-H$_2$ conversion factor. Our key science results are:

\begin{enumerate}
\setcounter{enumi}{4}
\item The clouds in our targets show $\langle \alpha_\mathrm{vir}\rangle = 1.5$ with a scatter of $0.4$~dex but only weak galaxy-to-galaxy variations (but see below). The typical cloud surface density, $\Sigma_\mathrm{mol}$, and normalized velocity dispersions, $\sigma_0 \equiv \sigma_v / \sqrt{R_\mathrm{3D}}$, vary among galaxies, but scale with one another in the sense expected for approximately virialized clouds. 

\item The above result implies good agreement between mass estimates from the virial method and from the CO emission using a spatially varying CO-to-H$_2$ conversion factor. This provides an independent validation of our adopted $\alpha_\mathrm{CO}$ prescription.

\item The typical cloud identified at 90~pc resolution in our targets has molecular gas mass surface density $\Sigma_\mathrm{mol} \approx 100$~M$_\odot$~pc$^{-2}$, virial parameter $\alpha_\mathrm{vir} \approx 1.5$, internal pressure $P_\mathrm{int}/k_{\rm B} \approx 3\! \times\! 10^5$ cm$^{-3}$~K, and gravitational free-fall time $t_\mathrm{ff} \approx 6.4$~Myr.  Again, there are  variations in these mean parameters between galaxies that are comparable to the range of GMC parameters within an individual galaxy, with cloud populations in NGC~4826 and NGC~5068 showing the largest shifts from the typical values.

\item We observe variations in surface density, line width, and cloud internal pressure from galaxy to galaxy. The $\pm 1\sigma$ of these parameters are roughly a factor of four for the surface density and an order of magnitude for the internal pressure. 

\item The mass distributions of clouds vary within and among galaxies.  We fit mass spectra to the cloud catalogues of individual galaxies and different galactic environments accounting for completeness. The fits imply that blending is a significant factor in our ability to recover clouds. In most cases, a pure power-law and a truncated Schechter-like function work equally well. The pure power-law fits show power law indices $\beta \lesssim -2$, with the shallower cases consistent with blending of distinct environments.

\item We often find high virial parameters, high line widths, and high surface densities in the centres of galaxies. This may partially reflect CO-to-H$_2$ conversion factor effects in which molecular gas tends to be overluminous in CO in galaxy centres. However, the broad line widths that we observe suggest the presence of unresolved streaming motions or diffuse gas in these systems. 

\item Clouds in Bar regions, including the centres of barred galaxies, show high line width at fixed surface density compared to Spiral and Interarm regions. Meanwhile clouds in Spiral regions show high surface density, lower virial parameter, and shorter gravitational free-fall time compared to Interarm clouds.

\end{enumerate}

Overall our scientific results are in good agreement with results from the non-parametric, fixed-scale analysis of \citet{sun_18,sun_20b} also using the PHANGS-ALMA data. The improved methods described in this paper will be applied to the full PHANGS-ALMA sample with catalogues presented in A.~Hughes et al. (in preparation).

\medskip

This work was carried out as part of the PHANGS collaboration.  We are grateful for the comments of an anonymous referee whose review improved this work.  This paper makes use of the following ALMA data: ADS/JAO.ALMA\#2015.1.00956.S and ADS/JAO.ALMA\#2017.1.00886.S ALMA is a partnership of ESO (representing its member states), NSF (USA) and NINS (Japan), together with NRC (Canada) , NSC and ASIAA (Taiwan), and KASI (Republic of Korea), in cooperation with the Republic of Chile. The Joint ALMA Observatory is operated by ESO, AUI/NRAO and NAOJ.  The National Radio Astronomy Observatory is a facility of the National Science Foundation operated under cooperative agreement by Associated Universities, Inc. This work made use of the {\sc astropy} \citep{Astropy}, {\sc spectral-cube}, {\sc radio-beam} and {\sc astrodendro} packages for Python and we gratefully acknowledge the ongoing efforts of our community's developers. This work also made use of the NASA Astrophysics Data System and the NASA Extragalactic Database.  ER acknowledges the support of the Natural Sciences and Engineering Research Council of Canada (NSERC), funding reference number RGPIN-2017-03987 and a resource allocation from Compute Canada. AH was supported by the Programme National Cosmology et Galaxies (PNCG) of CNRS/INSU with INP and IN2P3, co-funded by CEA and CNES, and by the Programme National ``Physique et Chimie du Milieu Interstellaire'' (PCMI) of CNRS/INSU with INC/INP co-funded by CEA and CNES. The work of AKL and JS is partially supported by the National Science Foundation (NSF) under Grants No. 1615105, 1615109, and 1653300, as well as by the National Aeronautics and Space Administration (NASA) under ADAP grants NNX16AF48G and NNX17AF39G. AU acknowledges support from the Spanish funding grants AYA2016-79006-P (MINECO/FEDER), PGC2018-094671-B-I00 (MCIU/AEI/FEDER), and PID2019-108765GB-I00 (MICINN). DL, TW, TS, and ES acknowledge funding from the European Research Council (ERC) under the European Union's Horizon 2020 research and innovation programme (grant agreement No. 694343). FB and IB acknowledge funding from the European Union’s Horizon 2020 research and innovation programme (grant agreement No 726384/EMPIRE). MC and JMDK gratefully acknowledge funding from the Deutsche Forschungsgemeinschaft (DFG) in the form of an Emmy Noether Research Group (grant number KR4801/1-1) and the DFG Sachbeihilfe (grant number KR4801/2-1).  JMDK is also supported by the European Research Council (ERC) under the European Union’s Horizon 2020 research and innovation programme via the ERC Starting Grant MUSTANG (grant agreement number 714907).
RSK and SCOG acknowledge financial support from the German Research Foundation (DFG) via the collaborative research center (SFB 881, Project-ID 138713538) ``The Milky Way System'' (subprojects A1, B1, B2, and B8). They also acknowledge funding from the Heidelberg Cluster of Excellence STRUCTURES in the framework of Germany’s Excellence Strategy (grant EXC-2181/1 - 390900948) and from the European Research Council via the ERC Synergy Grant ECOGAL (grant 855130) and the ERC Advanced Grant STARLIGHT (grant 339177).

{\it Data Availability:} The data underlying this article are available in the article and in its online supplementary material. Supporting data sets from in preparation articles will be publicly released on acceptance of those manuscripts.

\bibliography{main.bib}

\begin{thebibliography}{}
\makeatletter
\relax
\def\mn@urlcharsother{\let\do\@makeother \do\$\do\&\do\#\do\^\do\_\do\%\do\~}
\def\mn@doi{\begingroup\mn@urlcharsother \@ifnextchar [ {\mn@doi@}
  {\mn@doi@[]}}
\def\mn@doi@[#1]#2{\def\@tempa{#1}\ifx\@tempa\@empty \href
  {http://dx.doi.org/#2} {doi:#2}\else \href {http://dx.doi.org/#2} {#1}\fi
  \endgroup}
\def\mn@eprint#1#2{\mn@eprint@#1:#2::\@nil}
\def\mn@eprint@arXiv#1{\href {http://arxiv.org/abs/#1} {{\tt arXiv:#1}}}
\def\mn@eprint@dblp#1{\href {http://dblp.uni-trier.de/rec/bibtex/#1.xml}
  {dblp:#1}}
\def\mn@eprint@#1:#2:#3:#4\@nil{\def\@tempa {#1}\def\@tempb {#2}\def\@tempc
  {#3}\ifx \@tempc \@empty \let \@tempc \@tempb \let \@tempb \@tempa \fi \ifx
  \@tempb \@empty \def\@tempb {arXiv}\fi \@ifundefined
  {mn@eprint@\@tempb}{\@tempb:\@tempc}{\expandafter \expandafter \csname
  mn@eprint@\@tempb\endcsname \expandafter{\@tempc}}}

\bibitem[\protect\citeauthoryear{{Accurso} et~al.,}{{Accurso}
  et~al.}{2017}]{accurso2017}
{Accurso} G.,  et~al., 2017, \mn@doi [\mnras] {10.1093/mnras/stx1556}, \href
  {https://ui.adsabs.harvard.edu/abs/2017MNRAS.470.4750A} {470, 4750}

\bibitem[\protect\citeauthoryear{{Anand} et~al.,}{{Anand}
  et~al.}{2020}]{anand20}
{Anand} G.~S.,  et~al., 2020, arXiv e-prints, \href
  {https://ui.adsabs.harvard.edu/abs/2020arXiv201200757A} {p. arXiv:2012.00757}

\bibitem[\protect\citeauthoryear{{Astropy Collaboration} et~al.,}{{Astropy
  Collaboration} et~al.}{2013}]{Astropy}
{Astropy Collaboration} et~al., 2013, \mn@doi [\aap]
  {10.1051/0004-6361/201322068}, \href
  {http://adsabs.harvard.edu/abs/2013A%26A...558A..33A} {558, A33}

\bibitem[\protect\citeauthoryear{{Benincasa}, {Tasker}, {Pudritz}  \&
  {Wadsley}}{{Benincasa} et~al.}{2013}]{benincasa13}
{Benincasa} S.~M.,  {Tasker} E.~J.,  {Pudritz} R.~E.,   {Wadsley} J.,  2013,
  \mn@doi [\apj] {10.1088/0004-637X/776/1/23}, \href
  {https://ui.adsabs.harvard.edu/abs/2013ApJ...776...23B} {776, 23}

\bibitem[\protect\citeauthoryear{Bertoldi \& McKee}{Bertoldi \&
  McKee}{1992}]{bertoldi-mckee}
Bertoldi F.,  McKee C.~F.,  1992, ApJ, 395, 140

\bibitem[\protect\citeauthoryear{{Beuther}, {Meidt}, {Schinnerer}, {Paladino}
  \& {Leroy}}{{Beuther} et~al.}{2017}]{beuther17}
{Beuther} H.,  {Meidt} S.,  {Schinnerer} E.,  {Paladino} R.,   {Leroy} A.,
  2017, \mn@doi [\aap] {10.1051/0004-6361/201526749}, \href
  {https://ui.adsabs.harvard.edu/abs/2017A&A...597A..85B} {597, A85}

\bibitem[\protect\citeauthoryear{Blitz}{Blitz}{1993}]{psp3}
Blitz L.,  1993, in Protostars and Planets III. pp 125--161

\bibitem[\protect\citeauthoryear{Blitz \& Rosolowsky}{Blitz \&
  Rosolowsky}{2006}]{blitzros06}
Blitz L.,  Rosolowsky E.,  2006, ApJ, 650, 933

\bibitem[\protect\citeauthoryear{Blitz, Fukui, Kawamura, Leroy, Mizuno  \&
  Rosolowsky}{Blitz et~al.}{2007}]{psp5}
Blitz L.,  Fukui Y.,  Kawamura A.,  Leroy A.,  Mizuno N.,   Rosolowsky E.,
  2007, Protostars and Planets V, pp 81--96

\bibitem[\protect\citeauthoryear{Bolatto, Leroy, Rosolowsky, Walter  \&
  Blitz}{Bolatto et~al.}{2008}]{bolatto08}
Bolatto A.~D.,  Leroy A.~K.,  Rosolowsky E.,  Walter F.,   Blitz L.,  2008,
  ApJ, 686, 948

\bibitem[\protect\citeauthoryear{Bolatto, Wolfire  \& Leroy}{Bolatto
  et~al.}{2013}]{Bolatto_2013}
Bolatto A.~D.,  Wolfire M.,   Leroy A.~K.,  2013, \mn@doi [Annual Review of
  Astronomy and Astrophysics] {10.1146/annurev-astro-082812-140944}, 51, 207

\bibitem[\protect\citeauthoryear{{Braine}, {Gratier}, {Contreras}, {Schuster}
  \& {Brouillet}}{{Braine} et~al.}{2012}]{braine12}
{Braine} J.,  {Gratier} P.,  {Contreras} Y.,  {Schuster} K.~F.,   {Brouillet}
  N.,  2012, \mn@doi [\aap] {10.1051/0004-6361/201220093}, \href
  {https://ui.adsabs.harvard.edu/abs/2012A&A...548A..52B} {548, A52}

\bibitem[\protect\citeauthoryear{{Braine}, {Rosolowsky}, {Gratier}, {Corbelli}
  \& {Schuster}}{{Braine} et~al.}{2018}]{braine18}
{Braine} J.,  {Rosolowsky} E.,  {Gratier} P.,  {Corbelli} E.,   {Schuster}
  K.~F.,  2018, \mn@doi [\aap] {10.1051/0004-6361/201732405}, \href
  {https://ui.adsabs.harvard.edu/abs/2018A&A...612A..51B} {612, A51}

\bibitem[\protect\citeauthoryear{Braun, Walterbos, Kennicutt  \& Tacconi}{Braun
  et~al.}{1994}]{counterrot_m64}
Braun R.,  Walterbos R.~A.~M.,  Kennicutt R.~C.,   Tacconi L.~J.,  1994, ApJ,
  420, 558

\bibitem[\protect\citeauthoryear{{Cald{\'u}-Primo} \&
  {Schruba}}{{Cald{\'u}-Primo} \& {Schruba}}{2016}]{calduprimo16}
{Cald{\'u}-Primo} A.,  {Schruba} A.,  2016, \mn@doi [\aj]
  {10.3847/0004-6256/151/2/34}, \href
  {https://ui.adsabs.harvard.edu/abs/2016AJ....151...34C} {151, 34}

\bibitem[\protect\citeauthoryear{{Chevance} et~al.,}{{Chevance}
  et~al.}{2020}]{chevance20}
{Chevance} M.,  et~al., 2020, \mn@doi [\mnras] {10.1093/mnras/stz3525}, \href
  {https://ui.adsabs.harvard.edu/abs/2020MNRAS.493.2872C} {493, 2872}

\bibitem[\protect\citeauthoryear{Cohen, Dame, Garay, Montani, Rubio  \&
  Thaddeus}{Cohen et~al.}{1988}]{cohen88}
Cohen R.~S.,  Dame T.~M.,  Garay G.,  Montani J.,  Rubio M.,   Thaddeus P.,
  1988, ApJL, 331, L95

\bibitem[\protect\citeauthoryear{{Colombo} et~al.,}{{Colombo}
  et~al.}{2014}]{colombo14}
{Colombo} D.,  et~al., 2014, \mn@doi [\apj] {10.1088/0004-637X/784/1/3}, \href
  {https://ui.adsabs.harvard.edu/abs/2014ApJ...784....3C} {784, 3}

\bibitem[\protect\citeauthoryear{{Colombo}, {Rosolowsky}, {Ginsburg},
  {Duarte-Cabral}  \& {Hughes}}{{Colombo} et~al.}{2015}]{scimes}
{Colombo} D.,  {Rosolowsky} E.,  {Ginsburg} A.,  {Duarte-Cabral} A.,   {Hughes}
  A.,  2015, \mn@doi [\mnras] {10.1093/mnras/stv2063}, \href
  {https://ui.adsabs.harvard.edu/abs/2015MNRAS.454.2067C} {454, 2067}

\bibitem[\protect\citeauthoryear{{Colombo} et~al.,}{{Colombo}
  et~al.}{2019}]{colombo19}
{Colombo} D.,  et~al., 2019, \mn@doi [\mnras] {10.1093/mnras/sty3283}, \href
  {https://ui.adsabs.harvard.edu/abs/2019MNRAS.483.4291C} {483, 4291}

\bibitem[\protect\citeauthoryear{{Dale}, {Kruijssen}  \& {Longmore}}{{Dale}
  et~al.}{2019}]{dale19}
{Dale} J.~E.,  {Kruijssen} J.~M.~D.,   {Longmore} S.~N.,  2019, \mn@doi
  [\mnras] {10.1093/mnras/stz888}, \href
  {https://ui.adsabs.harvard.edu/abs/2019MNRAS.486.3307D} {486, 3307}

\bibitem[\protect\citeauthoryear{{Davis}, {Bureau}, {Onishi}, {Cappellari},
  {Iguchi}  \& {Sarzi}}{{Davis} et~al.}{2017}]{wisdom-gmcs}
{Davis} T.~A.,  {Bureau} M.,  {Onishi} K.,  {Cappellari} M.,  {Iguchi} S.,
  {Sarzi} M.,  2017, \mn@doi [\mnras] {10.1093/mnras/stw3217}, \href
  {https://ui.adsabs.harvard.edu/abs/2017MNRAS.468.4675D} {468, 4675}

\bibitem[\protect\citeauthoryear{{Dobbs} \& {Pringle}}{{Dobbs} \&
  {Pringle}}{2013}]{dobbs13}
{Dobbs} C.~L.,  {Pringle} J.~E.,  2013, \mn@doi [\mnras]
  {10.1093/mnras/stt508}, \href
  {https://ui.adsabs.harvard.edu/abs/2013MNRAS.432..653D} {432, 653}

\bibitem[\protect\citeauthoryear{{Dobbs}, {Rosolowsky}, {Pettitt}, {Braine},
  {Corbelli}  \& {Sun}}{{Dobbs} et~al.}{2019}]{dobbs19}
{Dobbs} C.~L.,  {Rosolowsky} E.,  {Pettitt} A.~R.,  {Braine} J.,  {Corbelli}
  E.,   {Sun} J.,  2019, \mn@doi [\mnras] {10.1093/mnras/stz674}, \href
  {https://ui.adsabs.harvard.edu/abs/2019MNRAS.485.4997D} {485, 4997}

\bibitem[\protect\citeauthoryear{{Donovan Meyer} et~al.,}{{Donovan Meyer}
  et~al.}{2013}]{canon-gmcs}
{Donovan Meyer} J.,  et~al., 2013, \mn@doi [\apj]
  {10.1088/0004-637X/772/2/107}, \href
  {https://ui.adsabs.harvard.edu/abs/2013ApJ...772..107D} {772, 107}

\bibitem[\protect\citeauthoryear{{Duarte-Cabral} \& {Dobbs}}{{Duarte-Cabral} \&
  {Dobbs}}{2016}]{duartecabral16}
{Duarte-Cabral} A.,  {Dobbs} C.~L.,  2016, \mn@doi [\mnras]
  {10.1093/mnras/stw469}, \href
  {https://ui.adsabs.harvard.edu/abs/2016MNRAS.458.3667D} {458, 3667}

\bibitem[\protect\citeauthoryear{{Egusa}, {Hirota}, {Baba}  \&
  {Muraoka}}{{Egusa} et~al.}{2018}]{egusa18}
{Egusa} F.,  {Hirota} A.,  {Baba} J.,   {Muraoka} K.,  2018, \mn@doi [\apj]
  {10.3847/1538-4357/aaa76d}, \href
  {https://ui.adsabs.harvard.edu/abs/2018ApJ...854...90E} {854, 90}

\bibitem[\protect\citeauthoryear{Elmegreen}{Elmegreen}{1989}]{eg89-pressure}
Elmegreen B.~G.,  1989, ApJ, 338, 178

\bibitem[\protect\citeauthoryear{Engargiola, Plambeck, Rosolowsky  \&
  Blitz}{Engargiola et~al.}{2003}]{eprb03}
Engargiola G.,  Plambeck R.~L.,  Rosolowsky E.,   Blitz L.,  2003, ApJS, 149,
  343

\bibitem[\protect\citeauthoryear{{Faesi}, {Lada}  \& {Forbrich}}{{Faesi}
  et~al.}{2018}]{faesi-ngc300}
{Faesi} C.~M.,  {Lada} C.~J.,   {Forbrich} J.,  2018, \mn@doi [\apj]
  {10.3847/1538-4357/aaad60}, \href
  {https://ui.adsabs.harvard.edu/abs/2018ApJ...857...19F} {857, 19}

\bibitem[\protect\citeauthoryear{{Field}, {Blackman}  \& {Keto}}{{Field}
  et~al.}{2011}]{field11}
{Field} G.~B.,  {Blackman} E.~G.,   {Keto} E.~R.,  2011, \mn@doi [\mnras]
  {10.1111/j.1365-2966.2011.19091.x}, \href
  {https://ui.adsabs.harvard.edu/abs/2011MNRAS.416..710F} {416, 710}

\bibitem[\protect\citeauthoryear{{Foreman-Mackey}, {Hogg}, {Lang}  \&
  {Goodman}}{{Foreman-Mackey} et~al.}{2013}]{emcee}
{Foreman-Mackey} D.,  {Hogg} D.~W.,  {Lang} D.,   {Goodman} J.,  2013, \mn@doi
  [\pasp] {10.1086/670067}, \href
  {https://ui.adsabs.harvard.edu/abs/2013PASP..125..306F} {125, 306}

\bibitem[\protect\citeauthoryear{{Fujimoto}, {Tasker}, {Wakayama}  \&
  {Habe}}{{Fujimoto} et~al.}{2014}]{fujimoto14}
{Fujimoto} Y.,  {Tasker} E.~J.,  {Wakayama} M.,   {Habe} A.,  2014, \mn@doi
  [\mnras] {10.1093/mnras/stu014}, \href
  {https://ui.adsabs.harvard.edu/abs/2014MNRAS.439..936F} {439, 936}

\bibitem[\protect\citeauthoryear{{Fujimoto}, {Chevance}, {Haydon}, {Krumholz}
  \& {Kruijssen}}{{Fujimoto} et~al.}{2019}]{fujimoto19}
{Fujimoto} Y.,  {Chevance} M.,  {Haydon} D.~T.,  {Krumholz} M.~R.,
  {Kruijssen} J.~M.~D.,  2019, \mn@doi [\mnras] {10.1093/mnras/stz641}, \href
  {https://ui.adsabs.harvard.edu/abs/2019MNRAS.487.1717F} {487, 1717}

\bibitem[\protect\citeauthoryear{{Fukui} \& {Kawamura}}{{Fukui} \&
  {Kawamura}}{2010}]{fukui-kawamura}
{Fukui} Y.,  {Kawamura} A.,  2010, \mn@doi [\araa]
  {10.1146/annurev-astro-081309-130854}, \href
  {http://adsabs.harvard.edu/abs/2010ARA%26A..48..547F} {48, 547}

\bibitem[\protect\citeauthoryear{{Fukui} et~al.,}{{Fukui}
  et~al.}{1999}]{nanten-fukui}
{Fukui} Y.,  et~al., 1999, \mn@doi [\pasj] {10.1093/pasj/51.6.745}, \href
  {https://ui.adsabs.harvard.edu/abs/1999PASJ...51..745F} {51, 745}

\bibitem[\protect\citeauthoryear{Gratier et~al.,}{Gratier
  et~al.}{2010}]{gratier-m33}
Gratier P.,  et~al., 2010, A{\&}A, 522, 3

\bibitem[\protect\citeauthoryear{{Grisdale}}{{Grisdale}}{2020}]{grisdale20}
{Grisdale} K.,  2020, \mn@doi [\mnras] {10.1093/mnras/staa3524}, \href
  {https://ui.adsabs.harvard.edu/abs/2020MNRAS.tmp.3338G} {}

\bibitem[\protect\citeauthoryear{{Guszejnov}, {Grudi{\'c}}, {Offner},
  {Boylan-Kolchin}, {Faucher-Gigu{\`e}re}, {Wetzel}, {Benincasa}  \&
  {Loebman}}{{Guszejnov} et~al.}{2020}]{guszejnov20}
{Guszejnov} D.,  {Grudi{\'c}} M.~Y.,  {Offner} S. S.~R.,  {Boylan-Kolchin} M.,
  {Faucher-Gigu{\`e}re} C.-A.,  {Wetzel} A.,  {Benincasa} S.~M.,   {Loebman}
  S.,  2020, \mn@doi [\mnras] {10.1093/mnras/stz3527}, \href
  {https://ui.adsabs.harvard.edu/abs/2020MNRAS.492..488G} {492, 488}

\bibitem[\protect\citeauthoryear{{Heiles}}{{Heiles}}{1971}]{heiles71}
{Heiles} C.,  1971, \mn@doi [\araa] {10.1146/annurev.aa.09.090171.001453},
  \href {https://ui.adsabs.harvard.edu/abs/1971ARA&A...9..293H} {9, 293}

\bibitem[\protect\citeauthoryear{{Henshaw} et~al.,}{{Henshaw}
  et~al.}{2016}]{henshaw16}
{Henshaw} J.~D.,  et~al., 2016, \mn@doi [\mnras] {10.1093/mnras/stw121}, \href
  {https://ui.adsabs.harvard.edu/abs/2016MNRAS.457.2675H} {457, 2675}

\bibitem[\protect\citeauthoryear{{Herrera-Endoqui}, {D{\'\i}az-Garc{\'\i}a},
  {Laurikainen}  \& {Salo}}{{Herrera-Endoqui} et~al.}{2015}]{herrera-endoqui15}
{Herrera-Endoqui} M.,  {D{\'\i}az-Garc{\'\i}a} S.,  {Laurikainen} E.,   {Salo}
  H.,  2015, \mn@doi [\aap] {10.1051/0004-6361/201526047}, \href
  {https://ui.adsabs.harvard.edu/abs/2015A&A...582A..86H} {582, A86}

\bibitem[\protect\citeauthoryear{{Herrera} et~al.,}{{Herrera}
  et~al.}{2020}]{herrera20}
{Herrera} C.~N.,  et~al., 2020, \mn@doi [\aap] {10.1051/0004-6361/201936060},
  \href {https://ui.adsabs.harvard.edu/abs/2020A&A...634A.121H} {634, A121}

\bibitem[\protect\citeauthoryear{{Heyer} \& {Brunt}}{{Heyer} \&
  {Brunt}}{2004}]{heyer_2004}
{Heyer} M.~H.,  {Brunt} C.~M.,  2004, \mn@doi [\apjl] {10.1086/425978}, \href
  {http://adsabs.harvard.edu/abs/2004ApJ...615L..45H} {615, L45}

\bibitem[\protect\citeauthoryear{{Heyer} \& {Dame}}{{Heyer} \&
  {Dame}}{2015}]{heyerdame_2015}
{Heyer} M.,  {Dame} T.~M.,  2015, \mn@doi [\araa]
  {10.1146/annurev-astro-082214-122324}, \href
  {https://ui.adsabs.harvard.edu/abs/2015ARA&A..53..583H} {53, 583}

\bibitem[\protect\citeauthoryear{Heyer, Carpenter  \& Snell}{Heyer
  et~al.}{2001}]{hc01}
Heyer M.~H.,  Carpenter J.~M.,   Snell R.~L.,  2001, ApJ, 551, 852

\bibitem[\protect\citeauthoryear{{Heyer}, {Krawczyk}, {Duval}  \&
  {Jackson}}{{Heyer} et~al.}{2009}]{heyer_2009}
{Heyer} M.,  {Krawczyk} C.,  {Duval} J.,   {Jackson} J.~M.,  2009, \mn@doi
  [\apj] {10.1088/0004-637X/699/2/1092}, \href
  {http://adsabs.harvard.edu/abs/2009ApJ...699.1092H} {699, 1092}

\bibitem[\protect\citeauthoryear{{Hirota} et~al.,}{{Hirota}
  et~al.}{2018}]{hirota-m83}
{Hirota} A.,  et~al., 2018, \mn@doi [\pasj] {10.1093/pasj/psy071}, \href
  {https://ui.adsabs.harvard.edu/abs/2018PASJ...70...73H} {70, 73}

\bibitem[\protect\citeauthoryear{{Hopkins}}{{Hopkins}}{2012}]{hopkins12}
{Hopkins} P.~F.,  2012, \mn@doi [\mnras] {10.1111/j.1365-2966.2012.20730.x},
  \href {https://ui.adsabs.harvard.edu/abs/2012MNRAS.423.2016H} {423, 2016}

\bibitem[\protect\citeauthoryear{Hughes et~al.,}{Hughes
  et~al.}{2010}]{hughes-2010}
Hughes A.,  et~al., 2010, MNRAS, 406, 2065

\bibitem[\protect\citeauthoryear{{Hughes} et~al.,}{{Hughes}
  et~al.}{2013}]{hughes13}
{Hughes} A.,  et~al., 2013, \mn@doi [\apj] {10.1088/0004-637X/779/1/46}, \href
  {https://ui.adsabs.harvard.edu/abs/2013ApJ...779...46H} {779, 46}

\bibitem[\protect\citeauthoryear{{Imanishi}, {Nakanishi}  \&
  {Izumi}}{{Imanishi} et~al.}{2019}]{ulirg-molgas}
{Imanishi} M.,  {Nakanishi} K.,   {Izumi} T.,  2019, \mn@doi [\apjs]
  {10.3847/1538-4365/ab05b9}, \href
  {https://ui.adsabs.harvard.edu/abs/2019ApJS..241...19I} {241, 19}

\bibitem[\protect\citeauthoryear{{Imara}, {De Looze}, {Faesi}  \&
  {Cormier}}{{Imara} et~al.}{2020}]{imara2020}
{Imara} N.,  {De Looze} I.,  {Faesi} C.~M.,   {Cormier} D.,  2020, \mn@doi
  [\apj] {10.3847/1538-4357/ab8883}, \href
  {https://ui.adsabs.harvard.edu/abs/2020ApJ...895...21I} {895, 21}

\bibitem[\protect\citeauthoryear{{Izquierdo} et~al.,}{{Izquierdo}
  et~al.}{2020}]{izquierdo20}
{Izquierdo} A.~F.,  et~al., 2020, \mn@doi [\mnras] {10.1093/mnras/staa3470},
  \href {https://ui.adsabs.harvard.edu/abs/2020MNRAS.tmp.3276I} {}

\bibitem[\protect\citeauthoryear{{Jeffreson} \& {Kruijssen}}{{Jeffreson} \&
  {Kruijssen}}{2018}]{jeffreson18}
{Jeffreson} S. M.~R.,  {Kruijssen} J.~M.~D.,  2018, \mn@doi [\mnras]
  {10.1093/mnras/sty594}, \href
  {https://ui.adsabs.harvard.edu/abs/2018MNRAS.476.3688J} {476, 3688}

\bibitem[\protect\citeauthoryear{{Jeffreson}, {Kruijssen}, {Keller}, {Chevance}
   \& {Glover}}{{Jeffreson} et~al.}{2020}]{jeffreson20}
{Jeffreson} S. M.~R.,  {Kruijssen} J.~M.~D.,  {Keller} B.~W.,  {Chevance} M.,
  {Glover} S. C.~O.,  2020, \mn@doi [\mnras] {10.1093/mnras/staa2127}, \href
  {https://ui.adsabs.harvard.edu/abs/2020MNRAS.498..385J} {498, 385}

\bibitem[\protect\citeauthoryear{{Kass} \& {Raftery}}{{Kass} \&
  {Raftery}}{1995}]{bayes-factor}
{Kass} R.~E.,  {Raftery} A.~E.,  1995, Journal of the American Statistical
  Association, 90, 773

\bibitem[\protect\citeauthoryear{Keto, Ho  \& Lo}{Keto et~al.}{2005}]{keto-m82}
Keto E.,  Ho L.~C.,   Lo K.~Y.,  2005, ApJ, 635, 1062

\bibitem[\protect\citeauthoryear{{Khoperskov}, {Vasiliev}, {Ladeyschikov},
  {Sobolev}  \& {Khoperskov}}{{Khoperskov} et~al.}{2016}]{khoperskov16}
{Khoperskov} S.~A.,  {Vasiliev} E.~O.,  {Ladeyschikov} D.~A.,  {Sobolev} A.~M.,
    {Khoperskov} A.~V.,  2016, \mn@doi [\mnras] {10.1093/mnras/stv2366}, \href
  {https://ui.adsabs.harvard.edu/abs/2016MNRAS.455.1782K} {455, 1782}

\bibitem[\protect\citeauthoryear{{Koda} et~al.,}{{Koda} et~al.}{2009}]{koda09}
{Koda} J.,  et~al., 2009, \mn@doi [\apjl] {10.1088/0004-637X/700/2/L132}, \href
  {https://ui.adsabs.harvard.edu/abs/2009ApJ...700L.132K} {700, L132}

\bibitem[\protect\citeauthoryear{{Kruijssen} et~al.,}{{Kruijssen}
  et~al.}{2019a}]{kruijssen19orb}
{Kruijssen} J.~M.~D.,  et~al., 2019a, \mn@doi [\mnras] {10.1093/mnras/stz381},
  \href {https://ui.adsabs.harvard.edu/abs/2019MNRAS.484.5734K} {484, 5734}

\bibitem[\protect\citeauthoryear{{Kruijssen} et~al.,}{{Kruijssen}
  et~al.}{2019b}]{kruijssen19-ngc300}
{Kruijssen} J.~M.~D.,  et~al., 2019b, \mn@doi [\nat]
  {10.1038/s41586-019-1194-3}, \href
  {https://ui.adsabs.harvard.edu/abs/2019Natur.569..519K} {569, 519}

\bibitem[\protect\citeauthoryear{{Lang} et~al.,}{{Lang}
  et~al.}{2020}]{lang2020}
{Lang} P.,  et~al., 2020, \mn@doi [\apj] {10.3847/1538-4357/ab9953}, \href
  {https://ui.adsabs.harvard.edu/abs/2020ApJ...897..122L} {897, 122}

\bibitem[\protect\citeauthoryear{Larson}{Larson}{1981}]{larson}
Larson R.~B.,  1981, MNRAS, 194, 809

\bibitem[\protect\citeauthoryear{{Leroy} et~al.,}{{Leroy}
  et~al.}{2013}]{Leroy_2013}
{Leroy} A.~K.,  et~al., 2013, \mn@doi [\aj] {10.1088/0004-6256/146/2/19}, \href
  {http://adsabs.harvard.edu/abs/2013AJ....146...19L} {146, 19}

\bibitem[\protect\citeauthoryear{{Leroy} et~al.,}{{Leroy}
  et~al.}{2015}]{leroy15}
{Leroy} A.~K.,  et~al., 2015, \mn@doi [\apj] {10.1088/0004-637X/801/1/25},
  \href {https://ui.adsabs.harvard.edu/abs/2015ApJ...801...25L} {801, 25}

\bibitem[\protect\citeauthoryear{{Leroy} et~al.,}{{Leroy}
  et~al.}{2016}]{leroy_16}
{Leroy} A.~K.,  et~al., 2016, \mn@doi [\apj] {10.3847/0004-637X/831/1/16},
  \href {http://adsabs.harvard.edu/abs/2016ApJ...831...16L} {831, 16}

\bibitem[\protect\citeauthoryear{{Leroy} et~al.,}{{Leroy} et~al.}{2019}]{z0mgs}
{Leroy} A.~K.,  et~al., 2019, \mn@doi [\apjs] {10.3847/1538-4365/ab3925}, \href
  {https://ui.adsabs.harvard.edu/abs/2019ApJS..244...24L} {244, 24}

\bibitem[\protect\citeauthoryear{{Leroy} et~al.,}{{Leroy}
  et~al.}{2020a}]{PHANGSSurvey}
{Leroy} A.,  et~al., 2020a, PHANGS-ALMA: Arcsecond CO (2--1) Imaging of Nearby
  Star-Forming Galaxies, (ApJ in preparation)

\bibitem[\protect\citeauthoryear{{Leroy} et~al.,}{{Leroy}
  et~al.}{2020b}]{PHANGSPipeline}
{Leroy} A.,  et~al., 2020b, PHANGS-ALMA Data Processing and Pipeline, (ApJS
  submitted)

\bibitem[\protect\citeauthoryear{{Li}, {Vogelsberger}, {Marinacci}, {Sales}  \&
  {Torrey}}{{Li} et~al.}{2020}]{li20}
{Li} H.,  {Vogelsberger} M.,  {Marinacci} F.,  {Sales} L.~V.,   {Torrey} P.,
  2020, \mn@doi [\mnras] {10.1093/mnras/staa3122}, \href
  {https://ui.adsabs.harvard.edu/abs/2020MNRAS.499.5862L} {499, 5862}

\bibitem[\protect\citeauthoryear{{Maeda}, {Ohta}, {Fujimoto}  \&
  {Habe}}{{Maeda} et~al.}{2020}]{gmc-ngc1300}
{Maeda} F.,  {Ohta} K.,  {Fujimoto} Y.,   {Habe} A.,  2020, \mn@doi [\mnras]
  {10.1093/mnras/staa556}, \href
  {https://ui.adsabs.harvard.edu/abs/2020MNRAS.493.5045M} {493, 5045}

\bibitem[\protect\citeauthoryear{{Meidt} et~al.,}{{Meidt}
  et~al.}{2015}]{meidt15}
{Meidt} S.~E.,  et~al., 2015, \mn@doi [\apj] {10.1088/0004-637X/806/1/72},
  \href {https://ui.adsabs.harvard.edu/abs/2015ApJ...806...72M} {806, 72}

\bibitem[\protect\citeauthoryear{{Meidt} et~al.,}{{Meidt}
  et~al.}{2018}]{meidt18}
{Meidt} S.~E.,  et~al., 2018, \mn@doi [\apj] {10.3847/1538-4357/aaa290}, \href
  {https://ui.adsabs.harvard.edu/abs/2018ApJ...854..100M} {854, 100}

\bibitem[\protect\citeauthoryear{{Miville-Desch{\^e}nes}, {Murray}  \&
  {Lee}}{{Miville-Desch{\^e}nes} et~al.}{2017}]{miville17}
{Miville-Desch{\^e}nes} M.-A.,  {Murray} N.,   {Lee} E.~J.,  2017, \mn@doi
  [\apj] {10.3847/1538-4357/834/1/57}, \href
  {https://ui.adsabs.harvard.edu/abs/2017ApJ...834...57M} {834, 57}

\bibitem[\protect\citeauthoryear{Mizuno et~al.,}{Mizuno et~al.}{2001}]{nanten}
Mizuno N.,  et~al., 2001, \pasj, 53, 971

\bibitem[\protect\citeauthoryear{{Mo}, {van den Bosch}  \& {White}}{{Mo}
  et~al.}{2010}]{MvdBW}
{Mo} H.,  {van den Bosch} F.~C.,   {White} S.,  2010, {Galaxy Formation and
  Evolution}

\bibitem[\protect\citeauthoryear{{Mok}, {Chandar}  \& {Fall}}{{Mok}
  et~al.}{2019}]{Mok19}
{Mok} A.,  {Chandar} R.,   {Fall} S.~M.,  2019, \mn@doi [\apj]
  {10.3847/1538-4357/aaf6ea}, \href
  {https://ui.adsabs.harvard.edu/abs/2019ApJ...872...93M} {872, 93}

\bibitem[\protect\citeauthoryear{{Mok}, {Chandar}  \& {Fall}}{{Mok}
  et~al.}{2020}]{Mok20}
{Mok} A.,  {Chandar} R.,   {Fall} S.~M.,  2020, \mn@doi [\apj]
  {10.3847/1538-4357/ab7a14}, \href
  {https://ui.adsabs.harvard.edu/abs/2020ApJ...893..135M} {893, 135}

\bibitem[\protect\citeauthoryear{{Neubert} \& {Protzel}}{{Neubert} \&
  {Protzel}}{2014}]{watershed-compactness}
{Neubert} P.,  {Protzel} P.,  2014, in 2014 22nd International Conference on
  Pattern Recognition. pp 996--1001

\bibitem[\protect\citeauthoryear{Nieten, Neininger, Gu{\'e}lin, Ungerechts,
  Lucas, Berkhuijsen, Beck  \& Wielebinski}{Nieten et~al.}{2006}]{iram-m31-aa}
Nieten C.,  Neininger N.,  Gu{\'e}lin M.,  Ungerechts H.,  Lucas R.,
  Berkhuijsen E.~M.,  Beck R.,   Wielebinski R.,  2006, A{\&}A, 453, 459

\bibitem[\protect\citeauthoryear{Oka, Hasegawa, Sato, Tsuboi, Miyazaki  \&
  Sugimoto}{Oka et~al.}{2001}]{gmcs-galcen}
Oka T.,  Hasegawa T.,  Sato F.,  Tsuboi M.,  Miyazaki A.,   Sugimoto M.,  2001,
  ApJ, 562, 348

\bibitem[\protect\citeauthoryear{{Pan} \& {Kuno}}{{Pan} \&
  {Kuno}}{2017}]{pan-m100}
{Pan} H.-A.,  {Kuno} N.,  2017, \mn@doi [\apj] {10.3847/1538-4357/aa60c2},
  \href {https://ui.adsabs.harvard.edu/abs/2017ApJ...839..133P} {839, 133}

\bibitem[\protect\citeauthoryear{{Pan}, {Fujimoto}, {Tasker}, {Rosolowsky},
  {Colombo}, {Benincasa}  \& {Wadsley}}{{Pan} et~al.}{2015}]{pan15}
{Pan} H.-A.,  {Fujimoto} Y.,  {Tasker} E.~J.,  {Rosolowsky} E.,  {Colombo} D.,
  {Benincasa} S.~M.,   {Wadsley} J.,  2015, \mn@doi [\mnras]
  {10.1093/mnras/stv1843}, \href
  {https://ui.adsabs.harvard.edu/abs/2015MNRAS.453.3082P} {453, 3082}

\bibitem[\protect\citeauthoryear{{Pattle} et~al.,}{{Pattle}
  et~al.}{2015}]{pattle15}
{Pattle} K.,  et~al., 2015, \mn@doi [\mnras] {10.1093/mnras/stv376}, \href
  {https://ui.adsabs.harvard.edu/abs/2015MNRAS.450.1094P} {450, 1094}

\bibitem[\protect\citeauthoryear{{Penzias}}{{Penzias}}{1975}]{penzias75}
{Penzias} A.~A.,  1975, in {Balian} R.,  {Encrenaz} P.,   {Lequeux} J.,  eds,
  Atomic and Molecular Physics and the Interstellar Matter. pp 373--408

\bibitem[\protect\citeauthoryear{{Pettitt}, {Egusa}, {Dobbs}, {Tasker},
  {Fujimoto}  \& {Habe}}{{Pettitt} et~al.}{2018}]{pettitt18}
{Pettitt} A.~R.,  {Egusa} F.,  {Dobbs} C.~L.,  {Tasker} E.~J.,  {Fujimoto} Y.,
   {Habe} A.,  2018, \mn@doi [\mnras] {10.1093/mnras/sty2040}, \href
  {https://ui.adsabs.harvard.edu/abs/2018MNRAS.480.3356P} {480, 3356}

\bibitem[\protect\citeauthoryear{{Pettitt}, {Dobbs}, {Baba}, {Colombo},
  {Duarte-Cabral}, {Egusa}  \& {Habe}}{{Pettitt} et~al.}{2020}]{pettitt20}
{Pettitt} A.~R.,  {Dobbs} C.~L.,  {Baba} J.,  {Colombo} D.,  {Duarte-Cabral}
  A.,  {Egusa} F.,   {Habe} A.,  2020, \mn@doi [\mnras]
  {10.1093/mnras/staa2242}, \href
  {https://ui.adsabs.harvard.edu/abs/2020MNRAS.498.1159P} {498, 1159}

\bibitem[\protect\citeauthoryear{Pety et~al.,}{Pety et~al.}{2013}]{Pety_2013}
Pety J.,  et~al., 2013, \mn@doi [{ApJ}] {10.1088/0004-637x/779/1/43}, 779, 43

\bibitem[\protect\citeauthoryear{{Pineda}, {Rosolowsky}  \& {Goodman}}{{Pineda}
  et~al.}{2009}]{pineda_09}
{Pineda} J.~E.,  {Rosolowsky} E.~W.,   {Goodman} A.~A.,  2009, \mn@doi [\apjl]
  {10.1088/0004-637X/699/2/L134}, \href
  {http://adsabs.harvard.edu/abs/2009ApJ...699L.134P} {699, L134}

\bibitem[\protect\citeauthoryear{{Rebolledo}, {Wong}, {Leroy}, {Koda}  \&
  {Donovan Meyer}}{{Rebolledo} et~al.}{2012}]{rebolledo12}
{Rebolledo} D.,  {Wong} T.,  {Leroy} A.,  {Koda} J.,   {Donovan Meyer} J.,
  2012, \mn@doi [\apj] {10.1088/0004-637X/757/2/155}, \href
  {https://ui.adsabs.harvard.edu/abs/2012ApJ...757..155R} {757, 155}

\bibitem[\protect\citeauthoryear{{Rebolledo}, {Wong}, {Xue}, {Leroy}, {Koda}
  \& {Donovan Meyer}}{{Rebolledo} et~al.}{2015}]{rebolledo15}
{Rebolledo} D.,  {Wong} T.,  {Xue} R.,  {Leroy} A.,  {Koda} J.,   {Donovan
  Meyer} J.,  2015, \mn@doi [\apj] {10.1088/0004-637X/808/1/99}, \href
  {https://ui.adsabs.harvard.edu/abs/2015ApJ...808...99R} {808, 99}

\bibitem[\protect\citeauthoryear{{Reina-Campos} \& {Kruijssen}}{{Reina-Campos}
  \& {Kruijssen}}{2017}]{reinacampos17}
{Reina-Campos} M.,  {Kruijssen} J.~M.~D.,  2017, \mn@doi [\mnras]
  {10.1093/mnras/stx790}, \href
  {https://ui.adsabs.harvard.edu/abs/2017MNRAS.469.1282R} {469, 1282}

\bibitem[\protect\citeauthoryear{{Renaud} et~al.,}{{Renaud}
  et~al.}{2013}]{renaud13}
{Renaud} F.,  et~al., 2013, \mn@doi [\mnras] {10.1093/mnras/stt1698}, \href
  {https://ui.adsabs.harvard.edu/abs/2013MNRAS.436.1836R} {436, 1836}

\bibitem[\protect\citeauthoryear{{Rice}, {Goodman}, {Bergin}, {Beaumont}  \&
  {Dame}}{{Rice} et~al.}{2016}]{rice2016}
{Rice} T.~S.,  {Goodman} A.~A.,  {Bergin} E.~A.,  {Beaumont} C.,   {Dame}
  T.~M.,  2016, \mn@doi [\apj] {10.3847/0004-637X/822/1/52}, \href
  {https://ui.adsabs.harvard.edu/abs/2016ApJ...822...52R} {822, 52}

\bibitem[\protect\citeauthoryear{{Riener}, {Kainulainen}, {Henshaw}  \&
  {Beuther}}{{Riener} et~al.}{2020}]{riener20}
{Riener} M.,  {Kainulainen} J.,  {Henshaw} J.~D.,   {Beuther} H.,  2020,
  \mn@doi [\aap] {10.1051/0004-6361/202038479}, \href
  {https://ui.adsabs.harvard.edu/abs/2020A&A...640A..72R} {640, A72}

\bibitem[\protect\citeauthoryear{{Roman-Duval}, {Heyer}, {Brunt}, {Clark},
  {Klessen}  \& {Shetty}}{{Roman-Duval} et~al.}{2016}]{romanduval2016}
{Roman-Duval} J.,  {Heyer} M.,  {Brunt} C.~M.,  {Clark} P.,  {Klessen} R.,
  {Shetty} R.,  2016, \mn@doi [\apj] {10.3847/0004-637X/818/2/144}, \href
  {https://ui.adsabs.harvard.edu/abs/2016ApJ...818..144R} {818, 144}

\bibitem[\protect\citeauthoryear{Rosolowsky}{Rosolowsky}{2005}]{mspec}
Rosolowsky E.,  2005, PASP, 117, 1403

\bibitem[\protect\citeauthoryear{Rosolowsky \& Blitz}{Rosolowsky \&
  Blitz}{2005}]{m64-gmcs}
Rosolowsky E.,  Blitz L.,  2005, ApJ, 623, 826

\bibitem[\protect\citeauthoryear{{Rosolowsky} \& {Leroy}}{{Rosolowsky} \&
  {Leroy}}{2006}]{cprops}
{Rosolowsky} E.,  {Leroy} A.,  2006, \mn@doi [\pasp] {10.1086/502982}, \href
  {http://adsabs.harvard.edu/abs/2006PASP..118..590R} {118, 590}

\bibitem[\protect\citeauthoryear{Rosolowsky, Pineda, Kauffmann  \&
  Goodman}{Rosolowsky et~al.}{2008}]{dendrograms}
Rosolowsky E.~W.,  Pineda J.~E.,  Kauffmann J.,   Goodman A.~A.,  2008, ApJ,
  679, 1338

\bibitem[\protect\citeauthoryear{{Rubio}, {Lequeux}  \& {Boulanger}}{{Rubio}
  et~al.}{1993}]{rubio93}
{Rubio} M.,  {Lequeux} J.,   {Boulanger} F.,  1993, \aap, \href
  {https://ui.adsabs.harvard.edu/abs/1993A&A...271....9R} {271, 9}

\bibitem[\protect\citeauthoryear{{Salo} et~al.,}{{Salo}
  et~al.}{2015}]{salo-s4ggalfit}
{Salo} H.,  et~al., 2015, \mn@doi [\apjs] {10.1088/0067-0049/219/1/4}, \href
  {https://ui.adsabs.harvard.edu/abs/2015ApJS..219....4S} {219, 4}

\bibitem[\protect\citeauthoryear{{S{\'a}nchez} et~al.,}{{S{\'a}nchez}
  et~al.}{2014}]{sanchez14}
{S{\'a}nchez} S.~F.,  et~al., 2014, \mn@doi [\aap]
  {10.1051/0004-6361/201322343}, \href
  {https://ui.adsabs.harvard.edu/abs/2014A&A...563A..49S} {563, A49}

\bibitem[\protect\citeauthoryear{{S{\'a}nchez} et~al.,}{{S{\'a}nchez}
  et~al.}{2019}]{sanchez19}
{S{\'a}nchez} S.~F.,  et~al., 2019, \mn@doi [\mnras] {10.1093/mnras/stz019},
  \href {https://ui.adsabs.harvard.edu/abs/2019MNRAS.484.3042S} {484, 3042}

\bibitem[\protect\citeauthoryear{{Sanders}, {Scoville}  \& {Solomon}}{{Sanders}
  et~al.}{1985}]{sss85}
{Sanders} D.~B.,  {Scoville} N.~Z.,   {Solomon} P.~M.,  1985, \mn@doi [\apj]
  {10.1086/162897}, \href
  {https://ui.adsabs.harvard.edu/abs/1985ApJ...289..373S} {289, 373}

\bibitem[\protect\citeauthoryear{{Sanders}, {Clemens}, {Scoville}  \&
  {Solomon}}{{Sanders} et~al.}{1986}]{sanders86}
{Sanders} D.~B.,  {Clemens} D.~P.,  {Scoville} N.~Z.,   {Solomon} P.~M.,  1986,
  \mn@doi [\apjs] {10.1086/191086}, \href
  {https://ui.adsabs.harvard.edu/abs/1986ApJS...60....1S} {60, 1}

\bibitem[\protect\citeauthoryear{{Sandstrom} et~al.,}{{Sandstrom}
  et~al.}{2013}]{Sandstrom_2013}
{Sandstrom} K.~M.,  et~al., 2013, \mn@doi [\apj] {10.1088/0004-637X/777/1/5},
  \href {http://adsabs.harvard.edu/abs/2013ApJ...777....5S} {777, 5}

\bibitem[\protect\citeauthoryear{{Schinnerer} et~al.,}{{Schinnerer}
  et~al.}{2013}]{paws}
{Schinnerer} E.,  et~al., 2013, \mn@doi [\apj] {10.1088/0004-637X/779/1/42},
  \href {https://ui.adsabs.harvard.edu/abs/2013ApJ...779...42S} {779, 42}

\bibitem[\protect\citeauthoryear{{Schruba} et~al.,}{{Schruba}
  et~al.}{2017}]{schruba17}
{Schruba} A.,  et~al., 2017, \mn@doi [\apj] {10.3847/1538-4357/835/2/278},
  \href {https://ui.adsabs.harvard.edu/abs/2017ApJ...835..278S} {835, 278}

\bibitem[\protect\citeauthoryear{{Schruba}, {Kruijssen}  \& {Leroy}}{{Schruba}
  et~al.}{2019}]{schruba19}
{Schruba} A.,  {Kruijssen} J.~M.~D.,   {Leroy} A.~K.,  2019, \mn@doi [\apj]
  {10.3847/1538-4357/ab3a43}, \href
  {https://ui.adsabs.harvard.edu/abs/2019ApJ...883....2S} {883, 2}

\bibitem[\protect\citeauthoryear{{Schwarz}}{{Schwarz}}{1978}]{BIC}
{Schwarz} G.,  1978, Annals of Statistics, \href
  {https://ui.adsabs.harvard.edu/abs/1978AnSta...6..461S} {6, 461}

\bibitem[\protect\citeauthoryear{{Scoville} \& {Solomon}}{{Scoville} \&
  {Solomon}}{1975}]{ss75}
{Scoville} N.~Z.,  {Solomon} P.~M.,  1975, \mn@doi [\apjl] {10.1086/181859},
  \href {https://ui.adsabs.harvard.edu/abs/1975ApJ...199L.105S} {199, L105}

\bibitem[\protect\citeauthoryear{{Scoville}, {Yun}, {Clemens}, {Sand ers}  \&
  {Waller}}{{Scoville} et~al.}{1987}]{scoville87}
{Scoville} N.~Z.,  {Yun} M.~S.,  {Clemens} D.~P.,  {Sand ers} D.~B.,   {Waller}
  W.~H.,  1987, \mn@doi [\apjs] {10.1086/191185}, \href
  {https://ui.adsabs.harvard.edu/abs/1987ApJS...63..821S} {63, 821}

\bibitem[\protect\citeauthoryear{Seabold \& Perktold}{Seabold \&
  Perktold}{2010}]{statsmodels}
Seabold S.,  Perktold J.,  2010, in 9th Python in Science Conference.

\bibitem[\protect\citeauthoryear{{Sheth} et~al.,}{{Sheth} et~al.}{2010}]{s4g}
{Sheth} K.,  et~al., 2010, \mn@doi [\pasp] {10.1086/657638}, \href
  {https://ui.adsabs.harvard.edu/abs/2010PASP..122.1397S} {122, 1397}

\bibitem[\protect\citeauthoryear{{Shetty}, {Beaumont}, {Burton}, {Kelly}  \&
  {Klessen}}{{Shetty} et~al.}{2012}]{shetty12}
{Shetty} R.,  {Beaumont} C.~N.,  {Burton} M.~G.,  {Kelly} B.~C.,   {Klessen}
  R.~S.,  2012, \mn@doi [\mnras] {10.1111/j.1365-2966.2012.21588.x}, \href
  {https://ui.adsabs.harvard.edu/abs/2012MNRAS.425..720S} {425, 720}

\bibitem[\protect\citeauthoryear{{Smith}, {Glover}, {Clark}, {Klessen}  \&
  {Springel}}{{Smith} et~al.}{2014}]{smith14}
{Smith} R.~J.,  {Glover} S. C.~O.,  {Clark} P.~C.,  {Klessen} R.~S.,
  {Springel} V.,  2014, \mn@doi [\mnras] {10.1093/mnras/stu616}, \href
  {https://ui.adsabs.harvard.edu/abs/2014MNRAS.441.1628S} {441, 1628}

\bibitem[\protect\citeauthoryear{Solomon \& de Zafra}{Solomon \&
  de~Zafra}{1975}]{sol75}
Solomon P.~M.,  de Zafra R.,  1975, ApJL, 199, L79

\bibitem[\protect\citeauthoryear{Solomon, Sanders  \& Scoville}{Solomon
  et~al.}{1979}]{sss79}
Solomon P.~M.,  Sanders D.~B.,   Scoville N.~Z.,  1979, ApJL, 232, L89

\bibitem[\protect\citeauthoryear{Solomon, Rivolo, Barrett  \& Yahil}{Solomon
  et~al.}{1987}]{srby87}
Solomon P.~M.,  Rivolo A.~R.,  Barrett J.,   Yahil A.,  1987, ApJ, 319, 730

\bibitem[\protect\citeauthoryear{{Solomon}, {Downes}, {Radford}  \&
  {Barrett}}{{Solomon} et~al.}{1997}]{Solomon_1997}
{Solomon} P.~M.,  {Downes} D.,  {Radford} S.~J.~E.,   {Barrett} J.~W.,  1997,
  \mn@doi [\apj] {10.1086/303765}, \href
  {http://adsabs.harvard.edu/abs/1997ApJ...478..144S} {478, 144}

\bibitem[\protect\citeauthoryear{{Sormani} et~al.,}{{Sormani}
  et~al.}{2019}]{sormani19}
{Sormani} M.~C.,  et~al., 2019, \mn@doi [\mnras] {10.1093/mnras/stz2054}, \href
  {https://ui.adsabs.harvard.edu/abs/2019MNRAS.488.4663S} {488, 4663}

\bibitem[\protect\citeauthoryear{{Stanimirovi{\'c}} \&
  {Lazarian}}{{Stanimirovi{\'c}} \& {Lazarian}}{2001}]{snez_01}
{Stanimirovi{\'c}} S.,  {Lazarian} A.,  2001, \mn@doi [\apjl] {10.1086/319837},
  \href {http://adsabs.harvard.edu/abs/2001ApJ...551L..53S} {551, L53}

\bibitem[\protect\citeauthoryear{Stanimirovi{\'c}, Staveley-Smith, Dickey,
  Sault  \& Snowden}{Stanimirovi{\'c} et~al.}{1999}]{lincom}
Stanimirovi{\'c} S.,  Staveley-Smith L.,  Dickey J.~M.,  Sault R.~J.,   Snowden
  S.~L.,  1999, MNRAS, 302, 417

\bibitem[\protect\citeauthoryear{{Sun} et~al.,}{{Sun} et~al.}{2018}]{sun_18}
{Sun} J.,  et~al., 2018, \mn@doi [\apj] {10.3847/1538-4357/aac326}, \href
  {https://ui.adsabs.harvard.edu/abs/2018ApJ...860..172S} {860, 172}

\bibitem[\protect\citeauthoryear{{Sun} et~al.,}{{Sun} et~al.}{2020a}]{sun_20}
{Sun} J.,  et~al., 2020a, \mn@doi [\apj] {10.3847/1538-4357/ab781c}, \href
  {https://ui.adsabs.harvard.edu/abs/2020ApJ...892..148S} {892, 148}

\bibitem[\protect\citeauthoryear{{Sun} et~al.,}{{Sun} et~al.}{2020b}]{sun_20b}
{Sun} J.,  et~al., 2020b, \mn@doi [\apjl] {10.3847/2041-8213/abb3be}, \href
  {https://ui.adsabs.harvard.edu/abs/2020ApJ...901L...8S} {901, L8}

\bibitem[\protect\citeauthoryear{{Tasker} \& {Tan}}{{Tasker} \&
  {Tan}}{2009}]{tasker09}
{Tasker} E.~J.,  {Tan} J.~C.,  2009, \mn@doi [\apj]
  {10.1088/0004-637X/700/1/358}, \href
  {https://ui.adsabs.harvard.edu/abs/2009ApJ...700..358T} {700, 358}

\bibitem[\protect\citeauthoryear{{Tress}, {Sormani}, {Smith}, {Glover},
  {Klessen}, {Mac Low}, {Clark}  \& {Duarte-Cabral}}{{Tress}
  et~al.}{2020}]{tress20b}
{Tress} R.~G.,  {Sormani} M.~C.,  {Smith} R.~J.,  {Glover} S. C.~O.,  {Klessen}
  R.~S.,  {Mac Low} M.-M.,  {Clark} P.,   {Duarte-Cabral} A.,  2020, arXiv
  e-prints, \href {https://ui.adsabs.harvard.edu/abs/2020arXiv201205919T} {p.
  arXiv:2012.05919}

\bibitem[\protect\citeauthoryear{{Utomo}, {Blitz}, {Davis}, {Rosolowsky},
  {Bureau}, {Cappellari}  \& {Sarzi}}{{Utomo} et~al.}{2015}]{utomo15}
{Utomo} D.,  {Blitz} L.,  {Davis} T.,  {Rosolowsky} E.,  {Bureau} M.,
  {Cappellari} M.,   {Sarzi} M.,  2015, \mn@doi [\apj]
  {10.1088/0004-637X/803/1/16}, \href
  {https://ui.adsabs.harvard.edu/abs/2015ApJ...803...16U} {803, 16}

\bibitem[\protect\citeauthoryear{{Vogel}, {Wright}, {Plambeck}  \&
  {Welch}}{{Vogel} et~al.}{1984}]{feather-vogel}
{Vogel} S.~N.,  {Wright} M.~C.~H.,  {Plambeck} R.~L.,   {Welch} W.~J.,  1984,
  \mn@doi [\apj] {10.1086/162351}, \href
  {https://ui.adsabs.harvard.edu/abs/1984ApJ...283..655V} {283, 655}

\bibitem[\protect\citeauthoryear{Vogel, Boulanger  \& Ball}{Vogel
  et~al.}{1987}]{vbb87}
Vogel S.~N.,  Boulanger F.,   Ball R.,  1987, ApJL, 321, L145

\bibitem[\protect\citeauthoryear{{Wei}, {Keto}  \& {Ho}}{{Wei}
  et~al.}{2012}]{wei12}
{Wei} L.~H.,  {Keto} E.,   {Ho} L.~C.,  2012, \mn@doi [\apj]
  {10.1088/0004-637X/750/2/136}, \href
  {https://ui.adsabs.harvard.edu/abs/2012ApJ...750..136W} {750, 136}

\bibitem[\protect\citeauthoryear{{Whitmore} et~al.,}{{Whitmore}
  et~al.}{2014}]{alma-antennae}
{Whitmore} B.~C.,  et~al., 2014, \mn@doi [\apj] {10.1088/0004-637X/795/2/156},
  \href {https://ui.adsabs.harvard.edu/abs/2014ApJ...795..156W} {795, 156}

\bibitem[\protect\citeauthoryear{Williams, de Geus  \& Blitz}{Williams
  et~al.}{1994}]{clumpfind}
Williams J.~P.,  de Geus E.~J.,   Blitz L.,  1994, ApJ, 428, 693

\bibitem[\protect\citeauthoryear{Wilson}{Wilson}{1994}]{wilson_6822}
Wilson C.~D.,  1994, ApJL, 434, L11

\bibitem[\protect\citeauthoryear{Wilson \& Reid}{Wilson \& Reid}{1991}]{wr91}
Wilson C.~D.,  Reid I.~N.,  1991, ApJL, 366, L11

\bibitem[\protect\citeauthoryear{Wilson \& Scoville}{Wilson \&
  Scoville}{1990}]{ws90}
Wilson C.~D.,  Scoville N.,  1990, ApJ, 363, 435

\bibitem[\protect\citeauthoryear{Wilson, Scoville, Madden  \&
  Charmandaris}{Wilson et~al.}{2003}]{gma-wilson}
Wilson C.~D.,  Scoville N.,  Madden S.~C.,   Charmandaris V.,  2003, ApJ, 599,
  1049

\bibitem[\protect\citeauthoryear{Wong et~al.,}{Wong et~al.}{2011}]{magma}
Wong T.,  et~al., 2011, ApJS, 197, 16

\bibitem[\protect\citeauthoryear{{Yim}, {Wong}, {Xue}, {Rand}, {Rosolowsky},
  {van der Hulst}, {Benjamin}  \& {Murphy}}{{Yim} et~al.}{2014}]{yim14}
{Yim} K.,  {Wong} T.,  {Xue} R.,  {Rand} R.~J.,  {Rosolowsky} E.,  {van der
  Hulst} J.~M.,  {Benjamin} R.,   {Murphy} E.~J.,  2014, \mn@doi [\aj]
  {10.1088/0004-6256/148/6/127}, \href
  {https://ui.adsabs.harvard.edu/abs/2014AJ....148..127Y} {148, 127}

\bibitem[\protect\citeauthoryear{{de Vaucouleurs}, {de Vaucouleurs}, {Corwin},
  {Buta}, {Paturel}  \& {Fouque}}{{de Vaucouleurs} et~al.}{1991}]{rc3}
{de Vaucouleurs} G.,  {de Vaucouleurs} A.,  {Corwin} Herold~G. J.,  {Buta}
  R.~J.,  {Paturel} G.,   {Fouque} P.,  1991, {Third Reference Catalogue of
  Bright Galaxies}.
Springer

\bibitem[\protect\citeauthoryear{{den Brok} et~al.,}{{den Brok}
  et~al.}{2020}]{denBrok20}
{den Brok} J.,  et~al., 2020, (MNRAS submitted)

\bibitem[\protect\citeauthoryear{van~der Walt et~al.,}{van~der Walt
  et~al.}{2014}]{scikit-image}
van~der Walt S.,  et~al., 2014, \mn@doi [PeerJ] {10.7717/peerj.453}, 2, e453

\makeatother
\end{thebibliography}

\appendix

\section{Native Resolution Catalogue Analysis}
\label{sec:fullres}

\begin{figure*}
\includegraphics[width=\textwidth]{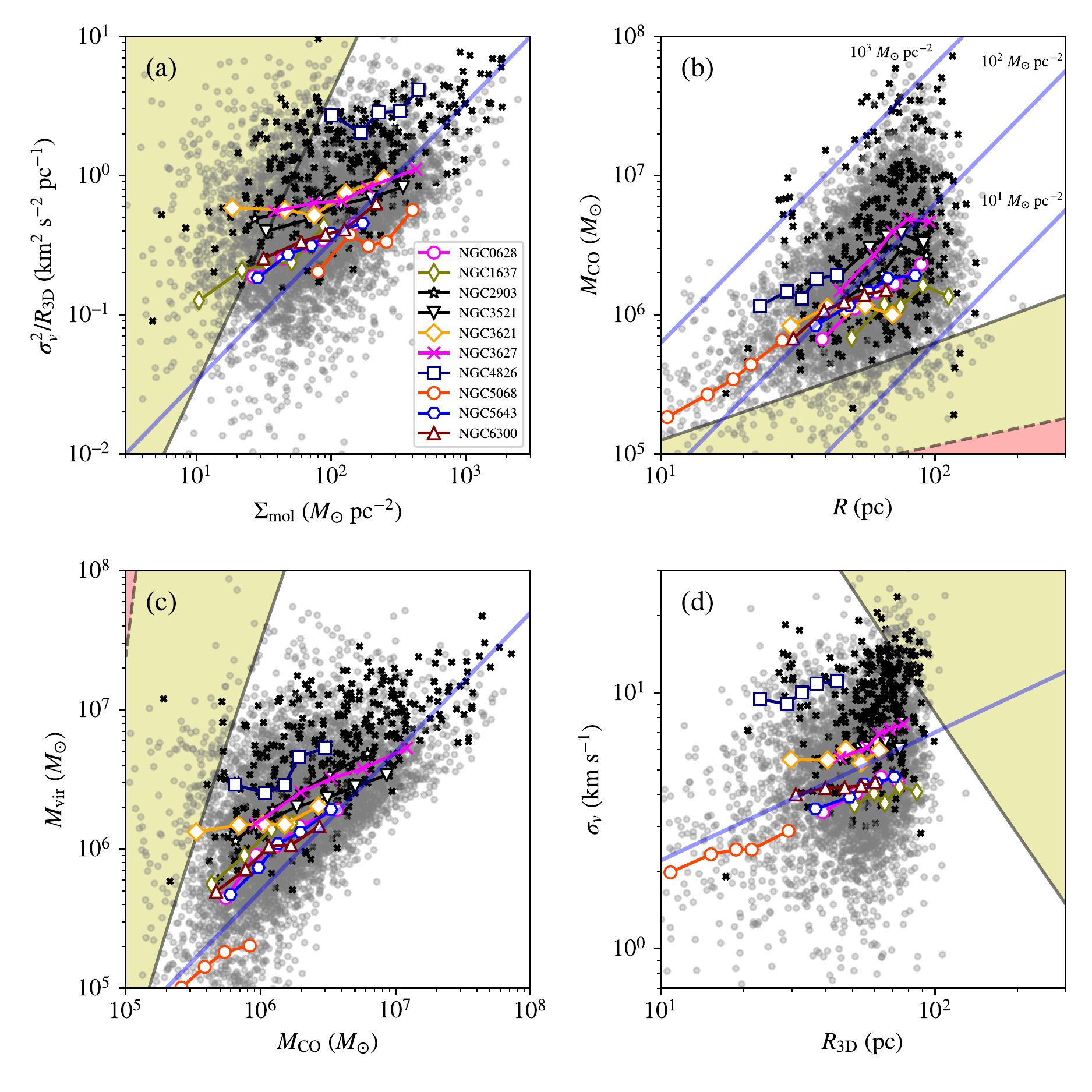}
\vspace*{-25px}
\caption{\label{fig:scaling} The scaling relationships between molecular cloud properties across the sample. In each panel, the individual clouds are presented as grey, translucent points and the curves with points show the relationship for the ten galaxies giving the median values in bins set by the quantiles of the $x$-axis. The shaded regions indicate where the false source insertion tests indicate $<50\%$ (yellow) and $<10\%$ (red) recovery of clouds in the test sample.  Red crosses indicate clouds found in the central kpc of their respective galaxies.  The solid blue line indicates a reference relationship. Panel (a) shows the relationship between surface density ($\Sigma_{\mathrm{mol}}$) and the turbulent line width on 1~pc scales ($\sigma_0$, plotted as $\sigma_0^2 = \sigma_v^2 / R$).  The blue line shows the locus of virialization. Panel (b) shows the mass--radius relationship for the recovered clouds.  The blue lines show lines of constant surface density.  Panel (c) shows the relationship between a virial mass estimate and the luminous mass estimate.  The blue line shows the locus of equality.  Panel (d) shows the size--line width relationship for GMCs with the blue line showing the scaling identified in the Milky Way (S87).}
\end{figure*}

In Figure~\ref{fig:scaling}, we present the scaling relationships for the GMC catalogs of the ten galaxies studied in this work without undergoing the homogenization process. In this analysis, there are 6567 total clouds but only 5697 clouds are included in the figures with peak signal-to-noise ratio $>6$. The completeness regions are established as before, namely as an average over the parameters ($c_i$) of the completeness fit across all galaxies.  There is a significantly larger range in completeness parameters owing to the different properties of the data (Table~\ref{tab:data}).

{\bf The \boldmath$\Sigma_\mathrm{mol}{-}\sigma_0$ relationship}  -- As in the homogenized population analysis, the population of molecular clouds appears consistent with the locus of virialization.  The binned medians for each of the galaxies show that there appear to be variations between the different systems.  Most follow a positive scaling relationship between the systems with the exception of NGC~4826, an abnormal system where most of the molecular mass is concentrated in a high surface density, nuclear disc with $R_\mathrm{gal}\approx 1.5~\mathrm{kpc}$.  While clearly distinct in the common-resolution analysis, the system appears even more abnormal here where the resolution of the data sets is also significantly different.

{\bf Mass--radius relationship} -- We show the native-resolution mass--radius relationship in Figure~\ref{fig:scaling}(b).  Compared to the homogenized data, GMCs show a wider range of catalogued properties but still with surface densities roughly consistent with $10^2~M_\odot~\mathrm{pc}^{-2}$.  The censoring at the low end of the mass and radius axes is more apparent in the native-resolution study.

{\bf Dynamical versus Luminous Mass} -- Figure~\ref{fig:scaling}(c) presents the native-resolution comparison of the dynamical ($M_{\mathrm{vir}}$) and luminous ($M_{\mathrm{CO}}$) masses of the GMCs.  Here, we see significantly wider ranges of mass scales and implied virial parameters for the clouds than we did in the homogenized data.  This illustrates the importance of the homogenization process, as a simple reading of these results would argue that the mass scales and dynamical state of these galaxies would be significantly different.  However, we see in the homogenized data that the galaxy populations all occupy a common section of parameter space.

{\bf Size--line width relationship} -- Figure~\ref{fig:scaling}(d) shows the size--line width relationship for the clouds studied at the native resolution.  The apparent differences between galaxies are clearly illustrated here where each galaxy occupies a distinct locus in parameter space, but all galaxy populations show good consistency with the slope of the relationship seen in the Milky Way. This shows that the implied index of $\beta=0.5$ that we adopt in the work is well justified but that catalogues created with heterogeneous noise levels and physical resolutions will necessarily lead to distinct populations.

\label{lastpage}

\end{document}